\documentclass[a4paper,11pt]{article}
\usepackage{imakeidx}
\makeindex[columns=2, title=Alphabetical Index, intoc]
\usepackage{jheppub} % for details on the use of the package, please see the JINST-author-manual
\usepackage{lineno}
\usepackage{amsmath}
\usepackage{dsfont} %mathds font
\usepackage{nicefrac}
\usepackage[dvipsnames]{xcolor}
\usepackage{braket}
\usepackage{enumitem}
\usepackage{mathtools}
\usepackage{cleveref}
\usepackage{bbold}
\usepackage{slashed}
\usepackage{multirow}
\usepackage{caption}
\usepackage{makecell}
\usepackage{amssymb}
\usepackage{verbatim}
\usepackage{physics}
\usepackage{booktabs}

\usepackage{tikz-feynman}
\tikzfeynmanset{warn luatex=false}

\usepackage{rotating}
\usepackage{cancel}
\usepackage{pifont}

\usepackage{longtable} 
\usepackage{textcomp}
\usepackage{textgreek}

\usepackage{mdframed}           
\mdfdefinestyle{theoremstyle}{
	linecolor=black,
	linewidth=1pt, 
	frametitlerule=true, 
	frametitlebackgroundcolor=gray!20, 
	innertopmargin=\topskip,
}
\mdtheorem[style=theoremstyle]{nb}{NB}

\usepackage{dashrule}

\makeatletter
\newcommand\ps@indexpagestyle{
  \renewcommand\@oddfoot{\hfill-- \thepage\ --\hfill}
  \renewcommand\@oddhead{}
}
\makeatother
\indexsetup{firstpagestyle=indexpagestyle}

\newcommand{\finalresult}{\texttt{FinalResult.m}}
\newcommand{\usefulfunctions}{\texttt{UsefulFunctions.m}}

\newcommand{\be}{\begin{equation}}
\newcommand{\ee}{\end{equation}}

\usepackage{dowith}

\newcommand{\UpperCase}[1]{
  \expandafter\newcommand\csname bb#1\endcsname{{\mathbb{#1}}}
  \expandafter\newcommand\csname cal#1\endcsname{{\mathcal{#1}}}   
  \expandafter\newcommand\csname rm#1\endcsname{{\mathrm{#1}}}
  \expandafter\newcommand\csname bf#1\endcsname{{\mathbf{#1}}}
  \expandafter\newcommand\csname bold#1\endcsname{{\boldsymbol{#1}}}
  \expandafter\newcommand\csname hat#1\endcsname{\hat{#1}}
  \expandafter\newcommand\csname tilde#1\endcsname{\widetilde{#1}}
  \expandafter\newcommand\csname bar#1\endcsname{\overline{#1}}
  \expandafter\newcommand\csname frak#1\endcsname{\mathfrak{#1}}
  }
\DoWith\UpperCase ABCDEFGHILJKMNOPQRSTUVWXYZ\StopDoing

\newcommand{\LowerCase}[1]{
  \expandafter\newcommand\csname rm#1\endcsname{{\mathrm{#1}}} 
  \expandafter\newcommand\csname bf#1\endcsname{{\mathbf{#1}}} 
  \expandafter\newcommand\csname bold#1\endcsname{{\boldsymbol{#1}}}
  \expandafter\newcommand\csname hat#1\endcsname{\hat{#1}}
  \expandafter\newcommand\csname tilde#1\endcsname{\tilde{#1}}
  \expandafter\newcommand\csname bar#1\endcsname{\bar{#1}}
  \expandafter\newcommand\csname frak#1\endcsname{\mathfrak{#1}}
  }
\DoWith\LowerCase abcdefghijklmnopqrstuvwxyz\StopDoing

\newcommand{\genBorn}{{\calA_n}}

\newcommand{\HP}{\calH}
\newcommand{\HPf}{{\calH_{\rmf}}}
\newcommand{\HPfg}{{\calH_{\rmf,g}}}
\newcommand{\HPfq}{{\calH_{\rmf,q}}}

\newcommand{\inotj}{(ij)}

\newcommand{\knotl}{(kl)}

\newcommand{\setg}[1]{\{g\}_{#1}}
\newcommand{\setq}[1]{\{q\}_{#1}}
\newcommand{\setqb}[1]{\{\qb\}_{#1}}
\newcommand{\setqp}[1]{\{\qp\}_{#1}}
\newcommand{\setqbp}[1]{\{\qbp\}_{#1}}

\newcommand{\Ng}{{N_g}}
\newcommand{\Nq}{{N_q}}
\newcommand{\Nqb}{{N_\qb}}

\newcommand{\DS}{{\rm DS}}
\newcommand{\noDS}{\cancel{{\rm DS}}}

\newcommand{\gqBorn}{{\calA_0}}

\newcommand{\proj}{\boldsymbol{\calP}_{\!\! \gqBorn}}
\newcommand{\FgqprocNLO}{{\calA_1}} 
\newcommand{\SgqprocNLO}{{\calA_2}}
\newcommand{\TgqprocNLO}{{\calA_3}} 
 
\newcommand{\FgqprocNNLO}{{\calA_4}} 
\newcommand{\SgqprocNNLO}{{\calA_5}}
\newcommand{\TgqprocNNLO}{{\calA_6}}

\newcommand{\barotimes}{\bar{\otimes}}
\newcommand{\colorprod}{\cdot}
\newcommand{\conv}{\otimes}

\newcommand{\dz}{\rmd z}

\newcommand{\eq}{Eq.~}
\newcommand{\eqdef}{=}

\newcommand{\oS}{\barS}
\newcommand{\oC}{\barC}

\newcommand{\qb}{{\bar{q}}}
\newcommand{\qp}{{q'}}
\newcommand{\qbp}{{\qb'}}

\newcommand{\zb}{{\barz}}

\newcommand{\fin}{{\mathrm{fin}}}

\newcommand{\gen}{{\mathrm{gen}}}

\newcommand{\LV}{{\rm LV}}

\newcommand{\NStilde}{\widetilde{\rm ns}}
\newcommand{\dc}{{\rm dc}}

\newcommand{\rest}{{\rm rest}}
\newcommand{\rhs}{right-hand side}
\newcommand{\R}{{\mathrm{R}}}
\newcommand{\RR}{{\mathrm{RR}}}
\newcommand{\RV}{{\mathrm{RV}}}

\newcommand{\Sec}{Sec.~}

\newcommand{\sing}{{\mathrm{sing}}}

\newcommand{\tc}{{\rm tc}}

\newcommand{\V}{\mathrm{V}}
\newcommand{\VV}{\mathrm{VV}}

\newcommand{\dsigmahat}{\rmd \hat{\sigma}}

\newcommand{\PAP}{\hat P^{(0)}}
\newcommand{\PAPone}{\hat P^{(1)}}
\newcommand{\CalPgen}{\calP^{\gen}}

\newcommand{\calPW}{\calP^{\calW}}

\newcommand{\PNLO}{\calP^{\mathrm{NLO}}}
\newcommand{\PNNLO}{\calP^{\mathrm{NNLO}}}
\newcommand{\PAPxPAP}[2]{\big[\PAP_{#1} \conv \PAP_{#2}\big]}
\newcommand{\PAPoxPAP}[2]{\big[\PAP_{#1} \, \barotimes \, \PAP_{#2}\big]}
\newcommand{\PAPoxPgen}[2]{\big[\PAP_{#1} \, \barotimes \, \CalPgen_{#2}\big]}
\newcommand{\PgenoxPgen}[2]{\big[\CalPgen_{#1} \, \bar{\otimes} \, \CalPgen_{#2}\big]}

\newcommand{\GFSR}[3]{G_{#1}\left|{\substack{ {#2} \\ {#3} }}\right.}

\newcommand{\SigmaDU}{\Sigma_{\rm DU}}
\newcommand{\SigmaFR}{\Sigma_{\rm FR}}
\newcommand{\SigmaSU}{\Sigma_{\rm SU}}

\newcommand{\db}{{\rm db}}
\newcommand{\inFsb}{{{\rm sb},\inF}}
\newcommand{\inSsb}{{{\rm sb},\inS}}
\newcommand{\el}{{\rm el}}

\newcommand{\du}{{\rm DU}}
\newcommand{\fr}{{\rm FR}}
\newcommand{\su}{{\rm SU}}

\newcommand{\Wacfin}[1]{\calW_{#1}^{#1 \parallel \Sp, \fin}}

\newcommand{\Wbdfin}[1]{\calW_{#1}^{\Fp \parallel \Sp, \fin}}
\newcommand{\Wr}[1]{\calW_\rmr^{(#1)}}

\newcommand{\lint}{\big\langle}
\newcommand{\rint}{\big\rangle}
\newcommand{\llint}{\Big\langle}
\newcommand{\rrint}{\Big\rangle}
\newcommand{\Lint}{\bigg\langle}
\newcommand{\Rint}{\bigg\rangle}

\DeclarePairedDelimiterX\mybraket[2]{\langle}{\rangle}{#1 \delimsize\vert #2}       % BRAKET  <M|M>  
\DeclarePairedDelimiterX\mybraketOp[3]{\langle}{\rangle}%
{#1\,\delimsize\vert\,\mathopen{}#2\,\delimsize\vert\,\mathopen{}#3} % BRAKET with OPERATOR   <M|O|M>

\newcommand{\Coll}[1]{C_{#1}}
\newcommand{\DoubSoft}[1]{S_{#1}}
\newcommand{\iden}{\mathbb{1}}

\newcommand{\ICatbar}{\overline{I}_1}

\newcommand{\IVirt}{I_{\rm V}}

\newcommand{\ISoft}{I_{\rm S}}

\newcommand{\ISofttilde}{\widetilde{I}_{\rm S}}
\newcommand{\IColl}{I_{\rm C}}

\newcommand{\ICollFour}{I_{\rm C}^{(4)}}
\newcommand{\ICollFournf}{I_{\rmC, \nf}^{(4)}}

\newcommand{\IColltilde}{\tildeI_{\rm C}}
\newcommand{\IColltildenf}{\tildeI_{\rm C, \nf}}

\newcommand{\GammaFour}{\Gamma^{(4)}}
\newcommand{\GammaLoop}{\Gamma^{\rm 1L}}
\newcommand{\ITot}{I_{\rm T}}

\newcommand{\Iccfin}{I_{\rm cc}^{\fin}}

\newcommand{\Itrifin}{I_{\mathrm{tri}}^\fin}
\newcommand{\Iuncfin}{I_{\rm unc}^\fin}

\newcommand{\LO}{\mathrm{LO}}

\newcommand{\NNLO}{\mathrm {NNLO}}
\newcommand{\ONLO}{\mathcal{O}_\text{NLO}}
\newcommand{\OColl}{\mathcal{O}_\text{NLO}} % I am calling O_C and O_NLO like the same

\newcommand{\CalP}{\mathcal{P}}

\newcommand{\PqqGen}{\CalPgen}
\newcommand{\convPgenPgen}[2]{\big[\PqqGen_{#1} \, \bar{\otimes} \, \PqqGen_{#2}\big]}

\newcommand{\CalPGenFour}{\mathcal{P}^{(4),\text{gen}}}

\newcommand{\CalPoneLgen}{\CalP^{\mathrm{1L,gen}}}

\newcommand{\Emax}{E_{\rm max}}
\newcommand{\EulerGamma}{\gamma_{\mathrm{E}}}
\newcommand{\Sep}{S_\ep}
\newcommand{\eps}{\epsilon}
\newcommand{\ep}{\epsilon}
\newcommand{\musq}{\mu^2}

\newcommand{\colsing}{X}

\newcommand{\fl}[1]{f_{#1}}

\newcommand{\TR}{T_\rmR}
\newcommand{\nf}{n_\rmf}

\newcommand{\Ca}{C_\rmA}
\newcommand{\Cf}{C_\rmF}
\newcommand{\Nc}{N_\rmc}

\newcommand{\partFuncNLOfp}[1]{\omega^{\Fp #1 }}

\newcommand{\partFuncNNLO}[2]{\omega^{\Fp #1, \Sp #2}}

\newcommand{\partFuncACfp}[1]{\omega^{\Fp #1, \Sp #1}_{#1 \parallel \Fp}}
\newcommand{\partFuncACsp}[1]{\omega^{\Fp #1, \Sp #1}_{#1 \parallel \Sp}}
\newcommand{\partFuncBD}[1]{\omega^{\Fp #1, \Sp #1}_{\Fp \parallel \Sp}}

\newcommand{\gsb}{g_{\rm s,b}}
\newcommand{\gs}{g_\rms}
\newcommand{\as}{\alpha_\rms}
\newcommand{\asbr}{[\alpha_\rms]}
\newcommand{\amu}{\frac{\alpha_\rms(\mu)}{2\pi}}

\newcommand{\Li}{\mathrm{Li}}
\newcommand{\Ltildei}{\widetilde{L}_i}
\newcommand{\Lmax}{L_{\rm max}}
\newcommand{\hypF}{{_2F_1}}
\newcommand{\THmn}{\Theta_{\Fp \Sp}}       % Heaviside step function Theta_mn
\newcommand{\THnm}{\Theta_{\Sp \Fp}}       % Heaviside step function Theta_nm
\newcommand{\T}{\boldsymbol{T}}

\newcommand{\Fp}{\mathfrak{m}}
\newcommand{\Sp}{\mathfrak{n}}

\newcommand{\xa}{\mathfrak{m}}
\newcommand{\yb}{\mathfrak{n}}

\newcommand{\hc}{h_\rmc}

\newcommand{\FLM}{F_{\mathrm{LM}}}
\newcommand{\FLMlo}[2]{\FLM^{#1}[#2]}
\newcommand{\FLMun}[3]{\FLM^{#1}[#2|#3]}
\newcommand{\FLMmunu}{F_{\mathrm{LM},\mu\nu}}
\newcommand{\FLMlomunu}[2]{\FLMmunu^{#1}[#2]}
\newcommand{\FLMunmunu}[2]{\FLMmunu^{#1}[#2]}

\newcommand{\FR}{F_\R}
\newcommand{\FRlo}[2]{\FR^{#1}[#2]}
\newcommand{\FLVfin}{F_{\LV,\fin}}
\newcommand{\FLVfinsq}{F_{\LV^2,\fin}}
\newcommand{\FVV}{F_\VV}
\newcommand{\FVVfin}{F_{\VV,\fin}}
\newcommand{\FRV}{F_\RV}
\newcommand{\FRVlo}[2]{\FRV^{#1}[#2]}

\newcommand{\FRVfin}{F_{\RV,\fin}}
\newcommand{\FRVfinlo}[2]{\FRVfin^{#1}{[#2]}}
\newcommand{\inF}{a} % labes the first parton in the initial state
\newcommand{\inS}{b} % labes the second parton in the initial state

\newcommand{\cTilde}{\tildec_\ep}

\newcommand{\Nscbd}{N_{\rm sc}^{(b,d)}}

\def\SLAC{SLAC National Accelerator Laboratory, Stanford University, Stanford, CA 94039, USA}

\def\KITA{Institute for Theoretical Particle Physics, KIT, Wolfgang-Gaede-Straße 1, 76131, Karlsruhe, Germany}

\def\TIF{Tif Lab, Dipartimento di Fisica, Università di Milano and
INFN, Sezione di Milano, Via Celoria 16, I-20133 Milano, Italy}
\def\MP{Max-Planck-Institut für Physik, Boltzmannstrasse 8, 85748 Garching, Germany}

\preprint{
\begin{flushright}
TIF-UNIMI-2025-8,
TTP25-006,
P3H-25-017,
MPP-2025-40,
SLAC-PUB-250317
\end{flushright}
}

\title{Towards a general subtraction formula for NNLO QCD corrections to processes at hadron colliders: final states with  quarks and gluons}

\author[a]{Federica Devoto,}
\author[b]{Kirill Melnikov,}
\author[c]{Raoul R{\"o}ntsch,}
\author[d]{Chiara Signorile-Signorile,}
\author[b,c]{Davide Maria Tagliabue,}
\author[b]{Matteo Tresoldi}

\emailAdd{federica@slac.stanford.edu}
\emailAdd{kirill.melnikov@kit.edu}
\emailAdd{raoul.rontsch@unimi.it}
\emailAdd{signoril@mpp.mpg.de}
\emailAdd{davide.tagliabue@kit.edu}
\emailAdd{matteo.tresoldi@partner.kit.edu}
\affiliation[a]{\SLAC}
\affiliation[b]{\KITA}
\affiliation[c]{\TIF}
\affiliation[d]{\MP}

\abstract{ 
We describe the calculation of integrated subtraction terms in the nested soft-collinear subtraction scheme for hadron collider processes with quarks and gluons, thereby extending the results presented in Ref.~\cite{Devoto:2023rpv}.
Although this extension eventually proves to be straightforward, it requires a more careful treatment of certain collinear limits to achieve a compact and physically-transparent final result.
We also show that the cancellation of infrared divergences can be organized in such a way that, once soft contributions are removed, it occurs 
 independently for each of the external partons.  We consider these results  to be  important stepping stones on the way to deriving finite remainders of the  integrated subtraction terms for fully-general hadron collider processes in the context of the nested soft-collinear subtraction scheme. 
}


\allowdisplaybreaks

\begin{document}
\maketitle
\flushbottom

%-----------------------------------------------------
%                     SECTIONS
%-----------------------------------------------------
%\input{responsetoreferee}
\section{Introduction}

The reliable theoretical description of particle production processes through high-energy collisions  
is an essential element  of the Large Hadron Collider (LHC)
physics program, which aims at probing the Standard Model (SM) of particle physics at the 
shortest distance scales. 
Providing these descriptions requires the development of  techniques to calculate multi-loop amplitudes, as well as those to organize the cancellation of infrared (IR) singularities which appear at intermediate stages of the computation. With the large and ever-increasing dataset from the LHC allowing us to probe ever more complicated processes, these techniques should be broadly applicable  -- i.e.\ they should be able to accommodate processes with a large number of final state particles, including jets --  as well as amenable to efficient implementation in a numerical code.

In this paper, we focus on the problem of IR singularities. 
While efficient, process-independent treatments of  IR singularities  at next-to-leading order (NLO) in QCD  were  developed  many years ago~\cite{Frixione:1995ms, Catani:1996vz, Nagy:2003qn, Bevilacqua:2013iha}, no fully general solution  to this problem 
is available at next-to-next-to-leading order (NNLO) in spite of the  significant progress achieved in the past twenty years. Indeed, during that time several NNLO subtraction and slicing schemes were  proposed~\cite{Frixione:2004is,Gehrmann-DeRidder:2005btv,Currie:2013vh,
Somogyi:2005xz,Somogyi:2006db,DelDuca:2016csb,DelDuca:2016ily,Czakon:2010td,
Czakon:2011ve,Czakon:2014oma,Anastasiou:2003gr,Caola:2017dug,Catani:2007vq,
Grazzini:2017mhc,Boughezal:2011jf,Gaunt:2015pea,Boughezal:2015dva,Sborlini:2016hat,
Herzog:2018ily,Magnea:2018hab,Magnea:2020trj,Chen:2022ktf,Bertolotti:2022aih,Capatti:2019ypt,Devoto:2023rpv,
Braun-White:2023sgd,Braun-White:2023zwd,Fox:2023bma,Gehrmann:2023dxm},
and applied to study many interesting processes at lepton and hadron colliders
(see e.g.\ Refs~\cite{Chen:2014gva,
Boughezal:2015dra,Caola:2015wna,Chen:2016zka,Campbell:2019gmd,Cacciari:2015jma,
Cruz-Martinez:2018rod,Gauld:2021ule,Catani:2022mfv,Chawdhry:2019bji,Chawdhry:2021hkp,
Czakon:2020coa,Gauld:2023zlv,Currie:2017eqf,Chen:2022tpk,Badger:2023mgf,
Czakon:2021mjy,Czakon:2015owf,Catani:2019hip,Buonocore:2023ljm,Brucherseifer:2014ama,
Berger:2016oht,Campbell:2020fhf,Bronnum-Hansen:2022tmr,Alvarez:2023fhi} for a selection of phenomenological papers employing  different theoretical methods).
 However, none solve the problem as comprehensively as at NLO. To illustrate this point, we emphasize  that, as of today, none of the fully-local NNLO subtraction schemes has been used to demonstrate the cancellation of IR 
singularities  and provide formulas for the finite remainders for processes with an arbitrary number of jets at a hadron collider. 

In view of this, many of the NNLO subtraction schemes are being actively developed and refined. In Ref.~\cite{Bertolotti:2022aih}, the \emph{local analytic sector subtraction} scheme~\cite{Magnea:2018hab,Magnea:2020trj} was used to demonstrate the cancellation of IR poles for an arbitrary final state produced in $e^+e^-$ collisions. Similarly, important recent advances have been made in the context of \emph{antenna subtractions}~\cite{Chen:2022ktf,Braun-White:2023sgd,Braun-White:2023zwd,Fox:2023bma,Gehrmann:2023dxm,Fox:2024bfp}, which streamline the computation of antenna functions as well as broaden their applicability to include sub-leading color effects. Improved treatments of power corrections have been proposed for both the $q_T$ slicing~\cite{Catani:2015vma,Ebert:2018gsn,Ebert:2019zkb,Ebert:2020dfc,Ferrera:2023vsw} and $N$-jettiness slicing~\cite{Ebert:2018lzn,Boughezal:2018mvf,Boughezal:2019ggi,Ebert:2019zkb,Vita:2024ypr,Campbell:2024hjq} methods. 
There has also been work towards extending local subtraction methods~\cite{Magnea:2024jqg, Jakubcik:2022zdi,Chen:2023fba,Chen:2023egx} and slicing methods~\cite{Melnikov:2018jxb,Melnikov:2019pdm,Behring:2019quf,Billis:2019vxg,Baranowski:2020xlp,Ebert:2020unb,Baranowski:2022khd,Baranowski:2022vcn,Bell:2023yso,Agarwal:2024gws,Baranowski:2024ene,Baranowski:2024vxg,Baranowski:2024ysi} to  N$^3$LO.

One of the main obstacles in constructing a generalized NNLO subtraction scheme is the proliferation of terms required at different stages of the calculation and the need to combine multiple contributions to obtain the physically-transparent structure of the final result. 
In the recent paper~\cite{Devoto:2023rpv}, we employed the so-called \emph{nested soft-collinear subtraction} scheme \cite{Caola:2017dug} to show how this problem can be overcome. 
The key element in the construction described in Ref.~\cite{Devoto:2023rpv} is the iterative nature of IR subtraction terms that emerges at the level of color-correlated matrix elements.  
Using this feature as a guiding principle allows us to treat the problem in an almost process-independent way, and significantly simplifies the intermediate steps required to demonstrate  the cancellation of IR poles. 
As a result, it became possible to show the cancellation of these poles for the process $q \qb \to \colsing + \Ng \, g$,  where the number of final-state gluons $\Ng$ is a parameter, and to produce relatively compact formulas for the finite remainder of the integrated subtraction terms.
 
Although the results reported in Ref.~\cite{Devoto:2023rpv} are very general, we did take advantage of the symmetries of the  final state and the simple structure of the collinear limits between  final-state gluons and incoming partons.  
Hence, to generalize the calculation of Ref.~\cite{Devoto:2023rpv} to an arbitrary process, we need to do two things: first, consider less symmetric final states and, second, account for flavor-changing  initial-state collinear emissions.
The goal of this paper is to address 
the first issue and we do this by considering a  process where an initial $gq$ state produces an arbitrary number of 
gluons and a quark. 
Furthermore, we also consider corrections in which an additional quark-antiquark pair is radiated,  and extract the contributions 
to the integrated subtraction terms that depend on the number of light quark flavors $\nf$.
We show that the general approach presented in Ref.~\cite{Devoto:2023rpv} can be 
applied to such processes with small modifications which mostly concern the interconnection of final-state collinear splittings 
 as these may be affected by the choice of unresolved partons in different ways.
We find that, if the subtraction terms described by these collinear limits are combined before integration, they lead to universal,  physically-transparent structures.
In fact, these features can already be seen at NLO, providing a guide as to how to solve this problem at NNLO. 

In general, we find that the method introduced in Ref.~\cite{Devoto:2023rpv} is sufficiently robust to provide the description of more complex final and initial states. Indeed, as we will see, the soft part is unchanged, while the collinear part can be treated leg-by-leg.  
Hence, in spite of the fact that we still work with a particular partonic process, we believe that the present paper is an important step towards providing integrated NNLO subtraction terms and  finite remainders for \emph{generic} multi-jet  processes at hadron and lepton colliders. 

The remainder of the paper is organized as follows. 
In \Sec\ref{sec:LO}, we introduce the process of interest and the notation. 
In \Sec\ref{sec_gq_NLO}, we present the treatment of NLO QCD corrections, and discuss our approach to the issues mentioned above.
In \Sec\ref{sec_corrections_NNLO}, we consider the NNLO corrections, and show that the relevant soft and collinear limits can be manipulated to identify contributions with a well-defined number of resolved partons. 
We discuss the simplification of these contributions in \Sec\ref{sec_simplifications_du}. 
Once this is done, we demonstrate the cancellation of IR singularities in \Sec\ref{sec:poles_cancell}, and present the finite remainder in \Sec\ref{sec_final_result}. 
We conclude in \Sec\ref{sec_conclusions}. We collect many of the constants, functions, and operators used throughout the paper in Appendix~\ref{sec:Splitting}.

\section{The leading order process}
\label{sec:LO}

We are interested in understanding higher-order QCD corrections to the inclusive production of $N$ jets at a hadron collider in association with a color-neutral system $\colsing$, $pp \to \colsing + N~{\rm jets}$.
This process receives contributions from many partonic channels. 
In this paper, we will focus on the following one
\begin{equation}
   \gqBorn: ~  \inF_g + \inS_q \to \colsing + \Ng \, g + q \,,
    \label{eq_NLO_gq_LO_process}
\end{equation}
where $\Ng = N - 1$, and $\inF$ and $\inS$ label initial state partons
with momenta $p_{a,b}$.
We write the cross section for process $\gqBorn$ as
\begin{equation}
\begin{split}
    2 s_{\inF\inS} \, \dsigmahat_\LO^{gq} 
    & = \Lint \frac{1}{\Ng!} \FLMlo{gq}{\setg{\Ng},\setq{1}} \Rint \\
    & = \frac{N_{\rm av}}{\Ng!} \int \rmd\Phi \, (2\pi)^4\delta^{(4)}(P_{\rm fin} - P_{\rm in}) \, \abs{\calM_0(P_{\rm fin}, P_{\rm in} )}^2 \,  
    \calO(P_{\rm fin}) \, ,
    \label{eq_FLM_LO_defn_gen}
\end{split}
\end{equation}
and we note that, in order to obtain its contribution to the hadronic process, we need to convolute the partonic cross section with parton distribution functions (PDF) $f_g$ and $f_q$. In  \eq\eqref{eq_FLM_LO_defn_gen}, 
$\calM_0$ is the matrix element of the process in 
\eq\eqref{eq_NLO_gq_LO_process}, $\rmd\Phi$ is the Lorentz-invariant phase space for the final-state particles, and $\calO$ is an IR-safe observable which  ensures that the final state contains at least $N$ resolved jets. 
The total momenta of the initial- and final-state particles are denoted as $P_{\rm in}$ and $P_\fin$, respectively, and the  delta function in \eq\eqref{eq_FLM_LO_defn_gen} enforces energy-momentum conservation. 
Summation over spins and colors of final-state partons, and averaging over spins and colors of initial-state partons are assumed in \eq\eqref{eq_FLM_LO_defn_gen}. 
The corresponding spin and color averaging factors are described by the factor 
$N_{\rm av}$.

Furthermore, 
$\setg{\Ng}$ and $\setq{1}$ 
in \eq\eqref{eq_FLM_LO_defn_gen}
denote lists of $\Ng$ final-state gluons,  and of  
a single final-state quark, respectively.
This notation is redundant for the process that we consider, but we introduce it with an eye on future generalizations. The factor $1/\Ng!$ corresponds to the symmetry factor arising from having $\Ng$ indistinguishable gluons in the final state.
It is convenient to absorb this symmetry factor and omit the initial-state partons and  lists of final-state 
partons by defining 
\begin{equation}
    \FLM^\gqBorn 
    \eqdef
    \frac{1}{\Ng!} \FLMlo{gq}{\setg{\Ng},\setq{1}} \,.
    \label{eq_FLM_gqBorn_def}
\end{equation}
Then, the differential cross section for 
the process ${\cal A}_0$ reads 
\begin{equation}
    2 s_{\inF\inS} \, \dsigmahat_\LO^{gq} 
    =
    \lint \FLM^\gqBorn \rint \,,
\end{equation}
and we will use this notation in what follows. 

\section{NLO QCD corrections  } 
\label{sec_gq_NLO}

Having fixed the notation, we continue by  discussing the  NLO QCD corrections to the partonic process $\gqBorn$ in \eq\eqref{eq_NLO_gq_LO_process}.
As usual, one has to compute virtual corrections to $\gqBorn$ and account for  real-emission contributions that are comprised of processes where the number 
of final-state partons is increased by one. For the complete calculation, many  real-emission processes need to be considered. 
However, since we would like to discuss the cancellation of IR  singularities in higher-order corrections to $\gqBorn$, we can omit contributions that do not lead to this process when soft or collinear limits are taken. 
There are also processes which lead to $\gqBorn$ under collinear limits but which are convoluted with different PDFs; we do not consider such processes either.

With this clarification, it is easy to convince oneself that the following three processes should be considered
\begin{equation}
\begin{split} 
    \FgqprocNLO: \quad & \inF_g + \inS_q \to X + (\Ng+1) g + q \,, \\
    \SgqprocNLO: \quad & \inF_g + \inS_q \to X + (\Ng-1) g + q + \qp \qbp \,, \quad \qp\neq q \,, \\
    \TgqprocNLO: \quad & \inF_g + \inS_q \to X + (\Ng-1) g + q  + q\qb \, .
\end{split} 
\label{eq:proc_A_NLO}
\end{equation}
However, further simplifications are possible.
Indeed, in order to provide a singular contribution to the process $\gqBorn$, each of the three processes ${\cal A}_{1,2,3}$ should ``lose'' a parton  in such a way that one gets both the initial and the final state of $\gqBorn$.
There are several ways in which this can happen in the process $\FgqprocNLO$, but in processes $\SgqprocNLO$ and $\TgqprocNLO$ this can only occur  if  
a  quark-antiquark pair, $q'\qb'$ or $q\qb$, respectively, is clustered into a gluon.
Since these collinear limits are identical for processes $\SgqprocNLO$ and $\TgqprocNLO$, we do not need to consider the latter  since we can retrieve all the required information from the former.

We begin with the discussion of process $\FgqprocNLO$ in \eq\eqref{eq:proc_A_NLO}, and write the corresponding differential cross section as 
\begin{equation}
    2s_{\inF\inS} \, \dsigmahat^{\FgqprocNLO}_\R 
    = \Lint \frac{1}{(\Ng+1)!} \FLMlo{gq}{\setg{\Ng+1},\setq{1}} \Rint \,. 
\end{equation}
Following Ref.~\cite{Devoto:2023rpv}, we introduce partitions  of unity  $\Delta^{(i)}$ and split the cross section into a sum of terms such that in each term only one of the partons can lead to a singular limit once it becomes unresolved. 
We find
\begin{equation}
    2s_{\inF\inS} \, \dsigmahat^{\FgqprocNLO}_\R  = \sum_{i} \Lint \frac{\Delta^{(i)}}{(\Ng+1)!} \FLMlo{gq}{\setg{\Ng+1},\setq{1}} \Rint  \,.
\label{eq_NLO_dsigma_R_first_writing}
\end{equation}
The index $i$ in the above formula runs over all final-state partons, and a possible choice of partitions $\Delta^{(i)}$ is described in Appendix B of Ref.~\cite{Devoto:2023rpv}.

The potentially-unresolved parton $i$ can be either a (anti)quark or a gluon.
Accounting for the symmetry properties of $\FLM$, we rename the final state momenta and call the unresolved parton $\Fp$.
There are $\Ng+1$ ways to choose an unresolved gluon and a single choice for an unresolved quark. 
Hence, we can write  
\begin{equation}
    2s_{\inF\inS} \, \dsigmahat^{\FgqprocNLO}_\R 
    = \lint \Delta^{(\Fp)} \FRlo{\FgqprocNLO}{\Fp} \rint \,,
\end{equation}
where 
\begin{equation}
    \FRlo{\FgqprocNLO}{\Fp} \eqdef \FLMlo{\FgqprocNLO}{\Fp_g} + \FLMlo{\FgqprocNLO}{\Fp_q} \,,
    \label{eq_NLO_FR_second_first_def}
\end{equation}
and the functions $F_{\rm LM}$ are defined as 
\begin{equation}
\begin{split}
    \FLMlo{\FgqprocNLO}{\Fp_g} & \eqdef \frac{1}{\Ng!} \FLMun{\inF\inS}{\setg{\Ng},\setq{1}}{\Fp_g} \,, \\
    \FLMlo{\FgqprocNLO}{\Fp_q} & \eqdef \frac{1}{(\Ng+1)!} \FLMun{\inF\inS}{\setg{\Ng+1}}{\Fp_q} \,.
    \label{eq_FLMlo_FgqprocNLO_Fp_g_and_q}
\end{split}
\end{equation}
We note that in the last $\FLM$ function the quark list has disappeared, and  that the symmetry factors that multiply various functions $\FLM$ are defined by the number of \emph{resolved} gluons. 

Following the discussion in Ref.~\cite{Devoto:2023rpv}, we 
extract soft and collinear divergences that affect the real emission contribution, and obtain 
\begin{equation}
    2s_{\inF\inS} \, \dsigmahat^{\FgqprocNLO}_\R 
    = \lint S_\Fp \FRlo{\FgqprocNLO}{\Fp} \rint 
    + \sum_{i \in \HP} \lint \oS_\Fp C_{i\Fp} \Delta^{(\Fp)} \FRlo{\FgqprocNLO}{\Fp} \rint
    + \lint \ONLO^{(\Fp)} \Delta^{(\Fp)} \FRlo{\FgqprocNLO}{\Fp} \rint \,.
    \label{eq_NLO_dsigma_R}
\end{equation}
We denote the list of $N$ final-state resolved partons contained in a given $\FLM$ as $\HPf$, and the list  that combines $\HPf$ and the initial-state particles as $\HP = \{\inF,\inS\} \cup \HPf$. 
The soft and collinear subtraction  operators in \eq\eqref{eq_NLO_dsigma_R} read
\begin{equation}
    \oS_\Fp = \iden - S_\Fp \,,
    \qquad
    \oC_{i\Fp} = \iden - C_{i\Fp} \,,
    \qquad
    \ONLO^{(\Fp)} = \sum_{i \in \HP} \oS_\Fp \oC_{i\Fp} \, \partFuncNLOfp{i} \,.
    \label{eq_ONLO_def}
\end{equation}
The partition functions $\partFuncNLOfp{i} $ are defined in Ref.~\cite{Devoto:2023rpv}, and allow us to treat one collinear singularity at a time.
We note that, in \eq\eqref{eq_NLO_dsigma_R}, the soft-limit operator $S_\Fp$ is only needed when $\Fp$ is a gluon.  
Hence, to apply \eq\eqref{eq_NLO_dsigma_R} for generic processes, we have to set the soft operator to zero when the unresolved parton is 
a quark or an antiquark, i.e.,
\begin{equation}
    S_q = S_\qb \eqdef 0 \,.
    \label{eq_single_soft_limit_quark_condition}
\end{equation} 
Upon integrating over the phase space of the soft gluon, 
the first term on the \rhs\ of \eq\eqref{eq_NLO_dsigma_R} can be written as\footnote{We use dimensional regularization with $d = 4-2\ep$ throughout the paper.}  
\begin{equation}
    \lint S_\Fp \FRlo{\FgqprocNLO}{\Fp} \rint 
    \equiv \lint S_\Fp \FLMlo{\FgqprocNLO}{\Fp_g} \rint
    = \asbr \lint \ISoft(\ep) \colorprod \FLM^\gqBorn \rint \,,
    \label{eq_NLO_soft_contribution}
\end{equation}
where $\ISoft$ is the \emph{soft operator} introduced in Ref.~\cite{Devoto:2023rpv} and defined explicitly in \eq\eqref{eq_ISoft_definition_appendix}.

Collinear limits require more attention.  
As stated above, we are interested in those limits that lead to singularities proportional to the matrix element of $\gqBorn$, and we isolate such limits using the \emph{projection operator} $\proj$.
Terms that involve the soft operator $S_\Fp$ are unaffected by $\proj$, 
but some non-vanishing collinear limits are removed, e.g.\ the limit where the final state quark $\Fp_q$ becomes collinear to the initial state gluon $a_g$; thus 
\begin{equation}
    \proj \lint  C_{\inF\Fp} \Delta^{(\Fp)} \FLMlo{\FgqprocNLO}{\Fp_q} \rint = 0 \, .
\end{equation}

This procedure allows us to select a subset of limits of the NLO matrix element that do not change the Born level configuration and whose combination is free of $1/\ep$ singularities. This subset of limits does not remove  \emph{all} the singularities of the hard N(N)LO matrix element; in particular, those singularities whose limits include a different Born-level matrix element to that of $\gqBorn$ are not subtracted. We do this because our main aim is to study the structure of final-state collinear limits in the presence of quarks rather than provide complete formulas for the physical real correction to the cross section.

With these preliminary remarks out of the way, 
we can analyze collinear singularities related to the initial-state radiation.
The only unresolved parton that can develop a collinear singularity with the incoming gluon $\inF_g$ and still contribute to $\gqBorn$ is the final-state gluon $\Fp_g$. 
Then, following the steps detailed in Ref.~\cite{Devoto:2023rpv}, we obtain
\begin{equation}
\begin{split}
    \proj \lint \oS_\Fp C_{\inF\Fp} \Delta^{(\Fp)} \FRlo{\FgqprocNLO}{\Fp} \rint 
    & = \lint \oS_\Fp C_{\inF\Fp} \Delta^{(\Fp)} \FLMlo{\FgqprocNLO}{\Fp_g} \rint \\
    & = \frac{\asbr}{\ep} \Big[\lint \Gamma_{\inF ,g} \, \FLM^\gqBorn \rint 
    + \lint \CalPgen_{gg} \conv \FLM^\gqBorn \rint \Big] \;, 
\end{split}  
\label{eq_NLO_hard_collinear_leg1}
\end{equation}
where we have used the fact that $C_{\inF \Fp} \Delta^{(\Fp)} = 1$, and introduced a particular notation for the left convolution
\begin{equation}
    \CalPgen_{\alpha\beta} \conv \FLM  
     \eqdef 
    \int_{0}^{1} \dz \, \CalPgen_{\alpha\beta}(z) \, \frac{\FLMlo{}{z \cdot p_\inF, p_\inS; \HPf}}{z}  \,.
\label{eq_convolution}
\end{equation}
Definitions of the generalized initial-state anomalous dimensions  $\Gamma_{\inF,f_\inF}$ arising from the soft-collinear limits and of the (azimuthally averaged) collinear splitting functions $\CalPgen_{\alpha\beta}$ are reported in Eqs \eqref{Eq:Gamma_1_2_definition_k_general_appendix} and \eqref{Eq:Paa_GEN_definition_k_general_appendix}, respectively. 
The computation of the soft-subtracted collinear limit for the initial-state parton $\inS_q$ is analogous. We
find 
\begin{equation}
   \proj \lint \oS_\Fp C_{b\Fp} \Delta^{(\Fp)} \FRlo{\FgqprocNLO}{\Fp} \rint
   =
   \frac{\asbr}{\ep} \Big[\lint \Gamma_{\inS,q} \, \FLM^\gqBorn \rint 
    + \lint \FLM^\gqBorn \conv \CalPgen_{qq} \rint \Big] \, , 
    \label{eq_NLO_hard_collinear_leg2}  
\end{equation} 
where the right convolution 
is defined as 
\begin{equation}
    \FLM \conv \CalPgen_{\alpha\beta}
     \eqdef 
    \int_{0}^{1} \dz \, \CalPgen_{\alpha\beta}(z) \, \frac{\FLMlo{}{p_\inF, z \cdot p_\inS; \HPf}}{z} \,.
\end{equation}

We continue with the discussion of the final-state collinear limits, which correspond to $i \in \HPf$ in \eq\eqref{eq_NLO_dsigma_R}.
We note that, in this case, $C_{i \Fp} \Delta^{(\Fp)} = E_{i}/(E_i+E_\Fp) = z_{i,\Fp}$, where $E_{i,\Fp}$ are energies of the corresponding partons.
The appearance of such factors makes the relation between the integrated  collinear limits and conventional splitting functions and anomalous dimensions less transparent. This is a new complication with respect to the calculation of Ref.~\cite{Devoto:2023rpv}. 

To address it, we will consider the two contributions to the function $\FRlo{\FgqprocNLO}{\Fp}$ in \eq\eqref{eq_NLO_FR_second_first_def} separately. 
We start with $\FLMlo{\FgqprocNLO}{\Fp_g}$, and compute the hard-collinear limits for final-state partons 
\begin{equation}
     \sum_{i \in \HPf} \lint \oS_\Fp C_{i\Fp} \Delta^{(\Fp)} \FLMlo{\FgqprocNLO}{\Fp_g} \rint
    = \frac{\asbr}{\ep} \sum_{i \in \HPf} \lint \Gamma_{i,f_i \to f_i g} \, \FLM^\gqBorn \rint \,,
    \label{eq_NLO_hard_collinear_unres_gluon}
\end{equation}
where we have used the fact that clustering  a collinear gluon with any parton does not change that parton's flavor.  
The \emph{generalized final-state anomalous dimension} $\Gamma_{i,f_i \to f_i g}$ can be found in \eq\eqref{eq_Gamma_a_to_bc_defs_appendix}. 
We note that these quantities generalize  the final-state anomalous dimensions $\gamma_{z,f_i \to f_i g}^{22}(L_i)$ defined in \eq\eqref{eq_usual_anom_dimens_defs} 
and used in Ref.~\cite{Devoto:2023rpv}, and that the quantity $\Gamma_{i,g \to gg}$ appeared in this reference where it was denoted by $\Gamma_{i,g}$.
Additionally, $\gamma_{z,f_i \to f_i g}^{22}(L_i)$ carries a subscript $z = z_{i,\Fp}$ to indicate that it is 
obtained from an integral of a particular splitting function weighted by a factor $z_{i,\Fp}$. 
We drop this subscript in the definition of $\Gamma_{i,f_i \to f_i f_\Fp}$ to lighten the notation. 

We emphasize that the generalized final-state anomalous dimensions differ from the conventional collinear anomalous dimensions that arise from integrals of the collinear splitting functions without additional factors $z_{i,\Fp}$. 
To see how conventional collinear anomalous dimensions emerge, we need to consider the other contribution to the function $\FRlo{\FgqprocNLO}{\Fp}$, where a
\emph{quark} is potentially-unresolved. 
Since in $\FLMlo{\FgqprocNLO}{\Fp_q}$ the only resolved final-state  partons are gluons, the  unresolved quark $\Fp_q$ has to be clustered with one of them. 
The clustered parton is a quark, so that the number of resolved gluons decreases by one and the number of resolved quarks increases by one. 
Working out the symmetry factors, we obtain 
\begin{equation}
\begin{split}
    \sum_{i\in\HPf} \lint C_{i\Fp} \Delta^{(\Fp)} \FLMlo{\FgqprocNLO}{\Fp_q} \rint 
    = \frac{\asbr}{\ep} \sum_{i \in \HPfq} \!\!  \lint 
      \Gamma_{i, q \to g q} \, \FLM^\gqBorn \rint \,,
    \label{eq_NLO_hard_collinear_unres_quark}
\end{split}
\end{equation}
where, on the \rhs, $\HPfq$ denotes the list of resolved final-state quarks.
In the current case, this notation is redundant as this list consists of a single parton, but we note that \eq\eqref{eq_NLO_hard_collinear_unres_quark} holds independently of the number of final-state quarks.

Combining Eqs \eqref{eq_NLO_hard_collinear_unres_gluon} and \eqref{eq_NLO_hard_collinear_unres_quark}, we obtain 
\begin{equation}
\begin{split}
    \sum_{i\in \HPf} \lint \oS_\Fp C_{i\Fp} \Delta^{(\Fp)} \FRlo{\FgqprocNLO}{\Fp} \rint
    = \frac{\asbr}{\ep} \Lint
    \bigg[\sum_{i \in \HPfg} \!\!  {\Gamma_{i,g \to gg}}
    + \sum_{i \in \HPfq} \!\! \Gamma_{i, q} \bigg]  \FLM^\gqBorn \Rint \,,
    \label{eq_IC_qTOqg_plus_qTOgq}
\end{split}
\end{equation}
where we define
\begin{equation}
    \Gamma_{i,q} \eqdef \Gamma_{i, q \to qg} + \Gamma_{i, q \to gq} \,.
    \label{eq_Gamma_i_q_FSR_def}
\end{equation}
We emphasize that the quantity $\Gamma_{i,q}$ in \eq\eqref{eq_Gamma_i_q_FSR_def} is directly related to the conventional quark anomalous dimension, since the weight factors  $z_{i,\Fp}$ disappear when combining the two splittings. 
To show this, we make use of the fact that  $\Gamma$ is proportional to $\gamma^{22}(L_i)$ (cf.~\eq\eqref{eq_Gamma_a_to_bc_defs_appendix}) and write\footnote{In the following equation, $S_z$ is analogous to $S_\Fp$ but refers specifically to the variable $z$, i.e.\ $\lim_{z \to 1}$.} 
\begin{equation}
\begin{split}
    \gamma_{z,q \to qg}^{22}(L_i) + \gamma_{z,q \to gq}^{22}
    & = - \int_{0}^{1} \dz \, (\iden - S_z) \, \frac{z P_{qq}(z) + z P_{qg}(z)}{[z(1-z)]^{2\ep}} + \Cf \frac{1 - e^{-2\ep L_i}}{\ep} \\
    & = - \int_{0}^{1} \dz \, (\iden - S_z) \, 
    \frac{P_{qq}(z)}{[z(1-z)]^{2\ep}}+ \Cf \frac{1 - e^{-2\ep L_i}}{\ep} \\
    & = \gamma_{1,q \to qg}^{22}(L_i) = \gamma_q + 2\T^2_q L_i + \order{\ep} \,,
\end{split}
\label{eq:dropping_z}
\end{equation}
where $\gamma_q = \frac{3}{2}\Cf$ is the quark anomalous dimension,  $L_i = \log(\Emax/E_i)$, and we have used the relation between splitting functions $P_{qg}(1-z) = P_{qq}(z)$ to remove the factors of $z$ from  the integrand.

Performing a similar calculation for $g^* \to gg$  splitting, we observe that the quantity $\Gamma_{i,g \to gg} = \frac{11}{6} \Ca + 2\Ca L_i + \order{\ep}$, i.e.\ to leading order in $\ep$, gives exactly the $\Ca$ term of the gluon anomalous dimension, but the $\nf$ term is missing.  This is easily remedied by considering the process with an additional
$\qp\qbp$ pair, which we referred to as $\SgqprocNLO$.
Its cross section reads
\begin{equation}
    2s_{\inF\inS} \, \dsigmahat_\R^{\SgqprocNLO}
    = 
    \Lint \frac{1}{(\Ng-1)!} \FLMlo{gq}{\setg{\Ng-1},\setq{1},\setqp{1},\setqbp{1}} \Rint \,.
\end{equation}
We can analyze this  contribution following what  we did for the process $\FgqprocNLO$.     
Since we are interested in contributions that contain singularities proportional to the matrix element for the process $\gqBorn$, only some collinear limits of $\SgqprocNLO$ need to be considered. 
In fact, the only way to arrive at $\gqBorn$ starting from $\SgqprocNLO$ is to cluster  
$\qp$ and $\qbp$ into a gluon. 
For this to happen, either $\qp$ or $\qbp$ should be designated as potentially-unresolved, and then the only collinear limit we need to consider is $\qp \parallel \qbp$. 
Making use of the projection operator $\proj$ to select  the relevant terms, we 
write 
\begin{equation}
    \proj \big[ 2s_{\inF\inS} \, \dsigmahat_\R^{\SgqprocNLO} \big] 
    = \lint C_{\qbp \Fp_\qp} \Delta^{(\Fp)} \FLMlo{\SgqprocNLO}{\Fp_\qp} \rint 
    + \lint C_{\qp \Fp_\qbp} \Delta^{(\Fp)} \FLMlo{\SgqprocNLO}{\Fp_\qbp} \rint \,,
    \label{eq_NLO_dsigmahat_SgqprocNLO_R_starting}
\end{equation}
where $\FLMlo{\SgqprocNLO}{\Fp_\qp}$ and $\FLMlo{\SgqprocNLO}{\Fp_\qbp}$ are defined analogously to \eq\eqref{eq_FLMlo_FgqprocNLO_Fp_g_and_q}, but using the list of partons in process $\SgqprocNLO$ instead of $\FgqprocNLO$.  
The two terms on the \rhs\ of \eq\eqref{eq_NLO_dsigmahat_SgqprocNLO_R_starting} give identical contributions since the splitting $g \to \qp \qbp$ is symmetric under the swapping  $\qp \leftrightarrow \qbp$. 
Furthermore, this splitting does not depend on the quark flavor. Thus, we  obtain the contribution of $\nf-1$ quark species from the process $\SgqprocNLO$, 
and add the contribution from the $\TgqprocNLO$ process, where $\qp = q$.
This is equivalent to taking 
one term in the right-hand side 
of \eq\eqref{eq_NLO_dsigmahat_SgqprocNLO_R_starting}, computing its contribution, and multiplying the  result by a factor of $2\nf$, obtaining 
\begin{equation}
    \sum_{n=2,3} \! \proj \big[2s_{\inF\inS} \, \dsigmahat_\R^{\genBorn} \big]
    = \frac{\asbr}{\ep} \sum_{i\in\HPfg} \!\! \lint 2\nf\, \Gamma_{i,g \to q\qb} \, \FLM^\gqBorn \rint \,.
    \label{eq_NLO_dsigmahat_STgqprocNLO}
\end{equation}
The anomalous dimension  $\Gamma_{i,g \to q {\bar q}}$ is given in \eq\eqref{eq_Gamma_a_to_bc_defs_appendix}. 
Note that we have ``distributed'' the clustered gluon over the list of gluons 
in $\FLM^\gqBorn$ using its symmetry under gluon permutations. 

Combining  the contributions from Eqs \eqref{eq_IC_qTOqg_plus_qTOgq} and \eqref{eq_NLO_dsigmahat_STgqprocNLO}, we find 
\begin{equation}
\begin{split}
    & \sum_{i\in \HPf} \lint \oS_\Fp C_{i\Fp} \Delta^{(\Fp)} \FRlo{\FgqprocNLO}{\Fp} \rint
    + \sum_{n=2,3} \! \proj \big[2s_{\inF\inS} \, \dsigmahat_\R^{\genBorn} \big] \\
    & = \frac{\asbr}{\ep} \Lint
    \bigg[\sum_{i \in \HPfg} \!\!  {\Gamma_{i,g}}
    + \sum_{i \in \HPfq} \!\! \Gamma_{i, q} \bigg]  \FLM^\gqBorn \Rint \,,
    \label{eq_NLO_dsigmahat_FSTgqprocNLO_appo}
\end{split}
\end{equation}
where, by analogy with \eq\eqref{eq_Gamma_i_q_FSR_def}, we introduced 
\begin{equation}
    \Gamma_{i,g} \eqdef \Gamma_{i,g \to g g} + 2\nf\, \Gamma_{i,g \to q\qb} \,.
  \label{eq_Gamma_i_fin_FSR_def}
\end{equation}
We note that, at $\order{\ep^0}$, $\Gamma_{i,g}$ corresponds to $\Gamma_{i,g} = \gamma_g + 2\T^2_g L_i + \order{\ep}$, where $\gamma_g$ is the gluon anomalous dimension. To see this, one can repeat  the manipulations performed in \eq\eqref{eq:dropping_z}   for $\Gamma_{i,g \to q\bar q}$, 
using the fact that $P_{gq}$ is symmetric under $z \leftrightarrow (1-z)$. As a result, the $z$ factor in the integrand can be replaced by a factor $1/2$ (see the discussion in Sec.\ 5.1 in Ref.~\cite{Devoto:2023rpv}). 

We can now use \eq\eqref{eq_NLO_dsigmahat_FSTgqprocNLO_appo} to extend the definition 
of the hard-collinear operator $\IColl$, introduced in Ref.~\cite{Devoto:2023rpv}, to
\begin{equation}
    \IColl(\ep) 
    \eqdef 
    \sum_{i \in \HP} \frac{\Gamma_{i,f_i}}{\ep} \,.
\label{eq_Ic_defs_gq_channel_first_process}
\end{equation}
Therefore, the whole hard-collinear contribution can be written as 
\begin{equation}
\begin{split}
    &\; \proj \sum_{i\in \HP} \lint \oS_\Fp C_{i\Fp} \Delta^{(\Fp)} \FRlo{\FgqprocNLO}{\Fp} \rint
    + \sum_{n=2,3} \proj \big[2s_{\inF\inS} \, \dsigmahat_\R^{\genBorn} \big] \\
    & = \asbr \lint \IColl(\ep) \colorprod \FLM^\gqBorn \rint 
    + \frac{\asbr}{\ep} \Big[ \lint \CalPgen_{gg} \conv \FLM^\gqBorn \rint + \lint \FLM^\gqBorn \conv \CalPgen_{qq} \rint \Big] \,.
    \label{eq_NLO_dsigmahat_FSTgqprocNLO_hard_collinear}
\end{split}
\end{equation}
The real-radiation contribution to the NLO cross section is obtained by combining the above expression with the soft contribution given in \eq\eqref{eq_NLO_soft_contribution}, and 
the fully-regulated $\ONLO$ term in \eq\eqref{eq_NLO_dsigma_R}.

To complete the computation, we need to account for the virtual corrections and perform the collinear renormalization of parton distribution functions. 
We use the operator $\IVirt$, introduced in Ref.~\cite{Devoto:2023rpv}, and defined in \eq\eqref{eq_IVirt_def_appendix}, to write the 
virtual corrections as 
\begin{equation}
    2s_{\inF\inS} \, \dsigmahat_\V^{\gqBorn} 
    =  \asbr \lint \IVirt(\ep) \colorprod \FLM^\gqBorn \rint 
    + \lint \FLVfin^\gqBorn \rint \, .
    \label{eq_NLO_dsigma^V}
\end{equation}
The function $\FLVfin^\gqBorn$ in the above equation is analogous to $\FLM$ in \eq\eqref{eq_FLM_LO_defn_gen},  but with $\abs{\calM_0}^2$ replaced by $2\Re[\bra{\calM_1^\fin}\ket{\calM_0}]$, where $\calM_1^\fin$ is the 
finite part of the one-loop virtual amplitude. 
The collinear renormalization contributions relevant for $\gqBorn$, which come 
from the PDF redefinition, read
\begin{equation}
    2s_{\inF\inS} \,  \dsigmahat_{\rm pdf}^{\gqBorn} 
    = \frac{\as(\mu)}{2\pi \ep} \Big[\lint \PAP_{gg} \conv \FLM^\gqBorn \rint 
    + \lint \FLM^\gqBorn \conv \PAP_{qq}\rint\Big] \,,
    \label{eq_NLO_dsigma^pdf}
\end{equation}
where the Altarelli-Parisi splitting functions $\PAP_{\alpha\beta}$ are defined in \eq\eqref{Eq_PAP_0_definition_appendix}.

Using these results, we can finally write the NLO corrections to the process $\gqBorn$. 
We combine Eqs \eqref{eq_NLO_dsigma_R}, \eqref{eq_NLO_soft_contribution}, \eqref{eq_NLO_dsigmahat_FSTgqprocNLO_hard_collinear}, \eqref{eq_NLO_dsigma^V}, and \eqref{eq_NLO_dsigma^pdf}, and find an expression that does not contain $\ep$ poles
\begin{equation}
\begin{split}
    &\; 2s_{\inF\inS} \Big[\dsigmahat_\V^\gqBorn + \sum_{n=1}^3 \proj \dsigmahat_\R^\genBorn  + \dsigmahat_{\rm pdf}^{\gqBorn} \Big]
    = \proj \lint \ONLO^{(\Fp)} \, \Delta^{(\Fp)} \FRlo{\FgqprocNLO}{\Fp} \rint
    + \lint\FLVfin^\gqBorn\rint \\
    & + \asbr \lint \ITot^{(0)} \colorprod \FLM^\gqBorn \rint 
    + \asbr \Big[\lint \PNLO_{gg} \conv \FLM^\gqBorn \rint 
    + \lint \FLM^\gqBorn \conv \PNLO_{qq} \rint \Big] \,.
    \label{eq_NLO_NgPlusq_channel_final_result}
\end{split}
\end{equation}
Here, $\ITot^{(0)}$ is the $\order{\ep^0}$ coefficient in the expansion of the IR-finite  operator
\begin{equation}
    \ITot(\ep) = \IVirt(\ep) + \ISoft(\ep) + \IColl(\ep) \,,
    \label{eq_NLO_IT_first_process_def}
\end{equation}
defined analogously to Ref.~\cite{Devoto:2023rpv}. 
Its expression is reported in \eq\eqref{eq_IT0_appendix}. 
Finally, for the boosted contribution, we used the relation $\alpha_\rms(\mu)/2\pi = \cTilde \asbr$, with $\cTilde = \Gamma(1-\ep)/e^{\ep \gamma_\rmE}$, and the following relation between generalized splitting functions and Altarelli-Parisi splitting kernels 
\begin{equation}
    \CalPgen_{\alpha \beta}(z,E_\inF) + \cTilde \PAP_{\alpha \beta}(z) = \ep \, \PNLO_{\alpha \beta}(z,E_\inF) + \order{\ep^2} \,, 
\label{eq_Pgen_PAP_PNLO_relation}
\end{equation}
valid for all $\alpha$ and $\beta$.
The explicit expressions of functions $\PNLO_{\alpha\beta}$ are reported in the ancillary file \finalresult\ provided with this paper, see Table \ref{table_ancillary_file}.
\\

Before concluding the NLO analysis, we briefly discuss the treatment of the singularities in $\FgqprocNLO$ that \emph{do not lead} to the partonic process $\gqBorn$. 
Considering   the function 
$\FLMlo{\FgqprocNLO}{\Fp_q}$
for the sake of example, and taking the collinear limit $\Fp_q \parallel \inF_g$, we derive    
\begin{equation}
    (f_g \conv f_q) \conv \lint  C_{\inF \Fp}\Delta^{(\Fp)}
    \FLMlo{\FgqprocNLO}{\Fp_q}
    \rint 
    = \frac{\asbr}{\ep} (f_g \conv f_q) \conv  \lint \CalPgen_{\qb g} \conv \FLMlo{\qb q}{\setg{\Ng+1}} \rint \,.
    \label{eq_NLO_dsigmahat_STgqprocNLOa}
\end{equation}
This singularity is cancelled by the collinear PDF renormalization in the Born process $\qb q \to \colsing + (\Ng+1) \, g$.
We emphasize that the use of the projection operator $\proj$ allows us to
discard such singularities, and  focus on the main goal of this paper, which is an understanding of intertwined
final-state collinear limits.

\section{NNLO QCD corrections: general framework} 
\label{sec_corrections_NNLO}

In this section, we study the NNLO QCD corrections to the process $\gqBorn$ in \eq\eqref{eq_NLO_gq_LO_process}.
Similarly to the NLO case, we discard all singular contributions 
that lead to partonic processes which  
differ from $\gqBorn$.
To compute the NNLO QCD corrections to $\gqBorn$, we have to account for the two-loop virtual corrections to it, the one-loop corrections to processes $\FgqprocNLO$ and $\SgqprocNLO$, and three double-real emission processes 
\begin{equation}
\begin{split}
    & \FgqprocNNLO : ~ \inF_g + \inS_q \to X + (\Ng+2) g + q \,,\\
    & \SgqprocNNLO : ~ \inF_g + \inS_q \to X + \Ng\, g + q + \qp\qbp \,, \\
    & \TgqprocNNLO : ~ \inF_g + \inS_q \to X + (\Ng - 2)  g + q + \qp\qbp + q''\qb'' \,, \label{eq_NNLO_gq_third_process} \, 
\end{split}
\end{equation}
where we consider both the cases $\qp \ne q$ and $\qp = q$, and the same for $q''$.
In order to cancel the $1/\ep$ poles at NNLO, we write 
\begin{equation}
 \dsigmahat_{\rm NNLO}^{\gqBorn} 
    = \dsigmahat_{\rm VV}^{\gqBorn} 
    + \proj \Big[
    \dsigmahat^{\FgqprocNLO}_\RV
    + \nf \dsigmahat^{\SgqprocNLO}_\RV
    + \dsigmahat^{\FgqprocNNLO}_\RR + \dsigmahat^{\SgqprocNNLO, {\rm int}}_\RR
    + \nf \dsigmahat^{\SgqprocNNLO}_\RR 
    + \nf^2 \dsigmahat^{\TgqprocNNLO}_\RR \Big] 
    + \dsigmahat^{\gqBorn}_{\rm pdf}
     \, .
     \label{eq_dsigmahat_NNLO_starting_point}
\end{equation}
We note that the only singular limits of the process ${\cal A}_{5}$ that will project onto 
the process $\gqBorn$ are those in which $q'$ and $\bar q'$ are either simultaneously soft, or are clustered together in the collinear regime (see the corresponding discussion after \eq\eqref{eq_NLO_dsigmahat_SgqprocNLO_R_starting}). In either case, they produce a factor of 
$\nf$, which we write  explicitly in \eq\eqref{eq_dsigmahat_NNLO_starting_point}. Similarly, the process ${\cal A}_6$ produces a factor $\nf^2$ when both $q'$ and $\bar q'$ as well as $q''$ and $\bar q''$ become collinear. There is also an interference term arising from process ${\cal A}_{5}$ if $q = q'$, which does not appear with a factor of $\nf$. This contribution is only singular in the triple-collinear limit and hence only produces simple poles in $\eps$. No other singular limits need to be considered as these would result in processes different from ${\cal A}_0$. In particular, we omit the discussion of IR divergences from flavor singlet configurations. 

To compute the double-real contributions to \eq\eqref{eq_dsigmahat_NNLO_starting_point}, we need to analyze quantities of the following type
\begin{equation}
    2s_{\inF\inS} \, \dsigmahat^{\genBorn}_{\RR}
    = \Lint \frac{\FLMlo{gq}{\setg{\Ng},\setq{\Nq},\setqb{\Nqb}, ... }}{\Ng! \times \Nq! \times \Nqb! \times ... } \Rint \,,
    \qquad n \in \{4,5,6\} \,,
    \label{eq_dsigmahat_RR_genBorn_starting}
\end{equation}
where integer numbers $\Ng, \Nq, \Nqb$, etc., indicate lengths of the relevant lists; they depend on the process $\genBorn$ under consideration.  
The first step consists of employing  damping factors to choose  the potentially-unresolved partons in  $\FLM$. 
Following Ref.~\cite{Devoto:2023rpv}, we will refer to such partons as $\Fp$ and $\Sp$, and to the relevant  damping factor as $\Delta^{(\Fp\Sp)}$.\footnote{The construction of damping factors is discussed in detail in Appendix B of Ref.~\cite{Devoto:2023rpv}.}
Writing  the \rhs\ in \eq\eqref{eq_dsigmahat_RR_genBorn_starting} as the sum of different contributions, each depending on a specific pair $(\Fp\Sp)$ of potentially unresolved partons, we obtain 
\begin{equation}
    2s_{\inF\inS} \, \dsigmahat^{\genBorn}_{\RR}
    = \sum_{(\Fp\Sp)} \lint \Delta^{(\Fp\Sp)} \THmn \FLMlo{\genBorn}{\Fp,\Sp} \rint \,,
    \qquad n \in \{4,5,6\} \,.
    \label{eq_dsigmahat_RR_genBorn_step_one}
\end{equation}
In the above equation, the sum runs over all unordered $(\Fp\Sp)$ pairs, except in the case of a $q_i \qb_i$ pair, where we consider both the pair $(q_i \qb_i)$ and $(\qb_i q_i)$.  
The function $\THmn$ will be specified shortly.  
For each $\genBorn$, labels $\Fp$ and $\Sp$ may refer to partons of different types, and one has to account for all the possible 
$(\Fp \Sp)$ pairings compatible with $\genBorn$.
In particular, for process $\FgqprocNNLO$, $(\Fp\Sp)$ can be a pair of gluons $(gg)$ or a quark-gluon pair $(q_i g)$; for 
$\SgqprocNNLO$, $(\Fp\Sp)$ can be a (anti)quark-gluon pair $(q_i g)$ or $(\qb_i g)$, or a quark-antiquark pair of the same $(q_i \bar{q}_i)$ or different flavor $(q_i \bar{q}_j)$; and for 
$\TgqprocNNLO$, $(\Fp\Sp)$ must be a fermion-fermion pair $(q_i q_j)$, $(\qb_i \qb_j)$, and $(q_i \qb_j)$. 

For concreteness, we explicitly write  the function $\FLMlo{\genBorn}{\Fp,\Sp}$ required for the process $\FgqprocNNLO$ in \eq\eqref{eq_dsigmahat_RR_genBorn_step_one}. 
The expression reads  
\begin{equation}
\begin{split}
    2s_{\inF\inS} \, \dsigmahat^{\FgqprocNNLO}_{\RR}
    = \lint \Delta^{(\Fp\Sp)} \THmn
    \big( \FLMlo{\FgqprocNNLO}{\Fp_g,\Sp_g}
    + \FLMlo{\FgqprocNNLO}{\Fp_q, \Sp_g} \big) \rint  \,.
    \label{eq_dsigmahat_FgqprocNNLO_RR_example}
\end{split}
\end{equation}
We note that symmetry factors depend on the list of resolved final-state  partons; they are included in the definition of $\FLMlo{\genBorn}{\Fp,\Sp}$
\begin{equation}
\begin{split}
        \FLMlo{\FgqprocNNLO}{\Fp_g,\Sp_g} & = \frac{1}{\Ng!} \FLMun{gq}{\setg{\Ng},\setq{1}}{\Fp_g,\Sp_g} \,, \\
        \FLMlo{\FgqprocNNLO}{\Fp_q,\Sp_g} & = \frac{1}{(\Ng+1)!} \FLMun{gq}{\setg{\Ng+1}}{\Fp_q,\Sp_g} \,.
        \label{eq_FLMA4}
\end{split}
\end{equation}

There are important differences in the singular limits of the $\FLM$ functions that appear in \eq\eqref{eq_dsigmahat_RR_genBorn_step_one}, depending on which partons are resolved and which are unresolved.
If the unresolved partons $(\Fp \Sp)$ are either two gluons or a $q_i \qb_i$ pair, then the double-soft  limit  is singular, whereas if they 
are $qg$, $\bar q g$, $q_i q_k$ etc.,  it is not. 
We will refer to the former combination as $\DS$ and the latter as $\noDS$, so that 
\begin{equation}
    \DS \eqdef \{(gg), (q_i \qb_i),(\qb_i q_i)\} \,,
    \label{eq_DS_def}
\end{equation}
and 
\begin{equation}
    \noDS \eqdef \{(q_i \qb_j), (q_i q_k), (\qb_i \qb_k), (q_i g), (\qb_i g), \text{ with } i\neq j\} \,.
    \label{eq_noDS_def}
\end{equation}
Then, if $(\Fp\Sp) \in \DS$, we introduce an energy ordering by identifying
$\THmn$ with the Heaviside function $\THmn = \Theta(E_\Fp - E_\Sp)$, where $E_{\xa,\yb}$ are the
energies of partons $\xa,\yb$.
Note that the presence of the energy ordering is the reason why we need to consider both $(q_i \qb_i)$ and $(\qb_i q_i)$ pairs.
If, on the other hand, $(\Fp\Sp) \in \noDS$, we do not introduce the energy ordering and take $\THmn = 1$.  
Thus, the proper reading of \eq\eqref{eq_dsigmahat_FgqprocNNLO_RR_example} 
requires associating $\THmn$ with the energy-ordering Heaviside function in the first term in brackets and with $1$ in the second term. 
Furthermore, we extend the convention of \eq\eqref{eq_single_soft_limit_quark_condition} so that $S_{\Fp \Sp} = 0$ for $(\Fp \Sp) \in \noDS$. 

Another important difference between various unresolved  partons concerns  single-soft singular limits, which only exist if at least one parton in the list  $(\Fp\Sp)$ is a gluon. 
To retain  generality of  the calculations, we keep the convention in \eq\eqref{eq_single_soft_limit_quark_condition} that the single-soft operator  makes an expression vanish if 
it is applied to a quark or an antiquark. 
Moreover, if the unresolved partons are a quark and a gluon, we always label them as $(\Fp_q \Sp_g)$, see  \eq\eqref{eq_FLMA4}.

Finally, since we are interested in unresolved contributions that lead to the process $\gqBorn$, 
we need to account for different  collinear limits depending  on the list of unresolved partons. 
For example, in case of the $q_i \qb_i$ final state, we exclude cases where $q_i$ and $\qb_i$ are collinear to different hard partons and only allow limits where they are collinear to each other, as discussed after \eq\eqref{eq_dsigmahat_NNLO_starting_point}. 
In addition, we note that (sequential) double-collinear limits may not be invariant under the exchange $\Fp \leftrightarrow \Sp$ for a generic unresolved final state $(\Fp\Sp)$. 
This possibility was pointed out in Ref.~\cite{Devoto:2023rpv}, although there (for the $gg$ unresolved final state) these limits did in fact commute. 
 
In spite of the differences between the various cases, the computational strategy is always the same and follows closely the discussion in Ref.~\cite{Devoto:2023rpv}.
In particular, as explained in Ref.~\cite{Devoto:2023rpv}, when computing the different contributions to the cross section in \eq\eqref{eq_dsigmahat_NNLO_starting_point}, it is beneficial to combine terms that have \emph{equal} number of \emph{resolved} partons.  
In the double-virtual contribution, the number of resolved partons is the same as in the Born process, i.e.\ $N$, in the real-virtual contribution this number can be $N+1$ or $N$, and in the double-real contribution it can be $N+2, N+1$ or $N$.

We proceed with the discussion  of the double-real contributions, starting from \eq\eqref{eq_dsigmahat_RR_genBorn_step_one}, and aiming at identifying the singular limits that result in a well-defined number of resolved partons. We begin by regularizing the double-soft and single-soft divergences, writing 
\begin{equation}
\begin{split}
    2s_{\inF\inS} \, \dsigmahat_\RR^{\genBorn} 
    = & \sum_{(\Fp\Sp)} \Big[ \lint S_{\Fp\Sp} \Delta^{(\Fp\Sp)} \THmn \FLMlo{\genBorn}{\Fp,\Sp} \rint 
    + \lint \oS_{\Fp\Sp} S_\Sp \Delta^{(\Fp\Sp)} \THmn \FLMlo{\genBorn}{\Fp,\Sp} \rint \\
    & + \lint \oS_{\Fp\Sp} \oS_\Sp \Delta^{(\Fp\Sp)} \THmn \FLMlo{\genBorn}{\Fp,\Sp} \rint \Big] \,,
\label{eq_dsigmahat_RR_genBorn}
\end{split} 
\end{equation}
with $n \in \{4,5,6\}$, and where $\oS_{\Fp\Sp} = 1 - S_{\Fp \Sp}$.
The above formula applies 
to all unresolved partons $(\Fp\Sp)$ and all processes $\genBorn$ due to the definitions of the double-soft operator $S_{\Fp \Sp}$ and the single-soft operator $S_{\Sp}$, explained  above.
The second and the third term on the \rhs\ of \eq\eqref{eq_dsigmahat_RR_genBorn} still contain collinear divergences that  must be regularized. The latter requires particular care since the single- and triple-collinear limits overlap. To overcome this, the nested soft-collinear scheme employs phase-space partitions and sectors~\cite{Czakon:2010td,Czakon:2011ve,Caola:2017dug}. However, these procedures lead to repeated computations of similar integrals, which result in an unwieldy number of  counterterms and obscures physical structures. Therefore, it is convenient to recombine the various limits from the different sectors, and to do so \emph{prior to} evaluating the effect of each limit on the $\FLM$ quantities and integrating over the unresolved phase space. 

The steps required to do so are somewhat tedious and are outlined in  Sec.\ 4 of Ref.~\cite{Devoto:2023rpv}. Extending them to make them applicable to a generic unresolved final state, we write \eq\eqref{eq_dsigmahat_RR_genBorn} as the sum of three contributions, 
\begin{equation}
    2s_{\inF\inS} \, \dsigmahat_\RR^{\genBorn} 
    = \sum_{(\Fp\Sp)} \Big[\SigmaFR^\genBorn[(\Fp\Sp)]_\RR
    + \SigmaSU^\genBorn[(\Fp\Sp)]_\RR
    + \SigmaDU^\genBorn[(\Fp\Sp)]_\RR \Big]\,,
    \qquad n \in \{4,5,6\} \,.
    \label{eq_dsigmahat_RR_genBorn_final}
\end{equation}
The three terms on the \rhs\ of the above equation correspond to the double-real fully-resolved, single-unresolved, and double-unresolved contributions, respectively. 
The fully-resolved term reads
\begin{equation}
    \SigmaFR^\genBorn[(\Fp\Sp)]_\RR
    = \lint \oS_{\Fp\Sp} \oS_\Sp \Omega_1 \Delta^{(\Fp\Sp)} \THmn \FLMlo{\genBorn}{\Fp,\Sp} \rint \,,
    \label{eq_SigmaFR_gqBorn}
\end{equation}
where the collinear regularization operator $\Omega_1$ is defined in \eq(D.5) of Ref.~\cite{Devoto:2023rpv}.
The double-unresolved and single-unresolved contributions in \eq\eqref{eq_dsigmahat_RR_genBorn_final} have slightly different expressions depending on whether $(\Fp\Sp) \in \DS$ or $(\Fp\Sp) \in \noDS$. 
For this reason, we present them separately.

Beginning with the double-unresolved terms, if $(\Fp\Sp) \in \DS$, we write\footnote{With nested angular brackets we indicate that the operations appearing within the innermost brackets
have to be performed first, and then we integrate over the corresponding unresolved degrees of freedom. This result is then acted upon by the outer operator.}
\begin{align}
    &\; \SigmaDU^\genBorn[(\Fp\Sp) \in \DS]_\RR \notag \\
    & = \lint S_{\Fp \Sp} \THmn \FLMlo{\genBorn}{\Fp,\Sp} \rint 
    + \sum_{i \in \HP} \lint \oS_\Fp C_{i\Fp} \Delta^{(\Fp)} \lint S_\Sp \THmn \FLMlo{\genBorn}{\Fp,\Sp} \rint \rint
    \notag \\
    %--------------------------------------------------------------------------------------------------
    & + \sum_{i\in\HP} \lint S_\Sp \lint \oS_\Fp C_{i \Fp} \Delta^{(\Fp\Sp)} \THnm \FLMlo{\genBorn}{\Fp,\Sp} \rint \rint
    + \frac{1}{2} \sum_{i,j \in \HP} \lint \oS_\Sp \oS_\Fp C_{j \Sp} C_{i \Fp} \Delta^{(\Fp\Sp)} \FLMlo{\genBorn}{\Fp,\Sp} \rint 
    \label{eq_SigmaDU_HP_sym_main_formula} \\
    %--------------------------------------------------------------------------------------------------
    & + \sum_{i\in\HP} \frac{N_{\Fp \parallel \Sp}(\ep)}{2} \lint \oS_{\Fp} C_{i\Fp} \Delta^{(\Fp)} \sigma_{i\Fp}^{-\ep} \lint (\iden - 2 \THmn S_\Sp) C_{\Fp \Sp} \FLMlo{\genBorn}{\Fp,\Sp} \rint \rint
    \notag \\
    & + \, \asbr^2 \, 2^{1+2\ep} 
    % c_\Fp 
    \delta_\Fp(\ep) \left(\frac{2\Emax}{\mu} \right)^{\!\!-2\ep} \! \bigg[-\lint \ISoft(\ep) \colorprod \FLM^\gqBorn \rint 
    + \frac{(2\Emax/\mu)^{-2\ep}}{2\ep^2} N_c(\ep) \sum_{i\in\HP} \T_i^2 \lint\FLM^\gqBorn\rint \bigg] 
    \notag \\
    & + \SigmaDU^{\genBorn,\rest}[(\Fp\Sp) \in \DS]_\RR \, , \notag
\end{align}
where the terms collectively denoted as $\SigmaDU^{\genBorn,\rest}[(\Fp\Sp) \in \DS]_\RR$ in the above equation can be found in Eqs (\ref{eq_SigmaDUrest_genBorn_ds} -- \ref{eq_SigmaDU_fin5_genBorn_ds}); they contain single $1/\ep$ poles at most. 
The quantities $N_{\Fp \parallel \Sp}$ and $N_c(\ep)$ are defined in \eq\eqref{eq:normalisation}, the quantities $\delta_\Fp(\ep)$ in \eq\eqref{eq_deltam_deltamperp_def}, and $\sigma_{ij} = \eta_{ij}/(1-\eta_{ij})$, where $\eta_{ij} = (1 -\cos\theta_{ij})/2$.
If $(\Fp\Sp) \in \noDS$, the double-unresolved contribution can be written as
\begin{align}
    &\; \SigmaDU^\genBorn[(\Fp\Sp) \in \noDS]_\RR \notag \\
    & = \sum_{i \in \HP} \lint C_{i\Fp} \Delta^{(\Fp)} \lint S_\Sp \FLMlo{\genBorn}{\Fp,\Sp}\rint \rint
    + \frac{1}{2} \sum_{i,j\in\HP} \lint \oS_\Sp \big(C_{j\Sp} C_{i\Fp} + C_{i\Fp} C_{j\Sp}\big) \Delta^{(\Fp \Sp)} \FLMlo{\genBorn}{\Fp, \Sp} \rint 
    \label{eq_SigmaDU_genSet_unsym_main_formula} \\
    %-------------------------------------------------------------------------------------
    & + \sum_{i \in \HP} N_{\Fp \parallel \Sp}(\ep) \lint C_{i\Fp} \sigma_{i\Fp}^{-\ep} \, \Delta^{(\Fp)} \lint \oS_\Sp C_{\Fp\Sp} \FLMlo{\genBorn}{\Fp,\Sp} \rint \rint 
    + \SigmaDU^{\genBorn,\rest}[(\Fp\Sp) \in \noDS]_\RR \,.
    \notag
\end{align}
Again, terms denoted as $\SigmaDU^{\genBorn,\rest}[(\Fp\Sp) \in \noDS]_\RR$ 
in the above equation have at most $1/\ep$ poles; they can be found in Eqs (\ref{eq_SigmaDUrest_genBorn_nods} -- \ref{eq_SigmaDU_fin3_genBorn_nods}). 
We emphasize once more that the derivation of the above formulas, as well as the notation that we employ, 
follows the discussion in 
Ref.~\cite{Devoto:2023rpv}
with minor modifications which account for the more general nature of the final state $(\xa \yb)$.

Moving on to  single-unresolved terms, if $(\Fp\Sp) \in \DS$, we write 
\begin{align}
    &\; \SigmaSU^\genBorn[(\Fp\Sp) \in \DS]_\RR \notag \\
    %%%%%%%%%%%%%%%%%%%%%%%%%%%%%%%%%%%%%%%%%%%%%%%%%%%%%%%%%%%%
    & = \lint \ONLO^{(\Fp)} \, \Delta^{(\Fp)} \lint S_\Sp \THmn \FLMlo{\genBorn}{\Fp,\Sp} \rint \rint
    + \sum_{i\in\HP} \lint \ONLO^{(\Fp)} (\iden - S_\Sp \THmn) C_{i\Sp} \Delta^{(\Fp \Sp)} \FLMlo{\genBorn}{\Fp,\Sp} \rint \notag \\
    %%%%%%%%%%%%%%%%%%%%%%%%%%%%%%%%%%%%%%%%%%%%%%%%%%%%%%%%%%%%
    & + \frac{1}{2} \lint \ONLO^{(\Fp)} \, \Delta^{(\Fp)} \lint (\iden - 2 \THmn S_\Sp)  C_{\Fp \Sp} \FLMlo{\genBorn}{\Fp,\Sp} \rint \rint \notag \\
    %%%%%%%%%%%%%%%%%%%%%%%%%%%%%%%%%%%%%%%%%%%%%%%%%%%%%%%%%%%%
    & + \sum_{i\in\HP} \llint \ONLO^{(i,\Fp)} \, \partFuncACsp{i} \, \big[(\eta_{i\Fp}/2)^{-\ep} -1\big] (\iden - S_\Sp \THmn) C_{i\Sp} \Delta^{(\Fp \Sp)} \FLMlo{\genBorn}{\Fp,\Sp}\rrint 
    \label{eq_SigmaSU_RR_genBorn_double_soft} \\
    %%%%%%%%%%%%%%%%%%%%%%%%%%%%%%%%%%%%%%%%%%%%%%%%%%%%%%%%%%%%
    & + \sum_{i\in\HP} \frac{1}{2} \llint \ONLO^{(i,\Fp)} \,  \partFuncBD{i} \Delta^{(\Fp)} \big[N_{\Fp \parallel \Sp}(\ep) \sigma_{i\Fp}^{-\ep} - 1\big] \lint (\iden - 2 \THmn S_\Sp) C_{\Fp \Sp} \FLMlo{\genBorn}{\Fp,\Sp} \rint \rrint \notag \\
    %%%%%%%%%%%%%%%%%%%%%%%%%%%%%%%%%%%%%%%%%%%%%%%%%%%%%%%%%%%%
    %
    & + \sum_{i\in\HP} \, \asbr\frac{N_\ep^{(b,d)}}{2} \gamma_{\bot, g \to \Fp\Sp}^{22} \lint \ONLO^{(i,[\Fp\Sp])} \, \partFuncBD{i} \sigma_{i[\Fp\Sp]}^{-\ep} (E_{[\Fp\Sp]}/\mu)^{-2\ep} (r_i^\mu r_i^\nu + g^{\mu\nu}) \Delta^{([\Fp\Sp])} \FLMunmunu{\genBorn}{\Fp\Sp} \rint \notag \\
    & + \sum_{i\in\HP} \, \asbr\frac{N_\ep^{(b,d)}}{2} \gamma_{\bot, g \to \Fp\Sp}^{22, \rmr} \lint \ONLO^{(i,[\Fp\Sp])} \, \partFuncBD{i} \sigma_{i[\Fp\Sp]}^{-\ep} (E_{[\Fp\Sp]}/\mu)^{-2\ep} \Delta^{([\Fp\Sp])} \FLMlo{\genBorn}{\Fp\Sp} \rint \,, \notag
\end{align}
whereas  if $(\Fp\Sp) \in \noDS$, the expression reads
\begin{align}
    &\; \SigmaSU^\genBorn[(\Fp\Sp) \in \noDS]_\RR \notag \\
    & = \lint \ONLO^{(\Fp)} \, \Delta^{(\Fp)} \lint S_\Sp \FLMlo{\genBorn}{\Fp,\Sp} \rint \rint 
    + \lint \OColl^{(\Fp)} \, \Delta^{(\Fp)} \lint \oS_\Sp C_{\Fp\Sp} \FLMlo{\genBorn}{\Fp,\Sp} \rint \rint \notag \\
    & + \sum_{i \in \HP} \llint \OColl^{(i,\Fp)} \, \partFuncBD{i} \big[N_{\Fp \parallel \Sp}(\ep) \sigma_{i\Fp}^{-\ep} - 1\big] \Delta^{(\Fp)} \lint \oS_\Sp C_{\Fp\Sp} \FLMlo{\genBorn}{\Fp,\Sp} \rint \rrint \notag \\
    & + \sum_{i \in \HP} \llint \Big[\ONLO^{(\Sp)} C_{i\Fp} + \OColl^{(\Fp)} \oS_\Sp C_{i\Sp} \Big]\Delta^{(\Fp \Sp)} \FLMlo{\genBorn}{\Fp,\Sp} \rrint
    \label{eq_SigmaSU_RR_genBorn_nodouble_soft} \\
    & + \sum_{i \in \HP} \Big\langle \Big[\ONLO^{(i,\Sp)} \, \partFuncACfp{i} \big[(\eta_{i\Sp}/2)^{-\ep} -1\big] C_{i\Fp} + \OColl^{(i,\Fp)} \, \partFuncACsp{i} \big[(\eta_{i\Fp}/2)^{-\ep} -1\big] \, \oS_\Sp  C_{i\Sp}\Big] \notag \\
    & \times \Delta^{(\Fp \Sp)} \FLMlo{\genBorn}{\Fp,\Sp} \Big\rangle \,. \notag
\end{align}
In Eqs \eqref{eq_SigmaSU_RR_genBorn_double_soft} and \eqref{eq_SigmaSU_RR_genBorn_nodouble_soft}, we have  used the notation 
\begin{equation}
    \ONLO^{(i,x)} = \oS_x \oC_{ix} \,,
    \qquad
    \ONLO^{(x)} = \sum_{i\in\HP} \ONLO^{(i,x)} \, \omega^{xi} \,,
\label{eq_ONLO_def_1}
\end{equation}
where $\omega^{xi}$ are NLO partition functions.  
The functions $\partFuncBD{i}, \partFuncACfp{i}, \partFuncACsp{i}$ originate from the NNLO partition functions $\partFuncNNLO{i}{i}$, evaluated in the limits $\Fp \parallel \Sp$, $i \parallel \Fp$, and $i \parallel \Sp$, respectively.  
The anomalous dimensions $\gamma_{\bot, g \to \Fp\Sp}^{22}$ and $\gamma_{\bot, g \to \Fp\Sp}^{22,\rmr}$ are defined in \eq\eqref{eq:gamma_tilde}, and the vector $r_i^\mu$ is described in Appendix E of Ref.~\cite{Devoto:2023rpv}.  

Although we  do not discuss the derivation of the above formulas for the double-unresolved and single-unresolved cases,  we believe it is useful to  present them to emphasize their proximity  to similar formulas presented in Ref.~\cite{Devoto:2023rpv}
for the $(gg)$ unresolved final state.  
It is this similarity and the appearance of universal structures  that will be key for deriving  formulas for  integrated NNLO subtraction terms for processes 
with an \emph{arbitrary number of jets} at  hadron colliders, a problem that we would like to investigate in the future. 
\\

We continue with the discussion of the real-virtual and double-virtual corrections in \eq\eqref{eq_dsigmahat_NNLO_starting_point}, starting from the former.  
They are analyzed in the same way as the real-emission contribution at NLO, see Sec.~\ref{sec_gq_NLO}.
A generic real-virtual contribution reads 
\begin{equation}
     2s_{\inF\inS} \, \dsigmahat_\RV^\genBorn 
     = \Lint \frac{\FRVlo{gq}{\setg{\Ng},\setq{\Nq},\setqb{\Nqb}, ...}}{\Ng! \times \Nq! \times \Nqb! \times ...} \Rint \,,
     \qquad
     n \in \{1,2,3\} \,,
\end{equation}
where $\FRV$ is defined similarly to $\FLM$ in \eq\eqref{eq_FLM_LO_defn_gen}.
We use damping factors $\Delta^{(\Fp)}$ and write
\begin{equation}
     2s_{ab} \, \dsigmahat_\RV^\genBorn 
     = \sum_\Fp \lint \Delta^{(\Fp)} \FRVlo{\genBorn}{\Fp} \rint \,,
\end{equation}
where  each $\FRVlo{\genBorn}{\Fp}$ function  depends  on a particular potentially-unresolved parton $\Fp$.
We extract soft and collinear singularities following steps discussed in the  NLO calculation, and obtain
\begin{equation}
    2s_{\inF\inS} \, \dsigmahat^{\genBorn}_\RV
    = \sum_\Fp \Big[\SigmaSU^\genBorn[\Fp]_\RV
    + \SigmaDU^\genBorn[\Fp]_\RV \Big]\,,
    \qquad n \in \{1,2,3\} \,,
    \label{eq_disigma_RV_genBorn}
\end{equation}
where
\begin{equation}
\begin{split}
    \SigmaSU^\genBorn[\Fp]_\RV 
    & = \lint \ONLO^{(\Fp)} \Delta^{(\Fp)} \FRVlo{\genBorn}{\Fp} \rint \,, \\ 
    %%%%%%%%%%%%%%%%%%%%%%%%%%%%%%%%%%%%%%%%%%%%%%%%%%%%%%%%%
    \SigmaDU^\genBorn[\Fp]_\RV 
    & = \lint S_\Fp \FRVlo{\genBorn}{\Fp} \rint 
    + \sum_{i\in\HP} \lint \oS_\Fp C_{i\Fp} \Delta^{(\Fp)} \FRVlo{\genBorn}{\Fp} \rint \,.
\end{split}
\label{eq_disigma_RV_genBorn_appo}
\end{equation}
We reiterate that, if parton $\Fp$ is a quark or an antiquark,  we drop the corresponding soft limit $S_{\Fp}$ from \eq\eqref{eq_disigma_RV_genBorn_appo} and
replace $\oS_\Fp$ with the identity operator.

Finally, we consider the double-virtual corrections.
Their IR singularities are universal~\cite{Catani:1998bh, Becher:2009qa, Becher:2013vva}, and the corrections to the Born process $\gqBorn$ are simply given by 
\begin{equation}
    2s_{\inF\inS} \, \dsigmahat_{\gqBorn}^\VV 
    = \lint \FVV^\gqBorn \rint 
    \equiv \SigmaDU^\gqBorn|_\VV \,.
    \label{eq_dsigmahat_VV_gqBorn}
\end{equation}
The function $\FVV$ is defined in a way that is similar to $\FLM$ in \eq\eqref{eq_FLM_LO_defn_gen}.
The explicit expression for \eq\eqref{eq_dsigmahat_VV_gqBorn} can be found in \eq(4.86) in Ref.~\cite{Devoto:2023rpv}. 

%At this point, we observe that 
At this point, we can combine the single-unresolved and double-unresolved terms  in Eqs \eqref{eq_disigma_RV_genBorn} and \eqref{eq_dsigmahat_VV_gqBorn}, originating from the real-virtual and double-virtual corrections, 
with the 
%will combine with the 
corresponding single-unresolved and double-unresolved contributions from the double-real corrections in \eq\eqref{eq_dsigmahat_RR_genBorn_final}.
Together, they should provide a physically-transparent description of ${\rm d} \hat \sigma_{\rm NNLO}^{\gqBorn}$ in \eq\eqref{eq_dsigmahat_NNLO_starting_point}. 
In the remaining sections of this paper, we discuss how  all these different contributions can be combined and simplified, demonstrate the  cancellation of the infrared poles, and derive compact results for  finite remainders.

\section{Simplifications of the double-unresolved contributions}
\label{sec_simplifications_du}

The formulas presented in the previous section 
express the contributions to the NNLO corrections in terms of soft and collinear operators acting on the functions $\FLM$, after the limits from different partitions and sectors have been combined. 
The action of these operators 
on the $\FLM$ quantities leads to universal structures (eikonal functions, collinear splitting functions) and reduced matrix elements that do not depend on the momenta of the unresolved partons.  
Integrating these universal functions, we obtain $1/\ep$ poles that have to disappear when double-virtual, real-virtual, and double-real contributions as well as the PDF renormalization are combined.  
In this section, we simplify the formulas obtained in the previous sections, focusing on  the double-unresolved contributions, which contain the strongest singularities. This preliminary work will enable us to dramatically simplify the demonstration of  the cancellation of 
singularities, which we discuss in Sec.~\ref{sec:poles_cancell}.

In our analysis below, we follow Ref.~\cite{Devoto:2023rpv}, where pure gluonic final states were considered. 
The terms involving virtual contributions and/or soft limits require only minor modifications, and we describe these in \Sec\ref{sec:color_corr}. 
However, as already seen in \Sec\ref{sec_gq_NLO}, the situation is more complex when considering the collinear limits of final states consisting of quarks and gluons, because certain combinations of such limits have to work in unison to produce such physical quantities as  splitting functions and collinear anomalous dimensions. We discuss this in \Sec\ref{sec:double_coll}. 
Understanding how this happens at NNLO is an important step required to extend the analysis of   Ref.~\cite{Devoto:2023rpv} to generic partonic processes.

%%%%%%%%%%%%%%%%%%%%%%%%%%%%%%%%%%%%%%%%%%%%%%%%%%%%%%%%%%%%%%%%%%%%%%%%%%%%%%%%%%%%%%%%%%%%%%%%%%%%%%%%%%%%%%%%%%%%%%%%%%%%%%%%%%%%%%%%%%%%%%%%%%%%%%%%%%%%%%%%%%%%%%

\subsection{Simplifying virtual and soft corrections}
\label{sec:color_corr}

In this section, we discuss the soft and virtual contributions. As stated above, these are very similar to those discussed in Ref.~\cite{Devoto:2023rpv}. There, the treatment of the singular triple-color correlated terms $\sim f_{abc} T_{i}^a T_{j}^b T_{k}^c $  was performed using generic representations of the color charges, and thus can be applied verbatim to the present case. We therefore focus on contributions containing products of the color-charge operators $\T_i \cdot \T_j$ or  $\{\T_i \cdot \T_j, \T_k \cdot \T_l\}$. These  appear in the double-virtual corrections, in the soft and collinear limits of the real-virtual corrections, as well as in the double-soft limit and a combination of the single-soft and hard-collinear limits in the double-real terms. 

The expressions for the double-virtual and the soft limit of the real-virtual contributions can be borrowed from Eqs (4.86, 4.102) in Ref.~\cite{Devoto:2023rpv} with obvious replacements  of $\FLM$ and $\FVV$ 
functions. 
This is possible because only an unresolved gluon contributes to the soft limit in the real-virtual case and because the double-virtual contributions are described by universal formulas that only depend on the momenta 
of hard partons, their color charges, and their collinear anomalous dimensions. 
The double-soft limit can again be borrowed from Ref.~\cite{Devoto:2023rpv}, but we have to supplement that result with the double-soft contribution from the  unresolved $q'\qb'$ pair,  which was calculated  for the nested subtraction scheme in Ref.~\cite{Caola:2018pxp}.  
These additional contributions are proportional to $\nf$, so it is easy to identify and track them. 
The full result  reads
\begin{equation}
\begin{split}
    &\; \lint \DoubSoft{\Fp \Sp} 
    \THmn \big(
    \FLMlo{\FgqprocNNLO}{\Fp_g,\Sp_g}
    + \nf \FLMlo{\SgqprocNNLO}{\Fp_{(q'},\Sp_{\qb')}} \big) \rint \\
    & = \asbr^2 
    \bigg\langle    
    \bigg[ \frac{1}{2}\ISoft^2(\ep)    
    +\left(\frac{\Ca}{\ep^2} c_1(\ep) + \frac{\beta_0}{\ep}  c_2(\ep)
    + {\beta_0} \,c_3(\ep)
    - \frac23 \nf \TR \, c_4(\ep)
    \right) \ISofttilde(2\eps) \bigg] \cdot \FLM^{\gqBorn} 
    \bigg\rangle \\
    & + \asbr^2 \sum_{\inotj} \lint \big[\big(S_{gg,T^2}^{\rm fin}\big)_{ij} + \nf \big(S_{q\qb,T^2}^{\rm fin}\big)_{ij}\big] (\T_i \cdot \T_j) \colorprod \FLM^\gqBorn \rint
     \,,
    \end{split}
\label{eq:DoubSoft_exp}
\end{equation}
where we have introduced the shorthand notation 
\begin{equation}
    \FLMlo{\SgqprocNNLO}{\Fp_{(q'},\Sp_{\qb')}}
    \eqdef
    \FLMlo{\SgqprocNNLO}{\Fp_{q'},\Sp_{\qb'}}
    + \FLMlo{\SgqprocNNLO}{\Fp_{\qb'},\Sp_{q'}} \,,
\end{equation}
to denote the $\qp\qbp$ final state. 
The constant factors $c_{1,2,3}$ in \eq\eqref{eq:DoubSoft_exp} have already been presented in Ref.~\cite{Devoto:2023rpv} and are reported in \eq\eqref{eq:constants_SS}. 
The new coefficient
\begin{equation}
     c_4(\ep) = - \frac{13}6 + \bigg(\frac{125}{18} - \frac{35}{3} \log2 - 12 \log^22 \bigg) \, \ep, 
\end{equation}
 has been extracted from Ref.~\cite{Caola:2018pxp}.
Finally, the last line of \eq\eqref{eq:DoubSoft_exp} contains finite remainders arising from the double soft configurations, which are extracted from Ref.~\cite{Caola:2018pxp}.
The explicit expression of $\big(S_{gg,T^2}^{\rm fin}\big)_{ij}$ and $\big(S_{q\qb,T^2}^{\rm fin}\big)_{ij}$ are given in the ancillary file \finalresult\ provided with  this paper (cf.\ Table \ref{table_ancillary_file}). 
Furthermore, $\beta_0$ in 
Eq.~(\ref{eq:DoubSoft_exp}) 
stands for the  leading-order QCD $\beta$-function. 
In contrast, in Ref.~\cite{Devoto:2023rpv}, $\beta_0$ was always assumed to be evaluated with $\nf=0$ since only 
final states with  gluons were accounted for  in that reference. 

Color-correlated terms in the double-unresolved contributions also arise when soft-limit operators or virtual corrections appear together with the collinear operators. 
The second term on the right-hand side of $\SigmaDU^\genBorn[\Fp]_\RV$ in \eq\eqref{eq_disigma_RV_genBorn_appo}
is the first example of the latter; we will refer to it as the 
hard-collinear limit of the real-virtual contribution.
Similarly to the NLO case, we are only interested in contributions that are relevant for cancelling IR divergences proportional to the matrix element squared for $\gqBorn$; this restricts the number of collinear limits that we have to consider.  
Indeed, for the initial-state collinear limits, we only need to account for the splitting $a_g \rightarrow \Fp_g \, g^*$ and $b_q\rightarrow \Fp_g \, q^*$. 
For the final-state collinear limits, we need to consider the $g^* \rightarrow \Fp_{g} i_g$, $q^* \rightarrow \Fp_{g} i_q$ and $q^* \rightarrow \Fp_{q} i_{g}$ splittings, as well as the branching $g^* \rightarrow \Fp_\qp i_\qbp$ and $g^* \rightarrow \Fp_\qbp i_\qp$. The latter splittings  are relevant for the process $\SgqprocNLO$, which we will collectively indicate as
\begin{equation}
    \FRVlo{\SgqprocNLO}{\Fp_{(q'\qb')}}
    \eqdef
    \FRVlo{\SgqprocNLO}{\Fp_{q'}} + \FRVlo{\SgqprocNLO}{\Fp_{\qb'}} \,.
\end{equation}
In full analogy with Ref.~\cite{Devoto:2023rpv}, we write 
\begin{align}
    &\; \proj \sum_{i \in \HP} \llint \oS_\Fp C_{i \Fp} \Delta^{(\Fp)} 
    \Big(\FRVlo{\FgqprocNLO}{\Fp_{g}}
    + \FRVlo{\FgqprocNLO}{\Fp_{q}}
    + \nf \FRVlo{\SgqprocNLO}{\Fp_{(q'\qb')}} \Big) \rrint  
    \notag \\
    %----------------------------------------------------------------------------------------------------
    & = \asbr^2 \bigg[\lint \IColl(\ep) \IVirt(\ep) \colorprod \FLM^\gqBorn
    \rint 
    + \frac{1}{\ep} \lint \CalPgen_{gg} \conv \big[\IVirt(\ep) \colorprod \FLM^\gqBorn \big] 
    + \big[\IVirt(\ep) \colorprod \FLM^\gqBorn \big] \conv \CalPgen_{qq} \rint \bigg] 
    \notag \\
    %----------------------------------------------------------------------------------------------------
    & + \asbr \bigg[ \lint \IColl(\ep) \colorprod \FLVfin^\gqBorn \rint 
    + \frac{1}{\ep} \lint \CalPgen_{gg} \conv \FLVfin^\gqBorn 
    + \FLVfin^\gqBorn \conv \CalPgen_{qq} \rint \bigg]
    \label{eq_RV_hard_coll_tot} \\
    %----------------------------------------------------------------------------------------------------
    & - \asbr^2 \frac{\Gamma(1-\eps)}{e^{\eps \gamma_E}} \frac{\beta_0}{\ep} 
    \bigg[\lint \IColl(\ep) \colorprod \FLM^\gqBorn \rint 
    + \frac{1}{\ep} \lint \CalPgen_{gg} \conv \FLM^\gqBorn
    + \FLM^\gqBorn \conv \CalPgen_{qq} \rint \bigg] 
    \notag \\
    %----------------------------------------------------------------------------------------------------
    & - \frac{\asbr^2}{\epsilon^2} \hc(\epsilon) 
    \llint \Big[\Ca \IColltilde(2\ep) 
    + 2 \big(\Cf - \Ca \big) \IColltildenf(2\ep) \Big] \colorprod \FLM^\gqBorn \rrint
    \notag \\
    %----------------------------------------------------------------------------------------------------
    & - \frac{\asbr^2}{2\epsilon^3} \Ca \hc(\epsilon) \lint \CalPoneLgen_{gg} \conv \FLM^\gqBorn
    + \FLM^\gqBorn \conv \CalPoneLgen_{qq} \rint \,, 
    \notag
\end{align}
where $\hc(\ep)$ is reported in \eq\eqref{eq:constants_RV}, and $\CalPoneLgen_{\alpha\alpha}$ are the one-loop splitting functions defined in  \eq\eqref{eq_Paa_1L_GEN_definition}.
The collinear operator $\IColltilde(2\epsilon)$ is the generalization of the operator $\IColl$ to the real-virtual case.
It is defined as follows
\begin{equation}
    \IColltilde(2\epsilon) 
    \eqdef
    \sum_{i \in \HP} \frac{\GammaLoop_{i,\fl{i}}(\eps)}{2\epsilon} \,, 
    \label{eq_IC_def}
\end{equation}
where $\GammaLoop_{i,f_i}$ represents the generalized initial- and final-state one-loop anomalous dimension, given in Eqs \eqref{eq_Gamma_1L_ISR_definition}, \eqref{eq_Gamma_1L_FSR_definition} and \eqref{eq_Gamma_1L_q_def_appendix}.  
We note that if previously-introduced quantities are written with the subscript $\nf$, such as $\IColltildenf$ in 
Eq.~(\ref{eq_RV_hard_coll_tot}), only the $\nf$-dependent final-state contributions should be retained in the corresponding functions.

The last contribution that we need to discuss arises from the action of one single-soft and one hard-collinear operator on the double-real cross section (see the second and third terms on the \rhs\ of \eq\eqref{eq_SigmaDU_HP_sym_main_formula}, and the first term on the \rhs\ of \eq\eqref{eq_SigmaDU_genSet_unsym_main_formula}). 
We find
\begin{align}
    &\; \proj \sum_{i \in \HP} \bigg[ \llint \oS_\Fp C_{i\Fp} \Delta^{(\Fp)} \llint S_\Sp \Big(\THmn \FLMlo{\FgqprocNNLO}{\Fp_{g}, \Sp_g}
    + \FLMlo{\FgqprocNNLO}{\Fp_{q}, \Sp_g}
    + \nf \FLMlo{\SgqprocNNLO}{\Fp_{(q'\qb')}, \Sp_g} \Big) \rrint \notag \\
    & + \llint S_\Sp \lint \oS_\Fp C_{i \Fp} \Delta^{(\Fp\Sp)} \THnm  \FLMlo{\FgqprocNNLO}{\Fp_{g}, \Sp_g} \rint \rrint  \bigg] \notag  \\
    %--------------------------------------------------------------------------------------------
    & =  \asbr^2 \bigg[\lint \ISoft(\ep) \IColl(\ep) \colorprod \FLM^\gqBorn \rint
    + \frac{1}{\ep} \lint \CalPgen_{gg} \conv \big[\ISoft(\ep) \colorprod \FLM^\gqBorn \big] 
    + \big[\ISoft(\ep) \colorprod \FLM^\gqBorn \big] \conv \CalPgen_{qq} \rint \bigg]
    \label{eq:single-soft_double-real}\\
    %---------------------------------------------------------------------------------------------
    & + \frac{\asbr^2}{2\ep^3} \Ca \hc(\ep) \lint \CalPGenFour_{gg} \conv \FLM^\gqBorn
    + \FLM^\gqBorn \conv \CalPGenFour_{qq} \rint \,
    \notag \\
    & + \frac{\asbr^2}{\ep^2} \hc(\ep) \Lint 
    \bigg[\Ca \Big(\IColl^{(4)}(\ep) + \!\! \sum_{i \in \HPfq} \!\! \sigma_{i,q \to gq} \Big)
    + (2\Cf-\Ca) \!\!\sum_{i \in \HPfg} \!\! \sigma_{i,g \to q\qb} \Big) \notag \\
    %---------------------------------------------------------------------------------------------
    & + 2(\Cf-\Ca)\ICollFournf(\ep) \bigg]\colorprod \FLM^\gqBorn \rrint \,,
\end{align}
where  
\begin{equation}
\begin{split}
    \sigma_{i,q \to gq}
    ={}& \frac{1}{2\ep} \bigg[e^{-2\ep \Lmax} \, \frac{\Gamma^2(1-\ep)}{\Gamma(1-2\ep)} \, \Gamma_{i,q \to gq} - \Gamma^{(4)}_{i, q \to gq} \bigg]  \, ,
    \\
     \sigma_{i,g \to q\qb}
    ={}& 2\nf \, \frac{1}{2\ep} \bigg[e^{-2\ep \Lmax} \, 
    \frac{\Gamma^2(1-\ep)}{\Gamma(1-2\ep)}  \, \Gamma_{i,g \to q\qb} 
    - \Gamma^{(4)}_{i,g \to q\qb} \bigg] 
    \,.
    \label{eq_sigma_i_def}
\end{split}
\end{equation}
As was pointed out 
in Ref.~\cite{Devoto:2023rpv},
one has to be careful with the 
order of soft and collinear operators  
when computing the hard-collinear limits of the 
double-real contributions. 
At the same time, the fact that in the present study we allow for 
more complex unresolved final states  does not impact the computation  significantly.  
The remaining double-unresolved terms in \eq\eqref{eq_SigmaDU_HP_sym_main_formula} and \eq\eqref{eq_SigmaDU_genSet_unsym_main_formula} that we have yet to analyze and which contribute to $\order{\ep^{-2}}$ involve two collinear limits; we discuss them in the following section.

%%%%%%%%%%%%%%%%%%%%%%%%%%%%%%%%%%%%%%%%%%%%%%%%%%%%%%%%%%%%%%%%%%%%%%%%%%%%%%%%%%%%%%%%%%%%%%%%%%%%%%%%%%%%%%%%%%%%%%%%%%%%%%%%%%%%%%%%%%%%%%%%%%%%%%%%%%%%%%%%%%%%%%

\subsection{Soft-regulated double-collinear contributions}
\label{sec:double_coll}

In this section, we discuss the double-unresolved terms that originate from collinear limits. 
Specifically, we consider the double-collinear limits present in the fourth and fifth terms on the \rhs\ of \eq\eqref{eq_SigmaDU_HP_sym_main_formula}, and in the second and third terms on the \rhs\ of \eq\eqref{eq_SigmaDU_genSet_unsym_main_formula}.  

We begin with the terms that do not involve the $C_{\Fp \Sp}$ limit. We write  
\begin{equation}
    \Sigma_{\rm DC} = 
    \Sigma_{\rm DC}^{\dc}
    + \Sigma_{\rm DC}^{\tc} \,,
\end{equation}
where $\Sigma_{\rm DC}^{\dc}$ contains limits where the unresolved partons $(\Fp\Sp)$ become collinear to two different partons, while in $\Sigma_{\rm DC}^{\tc}$ they become collinear to the same parton.\footnote{The notation ``dc'' and ``tc'' refers to their origin in double-collinear and triple-collinear partitions.} The expressions for these two terms read
\begin{equation}
\begin{split} 
    \Sigma_{\rm DC}^\dc
    = &\; \proj \sum_{\inotj} \llint \oS_\Sp \oS_\Fp C_{j \Sp} C_{i \Fp} \Delta^{(\Fp\Sp)} \Big( 
    \frac{1}{2} \FLMlo{\FgqprocNNLO}{\Fp_g,\Sp_g}
    + \FLMlo{\FgqprocNNLO}{\Fp_q,\Sp_g} \\ 
    & + \nf \FLMlo{\SgqprocNNLO}{\Fp_{(q'\qb')},\Sp_g}
    + \nf \FLMlo{\SgqprocNNLO}{\Fp_{(q'\qb')},\Sp_q}
    + \nf^2 \FLMlo{\TgqprocNNLO}{\Fp_{(q'\qb')},\Sp_{(q''\qb'')}}
    \Big) \rrint \, ,
    \\
    %-----------------------------------------------------------------------------------------
    \Sigma_{\rm DC}^\tc
    = &\; \proj 
    \sum_{i \in \HP} \frac{1}{2} \llint \oS_\Sp \oS_\Fp C_{i \Sp} C_{i \Fp} 
    \Delta^{(\Fp\Sp)} \Big(\FLMlo{\FgqprocNNLO}{\Fp_g,\Sp_g}
    + \nf \FLMlo{\SgqprocNNLO}{\Fp_{(q'},\Sp_{\qb')}} \Big) \rrint \\
    & + \proj \, \sum_{i \in \HP} \frac{1}{2} \llint \oS_\Sp \big( C_{i\Sp} C_{i\Fp} + C_{i\Fp} C_{i\Sp} \big) \Delta^{(\Fp \Sp)} \Big( \FLMlo{\FgqprocNNLO}{\Fp_q, \Sp_g} 
    + \nf \FLMlo{\SgqprocNNLO}{\Fp_{(q'\qb')},\Sp_g} \\
    & + \nf \FLMlo{\SgqprocNNLO}{\Fp_{(q'\qb')},\Sp_q} \Big) \rrint \,,
    \label{eq_NNLO_double_collinear}
\end{split}
\end{equation}
where  $\inotj$ stands for $i,j \in \HP$, with $i \neq j$.
In \eq\eqref{eq_NNLO_double_collinear}, the function $\FLMlo{\TgqprocNNLO}{\Fp_{(q'\qb')},\Sp_{(q''\qb'')}}$ is defined as
\begin{equation}
    \FLMlo{\TgqprocNNLO}{\Fp_{(q'\qb')},\Sp_{(q''\qb'')}} 
    \eqdef
    \FLMlo{\TgqprocNNLO}{\Fp_{(q'\qb')},\Sp_{q''}} 
    + \FLMlo{\TgqprocNNLO}{\Fp_{(q'\qb')},\Sp_{\qb''}} \,.
\end{equation}

We start with the discussion of $\Sigma_{\rm DC}^\dc$.  
Its calculation  follows Ref.~\cite{Devoto:2023rpv},  
and the final expression is equivalent to the product of two NLO-like expressions, since the two collinear limits $C_{i\Fp}$ and $C_{j\Sp}$ (with $i \neq j$)  are completely decoupled from each other.  
Combining the expressions that we obtain from the double-collinear limits of each $\FLM$ function contributing to $\Sigma_{\rm DC}^\dc$ to  reconstruct the quark and gluon anomalous dimensions, defined in Eqs \eqref{eq_Gamma_i_q_FSR_def} and \eqref{eq_Gamma_i_fin_FSR_def}, respectively, we obtain  
\begin{equation}
\begin{split}
    \Sigma_{\rm DC}^\dc
    = &\; \frac{\asbr^2}{\ep^2} 
    \bigg[ \frac12
    \sum \limits_{\inotj} \lint \Gamma_{i,f_i} \Gamma_{j,f_j} \colorprod \FLM^\gqBorn \rint
    + \lint \CalPgen_{gg} \conv \FLM^\gqBorn \conv \CalPgen_{qq} \rint \\
    %------------------------------------------------------------------------------
    & + \sum_{\substack{i \in \HP \\ i \not= \inF}}
    \lint \CalPgen_{gg} \conv \big(\Gamma_{i, f_i} \colorprod \FLM^\gqBorn \big)  \rint 
    + \sum_{\substack{i \in \HP \\ i \not= \inS}} 
    \lint \big(\Gamma_{i, f_i} \colorprod \FLM^\gqBorn \big) \conv \CalPgen_{qq} 
    \rint \bigg] \,. 
\end{split}
\label{eq_double_collinear_inotj}
\end{equation}
We note that the first term contains a factorized contribution of the anomalous dimensions of partons $i$ and $j$. 
To identify the universal operator $\IColl$ and its square, we need to supplement \eq\eqref{eq_double_collinear_inotj} with $i  =j$ double-collinear terms coming from $\Sigma_{\rm DC}^\tc$.

Although the analysis of $\Sigma_{\rm DC}^{\tc}$ is largely analogous to what was done in Ref.~\cite{Devoto:2023rpv}, it is useful to remind the reader that collinear limits for partons emitted off the same line are intertwined.  
However,  in order to write $\Sigma_{\rm DC}^\tc$ in terms of two factorized anomalous dimensions or two splitting functions convoluted in exactly the same way as in the perturbative solution of the Altarelli-Parisi equation, it is useful to disentangle them.

We start with the initial-state radiation.  In this case, only the terms in the first line on the \rhs\ of $\Sigma_{\rm DC}^\tc$ in \eq\eqref{eq_NNLO_double_collinear} give a nonzero contribution, since the remaining terms would involve an initial-state flavor change, and are thus projected out by $\proj$.  
Therefore, the computation of the double-collinear limit in the case $i = \inF$ or $i = \inS$ follows  the procedure described in Ref.~\cite{Devoto:2023rpv}, except that here we also consider the case where the initial-state parton is a gluon.  
As an example, we consider the function $\FLMlo{\FgqprocNNLO}{\Fp_g,\Sp_g}$ in $\Sigma_{\rm DC}^{\tc}$.  
Computing the collinear limit explicitly for $i = \inF$, we obtain  
\begin{equation}
\begin{split}
    \frac{1}{2} \lint \oS_\Fp \oS_\Sp C_{\inF \Fp} C_{\inF \Sp} \Delta^{(\Fp\Sp)} \FLMlo{\FgqprocNNLO}{\Fp_g,\Sp_g} 
    \rint 
    = &\; \frac{\asbr^2}{2\ep^2} \Big[\lint \Gamma_{\inF,g}^2 \colorprod \FLM^\gqBorn \rint 
    + 2 \lint \CalPgen_{gg} \conv \big(\Gamma_{\inF,g} \colorprod \FLM^\gqBorn\big) \rint \\
    & + \lint \PgenoxPgen{gg}{gg} \conv \FLM^\gqBorn \rint 
    + \lint G_{gg} \conv \FLM^\gqBorn \rint \Big] \,.
\end{split}
\label{eq_CanCam_FLMgg_Leg_a_final}
\end{equation}
We note that the first term on the \rhs\ of \eq\eqref{eq_CanCam_FLMgg_Leg_a_final} combines with the first term on the \rhs\ of \eq\eqref{eq_double_collinear_inotj} and contributes to the construction of $\IColl^2$.  
The second contribution in \eq\eqref{eq_CanCam_FLMgg_Leg_a_final} combines with the first term in the second line of \eq\eqref{eq_double_collinear_inotj}, leading to the operator $\IColl$ convoluted with $\CalPgen_{gg}$.  
Next, the term involving $\PgenoxPgen{gg}{gg}$, where the $\barotimes$ convolution is defined as  
\begin{equation}
    f_1(z_1,E) \,\barotimes\, f_2(z_2, E) 
    = 
    \int \dz_1 \dz_2 \, f_1(z_1, E) \, f_2(z_2, z_1 E) \, \delta(z - z_1 z_2) \,,
\label{eq_barotimes_convolution_def}
\end{equation}
is also expected, as it must combine with the contribution $\PAPxPAP{gg}{gg}$ arising from the PDF renormalization.  
Finally, the function $G_{gg}$ in \eq\eqref{eq_CanCam_FLMgg_Leg_a_final} is similar to the 
function $G_i$ introduced in Ref.~\cite{Devoto:2023rpv}. Here, we define it so that  it
can also be applied for an initial-state quark. We 
then write 
\begin{equation}
    G_{\alpha\alpha}(z,E_i) 
    \eqdef 
    \big(\Gamma_{i, \alpha} - \Gamma_{z \cdot i, \alpha} \big) \CalPgen_{\alpha\alpha}(z,E_i) \,,
    \qquad \text{with } ~
    \Gamma_{z \cdot i, \alpha} \eqdef \Gamma_{i,\alpha}\big|_{E_i \mapsto z E_i} \,,
    \label{eq_Gi_IS}
\end{equation}
where $(i,\alpha) \in \{(\inF,g), (\inS,q)\}$. The function in the above equation 
incorporates the modifications required by the non-trivial dependence of the double-collinear phase space of two unresolved partons on their energies.  

A similar calculation has to be performed for the final-state splittings but it is slightly more intricate because, in addition to the phase-space factors that appear in the collinear limits (discussed above), applying those limits to the  damping factors $\Delta^{(\Fp\Sp)}$  leads to energy fractions of final-state unresolved partons, and one needs to keep track of them to see simplified structures appearing in the final result. Although this complication was already present in the calculation reported in Ref.~\cite{Devoto:2023rpv}, it becomes more involved in the current case, as there are now different types of partons in the final state. 
To illustrate this, we consider the double-collinear limit 
\begin{equation}
    \lint \oS_\Fp \oS_\Sp C_{i\Sp} C_{i\Fp} \Delta^{(\Fp\Sp)} \FLMun{\inF\inS}{..., i, ...}{\Fp,\Sp} \rint \,, 
    \qquad
    i \in \HPf \,.
\end{equation}
The action of the first collinear operator clusters partons $i$ and $\Fp$ into a parton $[i\Fp]$, while the second operator combines  $[i\Fp]$ and $\Sp$ to produce parton $[i \Fp \Sp]$. We find
\begin{equation}
    C_{i\Sp} C_{i\Fp} \Delta^{(\Fp \Sp)} \FLMun{\inF\inS}{..., i,...}{\Fp,\Sp}
    = 
    \frac{\gsb^4}{\rho_{i\Fp} \rho_{i\Sp}} \
      \frac{  P_{[i\Fp\Sp][i\Fp]}(z_\Sp) \;
    P_{[i\Fp]i}(z_\Fp) 
    }{E_\Fp E_\Sp E_{[i\Fp]} E_{[i\Fp\Sp]}} \FLMlo{\inF\inS}{... , [i\Fp\Sp], ...} \,,
    \label{eq_double_coll_CinCim_FSR_collinear_limits}
\end{equation}
where 
$E_{[i\Fp]} = E_i + E_\Fp $ and $E_{[i \Fp\Sp]} = E_i + E_\Fp + E_\Sp$ are  the energies of the clustered partons $[i\Fp]$ and $[i \Fp \Sp]$, respectively, 
and we have used  
\begin{equation}
    z_\Fp = \frac{E_i}{E_{[i\Fp]}} \,,
    \qquad
    z_\Sp = \frac{E_{[i\Fp]}}{E_{[i\Fp\Sp]}} \,,
    \qquad
    C_{i\Sp} C_{i\Fp} \Delta^{(\Fp \Sp)} = z_\Sp z_\Fp = \frac{E_i}{E_{[i\Fp\Sp]}} \,,
    \label{eq5.18a}
\end{equation}
in addition to the collinear limits described by the   splitting functions $P_{[i\Fp]i}$ and $P_{[i\Fp\Sp][i\Fp]}$.

The expression 
in Eq.~(\ref{eq_double_coll_CinCim_FSR_collinear_limits})
needs to be integrated over the collinear phase space of the unresolved partons $\Fp, \Sp$, and the resolved parton $i$, keeping the sum of the three energies fixed. We would like to write the result as the product of the splittings 
$\Gamma_{i, \alpha \to \rho\sigma}$ defined in \eq\eqref{eq_Gamma_a_to_bc_defs_appendix}, which will allow us to combine these expressions with those in \eq\eqref{eq_double_collinear_inotj}. 
This does not happen automatically, because the integration over the variable $z_\Fp$, which describes the \emph{internal} splitting, introduces an \emph{additional} factor $z_\Sp^{-2\ep}$ into the integral over the \emph{external} splitting variable $z_\Sp$.
Writing this additional factor as  
\begin{equation}
    z_\Sp^{-2\ep} = 1 + (z_\Sp^{-2\ep} - 1) \,, 
\end{equation}
and integrating over $z_\Sp$, we can obtain the desired product of generalized anomalous dimensions from the first term on the \rhs\ in the above equation, and a new contribution from the second which has a stronger $\ep$-suppression.   
Generalizing the discussion in Ref.~\cite{Devoto:2023rpv}, we write the double-collinear soft-subtracted limit as 
\begin{equation}
\begin{split}
    &\; \lint \oS_\Fp \oS_\Sp C_{i\Sp} C_{i\Fp} \Delta^{(\Fp\Sp)} \FLMun{\inF\inS}{..., i,...}{\Fp,\Sp} \rint \\
    & = \asbr^2 \llint \Big[ \Gamma_{[i\Fp\Sp], [i\Fp] \to i\Fp} \, \Gamma_{[i\Fp\Sp], [i\Fp\Sp] \to [i\Fp]\Sp}
    + \GFSR{[i\Fp\Sp]}{f(z), [i\Fp] \to i\Fp ~~~\;}{\tildef(z), [i\Fp\Sp] \to [i\Fp]\Sp} \Big]  
    \FLMlo{\inF\inS}{..., [i\Fp\Sp],...} \rrint \,,
\end{split}
\label{eq_FSR_double_collinear_general_formula}
\end{equation}
where
\begin{equation}
\begin{split}
    &\; \GFSR{[i\Fp\Sp]}{f(z), [i\Fp] \to i\Fp ~~~\;}{\tildef(z), [i\Fp\Sp] \to [i\Fp]\Sp} 
    \eqdef \; \bigg[\bigg(\frac{2E_{[i\Fp\Sp]}}{\mu}\bigg)^{\!\! - 2\ep} \frac{\Gamma^2(1-\ep)}{\Gamma(1-2\ep)} \bigg]^2 \bigg[\gamma^{22}_{f(z), [i\Fp] \to i \Fp}(L_{[i\Fp\Sp]}) \\ 
    & + \delta_{[i\Fp]i} \frac{\T_{[i\Fp]}^2}{\ep} e^{-2\ep L_{[i\Fp\Sp]}} \bigg] 
    \bigg [\gamma^{42}_{\tildef(z),[i\Fp\Sp] \to [i\Fp]\Sp}(L_{[i\Fp\Sp]}) -  \gamma^{22}_{\tildef(z),[i\Fp\Sp] \to [i\Fp]\Sp}(L_{[i\Fp\Sp]})\bigg ] \,.
\end{split}
\label{eq_GFSR_general_def}
\end{equation}
Definitions relevant for  the two  equations above are collected in Appendix \ref{subsubsec_app_tree_level}.
Note that we introduced two different functions $f(z)$ and $\tildef(z)$ 
to describe the weights that appear in the integrands over energy fractions of the external and internal collinear splittings.  Since these weights arise from 
the collinear limits of $\Delta^{(\xa \yb)}$ factors, they are always equal 
to the relevant energy fraction $z$, c.f.  Eq.~(\ref{eq5.18a}).
However, 
as we pointed out several times, we benefit from combining 
various collinear limits before integrating over energy fractions since in some cases  the weights may disappear. Because of this, we find it convenient to introduce two different weights $f(z),\tilde f(z)$ 
in the definition of $G_\alpha$ in 
Eq.~(\ref{eq_GFSR_general_def}). 
Additionally, we note that in the case where the collinear limits in \eq\eqref{eq_FSR_double_collinear_general_formula} are taken in the order $C_{i\Fp} C_{i\Sp}$ instead of $C_{i\Sp} C_{i\Fp}$, \eq\eqref{eq_FSR_double_collinear_general_formula} remains valid as long as we exchange $\Fp \leftrightarrow \Sp$. 

To account for suitable combinations of the collinear limits, we
introduce the shorthand notation
\begin{equation}
    \GFSR{i}{g \hspace{9mm}}{z, \alpha \to \beta \gamma}
    \eqdef
    \GFSR{i}{z, g \to gg \;}{z, \alpha \to \beta \gamma}
    + 2\nf\, \GFSR{i}{z, g \to q\qb \;}{z, \alpha \to \beta \gamma} \,,
    \label{eq_GFSR_gluon}
\end{equation}
to describe the gluon splittings.
Furthermore, using  the properties of the generalized quark anomalous dimensions, we write
\begin{equation}
    \GFSR{i}{q \hspace{9mm}}{z, \alpha \to \beta \gamma}
    \eqdef  \GFSR{i}{z, q\to qg \;} {z,\alpha \to \beta\gamma}
    +
    \GFSR{i}{z, q\to gq \;} {z,\alpha \to \beta\gamma}
    \equiv 
    \GFSR{i}{1, q\to qg \;}{z,\alpha \to \beta\gamma},
    \quad 
    \label{eq_GFSR_quark}
\end{equation}
where the second equality follows from \eq\eqref{eq:dropping_z}.

 Each contribution to  $\Sigma_{\rm DC}^\tc$ in \eq\eqref{eq_NNLO_double_collinear} can be described using \eq\eqref{eq_CanCam_FLMgg_Leg_a_final} for initial-state radiation and \eq\eqref{eq_FSR_double_collinear_general_formula} for final-state radiation.  
All these terms can then be combined with \eq\eqref{eq_double_collinear_inotj} to obtain the final result for the double-collinear contribution.
It reads 
\begin{align}
    &\; \Sigma_{\rm DC} 
    = \frac{\asbr^2}{2} \Lint \bigg\{\IColl^2 
    + \frac{1}{\ep^2} \bigg[\sum_{i \in \HPfq} \!\! \Gamma_{i, q \to gq} \big(\Gamma_{i,g}-\Gamma_{i,q} \big)    
    + \!\! \sum_{i \in \HPfg} \!\! 2\nf\, \Gamma_{i, g \to q\qb} \big(\Gamma_{i,q}-\Gamma_{i,g}\big) \notag \\
    %%%%%%%%%%%%%%%%%%%%%%%%%%%%%%%%%%%%%%%%%%%%%%%%%%%%%%%%%%%%%%%%%%%%%%%%%%%%%%%%%%%%%
    & + \!\! \sum_{i \in \HPfg} \!\! \big(\GFSR{i}{g \hspace{8mm}}{z, g \to gg} + 2\nf\,\GFSR{i}{q \hspace{8mm}}{z,g \to q\qb} \big)
    + \!\! \sum_{i \in \HPfq} \!\! \big(\GFSR{i}{q \hspace{8mm}}{z,q \to qg} + \GFSR{i}{g \hspace{8mm}}{z,q \to gq} \big) \colorprod \FLM^\gqBorn \bigg\rangle \notag \\
    %%%%%%%%%%%%%%%%%%%%%%%%%%%%%%%%%%%%%%%%%%%%%%%%%%%%%%%%%%%%%%%%%%%%%%%%%%%%%%%%%%%%%
    & + \frac{\asbr^2}{2\eps^2} \Big[
    \lint \convPgenPgen{gg}{gg} \conv \FLM^\gqBorn
    + \FLM^\gqBorn \conv \convPgenPgen{qq}{qq} \rint 
    \label{eq_duouble_collinear_final_expression} \\
    %%%%%%%%%%%%%%%%%%%%%%%%%%%%%%%%%%%%%%%%%%%%%%%%%%%%%%%%%%%%%%%%%%%%%%%%%%%%%%%%%%%%%
    & + \frac{\asbr^2}{2\eps^2} 
    \lint G_{gg} \conv \FLM^\gqBorn
    + \FLM^\gqBorn \conv G_{qq} \rint  \notag \\
    %%%%%%%%%%%%%%%%%%%%%%%%%%%%%%%%%%%%%%%%%%%%%%%%%%%%%%%%%%%%%%%%%%%%%%%%%%%%%%%%%%%%%
    & + \frac{\asbr^2}{\eps} \Big[
    \lint \CalPgen_{gg} \conv \big[\IColl(\ep) \colorprod \FLM^\gqBorn \big] \rint 
    + \lint \big[\IColl(\ep) \colorprod \FLM^\gqBorn \big] \conv \CalPgen_{qq} \rint 
    + \lint \CalPgen_{gg} \conv \FLM^\gqBorn \conv \CalPgen_{qq} \rint \Big] \,. \notag
\end{align}

Finally, we consider those double-collinear terms that involve the operator $C_{\Fp \Sp}$,  i.e.\ the limit where the two unresolved partons $\Fp$ and $\Sp$ become collinear to each other, forming a clustered parton $[\Fp\Sp]$, which then goes collinear to a hard parton $i \in \HP$. Such contributions are found in the fifth term on the \rhs\ of \eq\eqref{eq_SigmaDU_HP_sym_main_formula}, and the third term on the \rhs\ of \eq\eqref{eq_SigmaDU_genSet_unsym_main_formula}.   
Their computation closely follows the discussion in Ref.~\cite{Devoto:2023rpv}. For this reason, we limit ourselves to reporting the final expressions. They  read  
\begin{align}
    &\; \Sigma_{\Fp \parallel \Sp}^{(1)} \notag \\
    & = \proj \sum_{i \in \HP} \frac{N_{\Fp \parallel \Sp}(\ep)}{2} 
    \llint \oS_{\Fp} C_{i\Fp} \, \sigma_{i\Fp}^{-\ep} \Big[(\iden - 2 \THmn S_\Sp) \Delta^{(\Fp)} C_{\Fp \Sp} 
    \big(\FLMlo{\FgqprocNNLO}{\Fp_g,\Sp_g} + \FLMlo{\SgqprocNNLO}{\Fp_{(q'},\Sp_{\qb')}} \big) \Big] \rrint \notag \\
    & = \frac{\asbr^2}{\ep} \Nscbd \Lint \gamma_{g}^{22}(0)
    \bigg(\IColl^{(4)}(\ep) 
    - \!\! \sum_{i \in \HPfq} \!\! \frac{\Gamma_{i,q\to gq}^{(4)}}{2\ep} 
    - \!\! \sum_{i \in \HPfg} \!\! \frac{2\nf \, \Gamma_{i,g\to q\qb}^{(4)}}{2\ep} \bigg) \colorprod \FLM^\gqBorn \Rint
    \label{eq:Sigma1_mn} \\
    & + \frac{\asbr^2}{2\ep^2} \Nscbd 
    \Big\langle \gamma_{g}^{22}(0) \Big[\CalPGenFour_{gg} \conv \FLM^\gqBorn + \FLM^\gqBorn \conv \CalPGenFour_{qq} \Big] \Big\rangle \, , \notag
\end{align}
and  
\begin{align}
    &\; \Sigma_{\Fp \parallel \Sp}^{(2)}
    = \proj \sum_{i \in \HP}  N_{\Fp \parallel \Sp}(\ep) \llint  C_{i\Fp} \sigma_{i\Fp}^{-\ep} \, \Delta^{(\Fp)} \oS_{\Sp} C_{\Fp \Sp} \big(\FLMlo{\FgqprocNNLO}{\Fp_q, \Sp_g} + \nf \FLMlo{\SgqprocNNLO}{\Fp_{(q'\qb')},\Sp_g} \big) \rrint \notag
    \\
    & = \frac{\asbr^2}{\ep} \Nscbd \!\! \sum_{i \in \HPfg} \!\!
    \Lint \bigg[\gamma_{q}^{22}(L_i) \frac{2\nf \, \Gamma_{i,g  \to q\qb}^{(4)}}{2\ep} + \frac{1}{2\ep} H_{i,g \to q\qb} \bigg] \colorprod \FLM^\gqBorn \Rint \label{eq:Sigma2_mn} \\
    & + \frac{\asbr^2}{\ep} \Nscbd \!\! \sum_{i \in \HPfq} \!\!
    \Lint \bigg[\gamma_{q}^{22}(L_i) \frac{\Gamma_{i,q \to gq}^{(4)}}{2\ep} + \frac{1}{2\ep} H_{i,q \to gq} \bigg] \colorprod \FLM^\gqBorn \Rint \,. \notag
\end{align}
In the above equations, we have used 
\begin{equation}
    \begin{aligned}
        \gamma_g^{22}(L_i) & \eqdef \gamma_{z,g\to gg}^{22}(L_i) + 2\nf\,\gamma_{z,g\to q\qb}^{22} \,, \\
        \gamma_q^{22}(L_i) & \eqdef \gamma_{z,q\to qg}^{22}(L_i) + \gamma_{z,q\to gq}^{22} \equiv \gamma_{1,q\to qg}^{22}(L_i) \,,
    \end{aligned}
    \qquad
    \text{with} ~ \gamma_{f_i}^{22}(0) \equiv \gamma_{f_i}^{22}(L_i = 0) \,,
    \label{eq_gamma_22_q_g_def}
\end{equation}
and 
\begin{equation}
\begin{split}
    H_{i,q \to gq} & \eqdef \frac{\T_q^2}{\ep} \bigg[e^{-2\ep L_i} \Gamma_{i,q\to gq}^{(4)} - e^{-2\ep \Lmax} \frac{\Gamma^2(1-\ep)}{\Gamma(1-2\ep)} \Gamma_{i,q\to gq} \bigg] \,, \\
    %%%%%%%%%%%%%%%%%%%%%%%%%%%%%%%%%%%%%%%%%%%%%%%%%%%%%%%%%%%%%%
     H_{i,g \to q\qb} & \eqdef 2\nf \, 
    \frac{\T_q^2}{\ep} \bigg[e^{-2\ep L_i} \Gamma_{i,g\to q\qb}^{(4)} - e^{-2\ep \Lmax} \frac{\Gamma^2(1-\ep)}{\Gamma(1-2\ep)} \Gamma_{i,g\to q\qb}
    \bigg] \,.
\end{split}
\label{eq_H_qTOgq_FSR_def}
\end{equation}

We have now analyzed all the double-unresolved terms in 
\eq\eqref{eq_SigmaDU_HP_sym_main_formula} and \eq\eqref{eq_SigmaDU_genSet_unsym_main_formula} which produce singularities of $\order{\ep^{-2}}$ or stronger, and have written the corresponding expressions in terms of the virtual, soft, and collinear operators $I_{\rm V,S,C}$. In the following section, we will employ these results to discuss the cancellation of infrared singularities.

\section{Cancellation of poles}
\label{sec:poles_cancell}

In this section, we use the results obtained in the previous section to demonstrate the cancellation of infrared singularities in the NNLO QCD corrections to the process $\gqBorn$. We have pointed out  in Ref.~\cite{Devoto:2023rpv} that this task is simplified significantly if  one combines virtual, soft, and collinear operators $\IVirt$, $\ISoft$ and $\IColl$ into a single color-correlated quantity $\ITot$ from \eq\eqref{eq_NLO_IT_first_process_def}, which is $\ep$-finite. 
Indeed, identifying this operator in the NNLO contribution to cross sections allowed us to cancel the poles through $1/\ep^2$ ``by eye''.
Furthermore, we  observed that combining the highest-order singularities into the $\ITot$ operator in two subsequent  orders of QCD perturbation theory lead to an iterative structure  consistent with the exponentiation 
\begin{equation}
    \dsigmahat \sim  \lint  e^{\asbr \ITot} \colorprod \FLM \rint \,.
    \label{eq:exponentiation}
\end{equation}
This, together with the fact that we were able to identify the $\ITot$ operator for the more complex process that we consider in this paper (cf.\ Section~\ref{sec_gq_NLO}), suggests that it would be beneficial to discuss the cancellation of $1/\ep$ poles by rewriting ${\rm d}\hat{\sigma}^{\rm NNLO}$ using the $\ITot^2$ and $\ITot$ operators, as far as possible.
We discuss this in the following sections, beginning with the single-unresolved contributions, and then turning to the double-unresolved ones.

%%%%%%%%%%%%%%%%%%%%%%%%%%%%%%%%%%%%%%%%%%%%%%%%%%%%%%%%%%%%%%%%%%%%%%%%%%%%%%%%%%%%%%%%%%%%%%%%%%%%%%%%%%%%%%%%%%%%%%%%%%%%%%%%%%%%%%%%%%%%%%%%%%%%%%%%%%%%%%%%%%%%%%

\subsection{Single-unresolved terms}
\label{sec:single_unr}

In this section, we discuss the cancellation of singularities for the single-unresolved contribution, and highlight the need to identify  the proper combinations of quark collinear splittings to simplify the calculation. The relevant expressions can be found in Eqs \eqref{eq_SigmaSU_RR_genBorn_double_soft}, \eqref{eq_SigmaSU_RR_genBorn_nodouble_soft}, and \eqref{eq_disigma_RV_genBorn_appo}. We begin by considering contributions 
which are proportional to the operator $\ONLO^{(\Fp)}$, and 
 first study those  terms which do not involve collinear operators, namely the real-virtual corrections (see $\SigmaSU^\genBorn[\Fp]_\RV$ in \eq\eqref{eq_disigma_RV_genBorn_appo}) and the soft limit of the double-real ones (see the first terms in Eqs.~\eqref{eq_SigmaSU_RR_genBorn_double_soft}
and~\eqref{eq_SigmaSU_RR_genBorn_nodouble_soft}). As before, the unresolved parton in the former contribution can be either a quark or a gluon, while the pair of unresolved partons in the latter contribution can be either $(\Fp_g\Sp_g)$ or $(\Fp_q\Sp_g)$.
Their sum evaluates to 
\begin{align}
    &\; \llint \ONLO^{(\Fp)} \Delta^{(\Fp)} \Big[
    \FRVlo{\FgqprocNLO}{\Fp_g}
    + \FRVlo{\FgqprocNLO}{\Fp_q} 
    + S_\Sp \Big(\THmn\FLMlo{\FgqprocNNLO}{\Fp_g,\Sp_g} 
    + \FLMlo{\FgqprocNNLO}{\Fp_q,\Sp_g} \Big) \Big] \rrint \notag \\
    & = \asbr \lint \ONLO^{(\Fp)} \Delta^{(\Fp)} \big[\IVirt + \ISoft(E_{\Fp}) 
    \big] \colorprod \FLMlo{\FgqprocNLO}{\Fp_g} \rint 
    \label{eq6.2} \\
    & + \asbr \lint \ONLO^{(\Fp)} \Delta^{(\Fp)} \big[\IVirt + \ISoft(\Emax) \big] \colorprod \FLMlo{\FgqprocNLO}{\Fp_q} \rint 
    + \lint \ONLO^{(\Fp)} \Delta^{(\Fp)} \big(\FRVfinlo{\FgqprocNLO}{\Fp_g} + \FRVfinlo{\FgqprocNLO}{\Fp_q}\big) \rint \,, \notag
\end{align}
where $\FRVfinlo{\FgqprocNLO}{\Fp}$ is defined analogously to $\FLVfin$.
We note that the $\IVirt$ and $\ISoft$ operators in \eq\eqref{eq6.2} act on the $(N+1)$ hard partons of functions $\FLMlo{\FgqprocNLO}{\Fp_g}$ and $\FLMlo{\FgqprocNLO}{\Fp_q}$.
Additionally, the energy dependencies of the two $\ISoft$ operators appearing on the right-hand side of the above equation are different.  
This is due to the presence of the energy-ordering function $\THmn$ in the case of the two-gluon unresolved final state, and its absence in the case of the quark-gluon one.  We note that 
$\ISoft(E_\Fp)$ and $\ISoft(\Emax)$ are defined in \eq\eqref{eq_ISoft_definition_appendix}, where in $\ISoft(E_\Fp)$ the substitution $\Emax \mapsto E_\Fp$ is applied.  

In order to combine the terms in \eq\eqref{eq6.2} into $\ITot$ operators, we 
require collinear operators $\IColl$.
These will arise from the second and the third term in \eq\eqref{eq_SigmaSU_RR_genBorn_double_soft} if $(\Fp\Sp) \in \DS$, and from the second and the fourth term in \eq\eqref{eq_SigmaSU_RR_genBorn_nodouble_soft} if $(\Fp\Sp) \in \noDS$.
Such  terms originate  from different  configurations of unresolved partons, but when combined they produce anomalous dimensions that appear in the collinear operator 
$\IColl$ for a  process ${\cal A}_0$  with an additional jet.

To show this, we first write the required contribution for 
the sum of the $gg$ and $q'\qb'$ unresolved final states, i.e., for $(\Fp\Sp) \in \DS$.
The result, which can be easily obtained by generalizing \eq(5.3) in Ref.~\cite{Devoto:2023rpv}, reads
\begin{equation}
\begin{split}
    &\; \proj \Lint \ONLO^{(\Fp)} 
    \bigg[ \sum_{i \in \HP}(\iden - S_\Sp \THmn) C_{i\Sp} \Delta^{(\Fp \Sp)} 
    + \frac{1}{2} \Delta^{(\Fp)} (\iden - 2 S_\Sp \THmn) C_{\Fp \Sp}
    \bigg] \FLMlo{\FgqprocNNLO}{\Fp_g,\Sp_g} \\
    & + \frac{\nf}{2} \ONLO^{(\Fp)} \Delta^{(\Fp)} C_{\Fp\Sp} \FLMlo{\SgqprocNNLO}{\Fp_{(q'},\Sp_{\qb')}} \Rint \\
    & = \frac{\asbr}{\ep} \lint \ONLO^{(\Fp)} \Delta^{(\Fp)} \big(
    \CalPgen_{gg} \conv \FLMlo{\FgqprocNLO}{\Fp_g} 
    + \FLMlo{\FgqprocNLO}{\Fp_g} \conv \CalPgen_{qq}\big) \rint \\
    & + \asbr \Lint \ONLO^{(\Fp)} \Delta^{(\Fp)} 
    \bigg(\IColl(E_{\Fp})
    - \!\! \sum_{i \in \HPfg} \!\! \frac{2\nf\, \Gamma_{i,g \to q\qb}}{\ep}
     - \!\! \sum_{i \in \HPfq} \!\! \frac{\Gamma_{i,q \to gq}}{\ep} 
     \bigg)
     \colorprod \FLMlo{\FgqprocNLO}{\Fp_g} \Rint \,. 
\end{split}
\label{eq_Sigma_N+1^div_2nd_3rd_lines_gg_def}
\end{equation}
We note that the collinear operator $\IColl(E_{\Fp})$ in the above formula 
includes the contribution of the parton $\Fp$ and, similarly to $\ISoft(E_\Fp)$ discussed above, is computed with the replacement $\Emax \mapsto E_\Fp$. 

For $(\Fp\Sp) \in \noDS$, we find
\begin{align}
    &\; \proj  \sum_{i \in \HP}
    \bigg\{
    \llint \Big[\ONLO^{(\Sp)} C_{i\Fp} + \OColl^{(\Fp)} \oS_\Sp C_{i\Sp} \Big]\Delta^{(\Fp \Sp)} \FLMlo{\FgqprocNNLO}{\Fp_q,\Sp_g}
    \rrint
    + \asbr \lint \OColl^{(\Fp)} \Delta^{(\Fp)} \lint C_{\Fp\Sp} \FLMlo{\FgqprocNLO}{\Fp_q} \rint \rint
    \nonumber \\
    & + \llint \nf \, \ONLO^{(\Sp)} \Coll{i\Fp} \Delta^{(\Fp \Sp)} 
    \FLMlo{\SgqprocNNLO}{\Fp_{(q'\qb')},\Sp_g}
    + \nf \, \ONLO^{(\Fp)} \Coll{i\Sp} \Delta^{(\Fp \Sp)}\, \FLMlo{\SgqprocNNLO}{\Fp_{(q'\qb')},\Sp_q} \rrint \bigg\} \nonumber \\
    & = \frac{\asbr}{\ep} \llint \ONLO^{(\Fp)} \, \Delta^{(\Fp)} \Big[ 
    \CalPgen_{gg} \conv \FLMlo{\FgqprocNLO}{\Fp_q} + \FLMlo{\FgqprocNLO}{\Fp_q} \conv \CalPgen_{qq} \Big] \rrint + \asbr \Lint \ONLO^{(\Fp)} \Delta^{(\Fp)}
    \label{eq_Sigma_N+1^div_2nd_3rd_lines_qg_def} \\
    & \times \bigg[
        \IColl(\Emax) \colorprod \FLMlo{\FgqprocNLO}{\Fp_q} 
        + \bigg(\sum_{i \in \HPfg} \!\! \frac{2\nf\, \Gamma_{i,g \to q\qb}}{\ep}
        + \!\! \sum \limits_{i \in \HPfq} \!\!
        \frac{\Gamma_{i,q \to gq}}{\ep}
        \bigg) \colorprod \FLMlo{\FgqprocNLO}{\Fp_g} \bigg] \Rint \,. \notag
\end{align}
Upon comparing Eqs \eqref{eq_Sigma_N+1^div_2nd_3rd_lines_gg_def} and \eqref{eq_Sigma_N+1^div_2nd_3rd_lines_qg_def}, it becomes apparent that \emph{in their sum}, the anomalous dimensions that do not combine into an $\IColl$ operator cancel.  
Furthermore, the two  collinear operators that appear carry the  appropriate energy dependence needed to reconstruct $\ITot(E_{\rm max}) $
and $\ITot(E_{\xa})$.  
  
The remaining $1/\ep$ divergences that reside in the generalized splitting functions $\CalPgen_{gg}$ and $\CalPgen_{qq}$ in Eqs~\eqref{eq_Sigma_N+1^div_2nd_3rd_lines_gg_def} and~\eqref{eq_Sigma_N+1^div_2nd_3rd_lines_qg_def} are removed by $\ONLO$-dependent terms that arise when PDF renormalization is applied to the finite remainder of the NLO cross section. 
Combining the two, we arrive at an $\ep$-finite expression
\begin{equation}
\begin{split}
    % \SigmaSU^{\fin,(4)}|_\gqBorn
    %
    &\; \asbr \llint  \ONLO^{(\Fp)} \Delta^{(\Fp)} \Big[ \ITot(E_\Fp) \colorprod \FLMlo{\FgqprocNLO}{\Fp_g} 
    + \ITot(\Emax) \colorprod \FLMlo{\FgqprocNLO}{\Fp_q} \\
    & + \PNLO_{gg} \conv \FRlo{\FgqprocNLO}{\Fp}   + \FRlo{\FgqprocNLO}{\Fp} \conv \PNLO_{qq} 
  + \FRVfinlo{\FgqprocNLO}{\Fp_g} + \FRVfinlo{\FgqprocNLO}{\Fp_q} 
    \Big ] \rrint \,. 
\end{split}
\end{equation}
In the above equation, $\FRlo{\FgqprocNLO}{\Fp}$ is defined in \eq\eqref{eq_NLO_FR_second_first_def}, and the generalized collinear 
splitting functions $\PNLO_{gg}$ and $\PNLO_{qq}$ have already appeared in the discussion of the NLO corrections.

The remaining single-unresolved terms can easily be identified as being $\ep$-finite. For example, the fourth and fifth term of \eq\eqref{eq_SigmaSU_RR_genBorn_double_soft}, which are acted on by the subtraction operator $\ONLO^{(i,\Fp)}$, contain a hard-collinear limit that produces a $1/\ep$ pole, but also a prefactor of $\order{\ep}$. The same argument applies to the third, fifth, and sixth terms of \eq\eqref{eq_SigmaSU_RR_genBorn_nodouble_soft}, while the final two lines of \eq\eqref{eq_SigmaSU_RR_genBorn_double_soft} are manifestly $\ep$-finite.
The manipulations required to bring such terms to their final form 
are therefore minimal. 
The complete expression for the single-unresolved contribution to $\dsigmahat_\NNLO^\gqBorn$ can be found in \eq\eqref{eq_dsigmahat_procgq_SU_final}.

%%%%%%%%%%%%%%%%%%%%%%%%%%%%%%%%%%%%%%%%%%%%%%%%%%%%%%%%%%%%%%%%%%%%%%%%%%%%%%%%%%%%%%%%%%%%%%%%%%%%%%%%%%%%%%%%%%%%%%%%%%%%%%%%%%%%%%%%%%%%%%%%%%%%%%%%%%%%%%%%%%%%%%

\subsection{Double-boosted contribution}
Having discussed the treatment of singularities arising in the single-unresolved contributions, we now turn to those terms in which two partons are unresolved. We find it convenient to organize them according to the number of initial-state partons which are boosted, and we begin with the double-boosted terms. These arise when soft-subtracted collinear operators act on \emph{both} incoming partons,  
 leading to double convolutions of generalized splitting functions with $\FLM^{\gqBorn}$. 
The relevant term appears as the last entry in \eq(\ref{eq_duouble_collinear_final_expression}).
Further terms of this type arise from the collinear renormalization of parton distribution functions.  
Combining such  contributions, we find 
\begin{equation}
\begin{split}
    \SigmaDU^\db = &\; \frac{\asbr^2}{\ep^2} \Big[\lint \CalPgen_{gg} \conv \FLM^\gqBorn \conv \CalPgen_{qq} \rint
    + \cTilde^2 \lint \PAP_{gg} \conv \FLM^\gqBorn \conv \PAP_{qq} \rint \\
    & +  \cTilde \lint \PAP_{gg} \conv \FLM^\gqBorn \conv \CalPgen_{qq} \rint 
    + \cTilde \lint \CalPgen_{gg} \conv \FLM^\gqBorn \conv \PAP_{qq}\rint\Big] \\
    %---------------------------------------------------------------------------------------------
    = &\; \asbr^2 \lint \PNLO_{gg} \conv \FLM^\gqBorn \conv \PNLO_{qq} \rint 
    + \order{\ep} \,,
\end{split}
\label{eq_double_boosted_expansion}
\end{equation}
where  $\cTilde = \Gamma(1-\ep)/e^{\ep \gamma_\rmE}$.
To derive the above result, we used the relation between $\CalPgen_{\alpha\beta}$, $\PAP_{\alpha\beta}$, and $\PNLO_{\alpha\beta}$, which we already presented in \eq\eqref{eq_Pgen_PAP_PNLO_relation} in the context of NLO corrections.

%%%%%%%%%%%%%%%%%%%%%%%%%%%%%%%%%%%%%%%%%%%%%%%%%%%%%%%%%%%%%%%%%%%%%%%%%%%%%%%%%%%%%%%%%%%%%%%%%%%%%%%%%%%%%%%%%%%%%%%%%%%%%%%%%%%%%%%%%%%%%%%%%%%%%%%%%%%%%%%%%%%%%%

\subsection{Single-boosted contributions}
\label{sec:NNLO-single-boosted}

We continue with the analysis of the single-boosted contributions. It is straightforward to identify terms proportional to a boost,  convoluted with the operators $\IVirt$, $\ISoft$, and $\IColl$ in Eqs~\eqref{eq_RV_hard_coll_tot}, \eqref{eq:single-soft_double-real}, and \eqref{eq:Sigma1_mn}, respectively, and to combine these into a boost acting on the IR-finite operator $\ITot$.  Similarly, terms proportional to the finite remainder of the one-loop amplitude $\FLVfin^\gqBorn$ in \eq\eqref{eq_RV_hard_coll_tot} can be combined with terms originating from the PDF renormalization to obtain an IR-finite result, as happens at NLO.
We note that this renders finite \emph{all} contributions with color correlations.

We therefore concentrate on the remaining terms proportional to $\FLM^{\gqBorn}$ that cannot be absorbed in $\ITot$, and consider those that exhibit $\order{\ep^{-3}}$ and $\order{\ep^{-2}}$ singularities. 
We combine the relevant terms from Eqs \eqref{eq_RV_hard_coll_tot}, \eqref{eq:single-soft_double-real}, \eqref{eq_duouble_collinear_final_expression}, and \eqref{eq:Sigma1_mn}, with terms coming from the renormalizations of the PDFs.\footnote{We point out that contributions proportional to one-loop splitting functions that appear in the PDF renormalization only affect the single $1/\ep$ pole, therefore we do not discuss them in this section.} For simplicity, we report below the contribution associated with the boost of the initial-state parton $\inF_g$, which reads
\begin{equation}
\begin{split}
    &\; \Sigma_{{\rm DU}, \order{\ep^{-2}}}^{\inFsb}
    = \frac{\asbr^2}{2\ep^3} \llint \Big[\Ca \hc(\ep) \big(\CalPGenFour_{gg} 
    - \CalPoneLgen_{gg} \big) 
    + \ep \, G_{gg} \Big] \conv \FLM^\gqBorn \rrint \\
    & + \frac{\asbr^2}{2\ep^2} \bigg\{\llint \Big[N^{(b,d)}_{\rm sc}
    \gamma_{g}^{22}(0) \, \CalPGenFour_{gg} 
    - 2\cTilde \beta_0 \, \CalPgen_{gg}
    - \cTilde^2 \beta_0 \PAP_{gg} \Big] \conv \FLM^\gqBorn \rrint \\
    & + \llint \Big[\convPgenPgen{gg}{gg} 
    + 2 \cTilde \PAPoxPgen{gg}{gg} 
    + \cTilde^2 \PAPxPAP{gg}{gg} \\
    & + 2\nf\Big(\convPgenPgen{qg}{gq}
    + 2\cTilde \PAPoxPgen{qg}{gq}
    + \cTilde^2 \PAPxPAP{gq}{qg} \Big) \Big] 
    \conv \FLM^\gqBorn \rrint \bigg\} \,.
\end{split}
\label{eq:Sigma_DU_sb}
\end{equation} 

To discuss the $1/\ep$-divergences in \eq\eqref{eq:Sigma_DU_sb}, we need to expand 
various quantities that appear there in  powers of $\ep$.
Using the definitions of $\CalPGenFour_{\alpha\alpha}$ and $\CalPoneLgen_{\alpha\alpha}$ in Eqs \eqref{Eq:Paa_GEN_definition_k_general_appendix} and \eqref{eq_Paa_1L_GEN_definition}, respectively, we obtain 
\begin{equation}
\begin{split}
    & \Ca \hc(\ep) \Big(\CalPGenFour_{\alpha \alpha}(z,E) - \CalPoneLgen_{\alpha \alpha}(z,E) \Big) = 2\T_{\alpha}^2 \ep \log(z) \PAP_{\alpha \alpha}(z) + \order{\ep^2} \,,
    \\
    & \ep \, G_{\alpha\alpha}(z, E) = - 2\T_{\alpha}^2 \ep \log(z) \PAP_{\alpha \alpha}(z) + \order{\ep^2} \,. 
    \label{eq_expansion_G_i_ISR}
\end{split}
\end{equation}
This equation, which is valid for both $\alpha = q$ and $g$, 
implies  that, in spite of its appearance,  the first  line in \eq\eqref{eq:Sigma_DU_sb} is only $\order{\ep^{-1}}$. 

To show that the second line in \eq\eqref{eq:Sigma_DU_sb} is at most $\order{\ep^{-1}}$, it is sufficient to use the expansions 
\begin{equation}
    \gamma_g^{22}(0) = \beta_0  + \order{\ep} \,,
    \qquad 
    \cTilde = 1 + \order{\ep^2} \,, 
    \qquad
    \Nscbd = 1 + \order{\ep} \,,
    \label{eq:relations}
\end{equation}
and also (cf.~\eq\eqref{Eq:Paa_GEN_definition_k_general_appendix})
\begin{equation}
    \calP_{\alpha \beta}^{(k),\text{gen}} = - \PAP_{\alpha \beta} + \order{\ep} \,,
    \qquad
    k = 2,4 \,,
    \label{eq:relations_appo}
\end{equation}
which holds for all indices $(\alpha\beta)$.
Turning to the last two lines in \eq\eqref{eq:Sigma_DU_sb}, we can use Eqs~\eqref{eq:relations} and~\eqref{eq:relations_appo} to write
%To show the cancellation of the $1/\ep^2$ pole, we replace $\cTilde$ with 
%$1$ in virtue of \eq\eqref{eq:relations}, and we use \eq\eqref{eq:relations_appo} to write
\begin{equation}
\begin{split}
    &\; \convPgenPgen{\alpha\beta}{\gamma\delta} + 2\cTilde \PAPoxPgen{\alpha\beta}{\gamma\delta} + \cTilde^2 \PAPxPAP{\alpha\beta}{\gamma\delta} \\
    = &\; \PAPoxPAP{\alpha\beta}{\gamma\delta} - 2 \PAPoxPAP{\alpha\beta}{\gamma\delta} + \PAPxPAP{\alpha\beta}{\gamma\delta} + \order{\ep} \\
    = &\; \order{\ep} \,,
\end{split}
\label{eq_convolution_cancellation}
\end{equation}
for any flavour-pair $\alpha\beta$ and $\gamma\delta$. 
Note that the last step in \eq\eqref{eq_convolution_cancellation} follows from  
the fact that $\PAP$ is energy-independent, which implies that the   
$\barotimes $ and $\otimes$ convolutions are identical (cf.\ \eq\eqref{eq_barotimes_convolution_def}).  
We have therefore demonstrated that the single-boosted term in \eq\eqref{eq:Sigma_DU_sb} has at most poles of $\order{\ep^{-1}}$, and we will return to the remaining poles of this order at the end of this section. Moreover, these simple poles do not contain color correlations; all singularities multiplying products of color charges have already been combined into $\ITot$, as discussed at the beginning of this section.

The contribution with the boost applied to the initial-state parton $b_q$ can be obtained from \eq\eqref{eq:Sigma_DU_sb} by replacing $q \leftrightarrow g$ in the indices of the initial-state splitting functions, and by removing the final line. 
This second action is required because, as mentioned earlier, we do not consider flavor-singlet configurations in this paper.
Since the relations in Eqs~\eqref{eq_expansion_G_i_ISR},~\eqref{eq:relations_appo}, and~\eqref{eq_convolution_cancellation} are applicable for all values of the indices $\alpha$ and $\beta$, the above discussion can be repeated  verbatim to demonstrate the cancellation of the poles through $\order{\ep^{-2}}$ for this piece too.

\subsection{Unboosted  terms }
Finally, we consider  the unboosted  terms. Many of these originate from virtual corrections and/or a soft gluon emission, and such contributions include color-correlated matrix elements squared with quadratic  $\T_i \cdot \T_j$,  triple  $T^a_i T^b_j T^c_k$ and quartic  $\{\T_i \cdot \T_j, \T_k \cdot \T_l\}$ products of color charges.

We first comment on the triple-color  correlators.  
In Ref.~\cite{Devoto:2023rpv}, we have shown that they originate from three distinct sources: 
i) the soft limit of the real-virtual contributions \cite{Catani:1999ss}, 
ii) commutators of the  soft $\ISoft$ and virtual $\IVirt$ operators, and iii) double-virtual corrections~\cite{Catani:1998bh,Becher:2009qa,Gardi:2009qi}. 
All divergent triple-color correlated contributions cancel when terms from the three sources are combined. 
Since the calculations in Ref.~\cite{Devoto:2023rpv} employed generic representation of 
color charges, they can be applied verbatim to the discussion 
of such contributions in the NNLO corrections to  $\gqBorn$; hence, we do not discuss them further.   
The remaining color-correlated terms arise from various virtual corrections and/or soft limits, and are captured in the operators $\ISoft$ and $\IVirt$ in Eqs~\eqref{eq:DoubSoft_exp},~\eqref{eq_RV_hard_coll_tot}, and~\eqref{eq:single-soft_double-real}.\footnote{We also need to include the double-virtual contribution and the soft limit of the real-virtual correction, the latter of which we borrow from Ref.~\cite{Devoto:2023rpv}.} 
The terms proportional to the finite remainder of the one-loop amplitude $\FLVfin^\gqBorn$ can easily be combined into the finite operator $\ITot$.  
We focus on the remaining terms proportional to $\FLM^{\gqBorn}$, and combine these with the collinear operators in Eqs~\eqref{eq_duouble_collinear_final_expression},~\eqref{eq:Sigma1_mn}, and~\eqref{eq:Sigma2_mn} to reconstruct, as far as possible, the IR finite operator $\ITot$.  
Again, we discard the terms that lead only to $\order{\ep^{-1}}$ poles for the time being, and write the unboosted contribution as a sum of two terms 
\be
    \Sigma_{\rm DU, \order{\ep^{-2}}}^{\rm ub}
     = \Sigma_{\rm DU}^{\rm ub, \, s}
+ \Sigma_{\rm DU}^{\rm ub, \, c}.
\label{eq:6.12}
\ee
The first term on the \rhs~of \eq\eqref{eq:6.12} contains $\ITot$ as well as left-over $\ISoft$ and $\IVirt$ operators (i.e., all color-correlated contributions), while the second term contains left-over terms arising from collinear limits, and hence no color-correlated matrix elements are present there. The expression for the first term reads 
\begin{align}
    &\; \Sigma_{\rm DU}^{\rm ub, \, s}
    = \asbr^2 \Lint \bigg[\frac{1}{2} \big(\ITot^{(0)}\big)^2 + K \ITot^{(0)} + \beta_0 \ITot^{(1)} \bigg]  \colorprod \FLM^\gqBorn \Rint \notag \\
    %-----------------------------------------------------------------------------------------------
    & + \asbr^2 \frac{\beta_0}{\epsilon} \llint \Big[(\cTilde-1) \IVirt(2\ep) + c_2(\ep) \ISofttilde(2\ep) - \ISoft(2\ep)\Big] \colorprod \FLM^\gqBorn \rrint \notag \\
    %-----------------------------------------------------------------------------------------------
    & + 2^{1+2\ep} \asbr^2 \,\delta(\epsilon) \Lint \bigg[\left(2\Emax/\mu \right)^{-2\epsilon} \bigg( \!\! - \ISoft(\epsilon) + \frac{(2\Emax/\mu)^{-2\epsilon}}{2\epsilon^2} N_c(\epsilon) \sum_{i \in \HP} \T_i^2 \bigg)  
    \label{eq_SigmaDU_ub_s} \\
    %-----------------------------------------------------------------------------------------------
    & + 2 \ISoft(2\ep)\bigg] \colorprod \FLM^\gqBorn \bigg\rangle 
    + \asbr \bigg[\Ca \bigg(\frac{c_1(\ep)}{\ep^2} - \frac{A_K(\ep)}{\ep^2}\bigg) - \cTilde K + \beta_0\,c_3(\ep) - \frac{2}{3} \nf\TR c_4(\ep) \notag \\
    & - 2^{2+2\ep} \delta(\ep) \bigg] \Big\langle \ISoft(2\ep) \colorprod \FLM^\gqBorn \Big\rangle \,, \notag
\end{align}
where $\delta(\ep) = \delta_g(\epsilon) + 2\nf\,\delta_q(\epsilon)$ (cf.\ Eqs \eqref{eq_deltam_deltamperp_def} and \eqref{eq_P_gg_Pgq_perp_defs}), and the remaining coefficients are collected in Appendix \ref{app_subsec_useful_constants}.
The first line of \eq\eqref{eq_SigmaDU_ub_s} is manifestly finite, with the first term having the expected form for the exponentiation given in \eq\eqref{eq:exponentiation}. In the second line, we can use the expansions of $\cTilde$ and $c_2$, as well as the relation
\be
\ISoft(\ep) - \ISofttilde(\ep) = \order{\ep} \,,
\ee
to demonstrate that only $\order{\ep^{-1}}$ poles are present, and that these are proportional to the residues of the double poles of $\ISoft$ and $\IVirt$ (i.e., sums over Casimir operators), and hence are not color-correlated. 
In the third line, the combination of $\ISoft(\ep)$, $\ISoft(2\ep)$, and the sum over the Casimirs cancels both the $1/\ep^2$ and the $1/\ep$ poles, leaving a finite result. 
Finally, it is simple to check that the terms in brackets in the final two lines are $\order{\ep}$, and this produces a color-uncorrelated simple pole when multiplied by the operator $\ISoft(2\ep)$. We therefore conclude that the expression in \eq\eqref{eq_SigmaDU_ub_s} has only $\order{\ep^{-1}}$ singularities, with all color-correlated terms being $\ep$-finite.

We now move on to   the unboosted component arising from collinear limits. It reads
\begin{align}
    &\; \SigmaDU^{\rm ub, \, c}
    = \frac{\asbr^2}{\ep^2} h_\rmc(\ep) \Ca \llint \Big[\ICollFour(\ep) - \IColltilde(2\ep) \Big] \colorprod \FLM^\gqBorn \rrint 
    + \frac{\asbr^2}{\ep} \llint \Big[ \Nscbd \gamma_g^{22}(0) \ICollFour(\ep) \notag \\ 
    & - \beta_0 \IColl(2\ep) \Big] \colorprod \FLM^\gqBorn \rrint + \frac{\asbr^2}{\ep^2} \sum_{i\in\HPf} \Lint\bigg\{ \frac{\Nscbd}{2} \Big[\gamma_q^{22}(L_i) \, c_{f_i} \GammaFour_{i,f_i \to \tildef_i q} + H_{i,f_i \to \tildef_i q} \Big] \label{eq_SigmaDU_ub_c} \\
    & + h_\rmc(\ep) \calX_{i,f_i} + \frac12 \Big[\GFSR{i}{f_i \hspace{9.5mm}}{z,f_i \to f_i g}
    + c_{f_i} \GFSR{i}{\tildef_i \hspace{9.5mm}}{z,f_i \to \tildef_i q}
    + c_{f_i} \Gamma_{i,f_i \to \tildef_i q} \big(\Gamma_{i,\tildef_i} - \Gamma_{i,f_i}\big) \Big] \bigg\} \colorprod \FLM^\gqBorn \Rint \,, \notag 
\end{align}
where the $H$-functions are defined in \eq\eqref{eq_H_qTOgq_FSR_def}. Moreover, $f_i$ labels the flavour of parton $i \in \HPf$, while $\tildef_i$ labels the ``complementary flavor" of $f_i$.
Therefore, if parton $i$ is a gluon, we have $f_i = g$ and $\tildef_i = q$, while if parton $i$ is a quark, then $f_i = q$ and $\tildef_i = g$.
In \eq\eqref{eq_SigmaDU_ub_c}, $c_{f_i}$ and $\calX_{i,f_i}$ are defined as
\begin{equation}
\begin{split}
    c_{f_i} & \eqdef
    \left\{
    \begin{aligned}
        & 2\nf, & & f_i = g, \\
        & 1,    & & f_i = q \,,
    \end{aligned}
    \right.\ \\
    %%%%%%%%%%%%%%%%%%%%%%%%%%%%%%%%%%%%%%%%%%%%%%%%%%%%%%%%%%%%%%%%%%%%%%%%%%%%%%%%%%%%
    \calX_{i,f_i} & \eqdef
    \left\{ 
    \begin{aligned}
        & (2\Cf - \Ca) \sigma_{i,g \to q\qb} + 4\nf(\Cf-\Ca) \bigg(\frac{\GammaFour_{i,g\to q\qb}}{2\ep} - \frac{\GammaLoop_{i,g\to q\qb}}{2\ep} \bigg), 
        & & \quad f_i = g, \\
        & \Ca \sigma_{i,q \to gq}, 
        & & \quad f_i = q \,.
    \end{aligned}
    \right.\
\end{split}
\end{equation}
Furthermore,  the $\sigma$-functions are defined in \eq\eqref{eq_sigma_i_def},
$\Nscbd$ and $h_\rmc(\ep)$ in Eqs \eqref{eq:normalisation}, and \eqref{eq:constants_RV}, respectively, the $G$-functions in Eqs \eqref{eq_GFSR_general_def}, \eqref{eq_GFSR_gluon} and \eqref{eq_GFSR_quark}, and the $H$-functions in \eq\eqref{eq_H_qTOgq_FSR_def}. 

\eq\eqref{eq_SigmaDU_ub_c} has two important features.  
The first one is that 
it is given by the sum of contributions over 
external partons, which 
suggests a straightforward generalization for even more complex processes than considered here. 
The second is that the contributions from each leg can be written using the universal quantities $c_{f_i}$ and $\calX_{i,f_i}$.  
As we will show below, these features allow for a simple discussion of the cancellation of $1/\ep$ poles, at least through $\order{\ep^{-2}}$, which is quite 
remarkable given the complexity of the process considered.  

 The cancellation of the $1/\ep^3$ pole takes place in a straightforward way.  
The only term of order $\order{\ep^{-3}}$ is the first on the \rhs\ of \eq\eqref{eq_SigmaDU_ub_c}, since both collinear operators are $\order{\ep^{-1}}$ (cf.\ \eq\eqref{eq_IColl_IColltilde_def_appendix}),  
while all the remaining quantities are $\order{\ep^0}$.  
However, $\ICollFour$ and $\IColltilde$ are related by the equation  
\begin{equation}
    \ICollFour(\ep) = \IColltilde(2\ep) + \order{\ep^0} \,,
\end{equation}
which immediately ensures the cancellation of the $1/\ep^3$ pole.

We move to the $1/\ep^2$ pole, distinguishing the case of initial-state radiation from that of final-state radiation.  
We begin with the former, which is contained in the first two terms on the \rhs\ of \eq\eqref{eq_SigmaDU_ub_c}, since the second and third lines pertain to final-state radiation.  
The objects describing the contribution of the $\inF$ and $\inS$ legs in the operators $\ICollFour$ and $\IColltilde$ correspond to the generalized initial-state anomalous dimensions $\GammaFour_{i,f_i}$ and $\GammaLoop_{i,f_i}$, defined in Eqs \eqref{Eq:Gamma_1_2_definition_k_general_appendix} and \eqref{eq_Gamma_1L_ISR_definition}, respectively.  
We have
\begin{equation}
    \frac{\GammaFour_{i,f_i}}{2\ep} = \frac{\GammaLoop_{i,f_i}}{2\ep} + \order{\ep} \,,
    \qquad
    i \in \{\inF,\inS\} \,,
\end{equation}
and also $h_\rmc(\ep) = 1 + \order{\ep^3}$. A similar argument holds for the quadratic poles due to the second term on the \rhs\ \eq\eqref{eq_SigmaDU_ub_c}, and we can conclude that the $1/\ep^2$ poles in this equation vanish for $i \in \{\inF,\inS\}$.

Next, we consider the final-state radiation, that is, $i \in \HPf$, where the second and third lines in \eq\eqref{eq_SigmaDU_ub_c} also contribute.  
The cancellation mechanism for the final-state gluon and quark legs is entirely analogous.  
In this case, the cancellation of the $1/\ep^2$ pole does not occur immediately by inspection, but still follows from elementary algebraic manipulations.  
To see this, it is sufficient to expand  the anomalous dimensions in \eq\eqref{eq_gamma_22_q_g_def} and \eq\eqref{eq_usual_anom_dimens_defs} in powers of $\ep$, obtaining  
\begin{equation}
    \gamma_{x}^{nk}(L_i) = \sum_{m=0}^{\infty} \gamma_{x}^{nk,(m)}(L_i) \, \ep^m \,,
    \qquad
    L_i = \log\bigg(\frac{\Emax}{E_i}\bigg) \,.
\end{equation}
Furthermore, a similar expansion for the one-loop anomalous dimension contained in $\GammaLoop_{i,f_i}$ (cf.\ Eqs (\ref{eq_Gamma_1L_FSR_definition} -- \ref{eq_gamma_nk_1L_appendix})) reads
\begin{equation}
    \gamma_{x}^{\rm 1L}(L_i)
    \eqdef - \frac{\ep^2 \cos(\pi\ep)}{C_x} \, \gamma_{x}^{3(k+1),\mathrm{1L}}(L_i) 
    = \sum_{m=0}^{\infty} \gamma_{x}^{{\rm 1L},(m)}(L_i) \, \ep^m \,, 
\end{equation}  
where $C_x = 2\Cf-\Ca$ if $x = (g \to q\qb)$ and $C_x = \Ca$ otherwise. 
We then observe that only the coefficients with $m=0,1$ contribute to the cancellation of the poles.  
For instance, in the case of the quark leg, we can 
use the expansions
\begin{equation}
\begin{split}
    \Gamma^{(k)}_{i,x} & = \gamma_{x}^{2k,(0)}(L_i) + \ep \, \Big(\gamma_{x}^{2k,(1)}(L_i) - k\, \Ltildei \gamma_{x}^{2k,(0)}(L_i) \Big) + \order{\ep^2} \,, \\
    \GammaLoop_{i,f_i} & = \gamma_{f_i}^{\rm 1L,(0)}(L_i) + \ep \, \Big(\gamma_{f_i}^{{\rm 1L},(1)}(L_i) - 4 \Ltildei \gamma_{f_i}^{\rm 1L,(0)}(L_i) \Big) + \order{\ep^2} \,,
\end{split}
\end{equation}
and noticing that $\gamma_{x}^{2k,(0)}(L_i) \equiv \gamma_{f_i}^{{\rm 1L},(0)}(L_i) = \gamma_i + 2\T_i^2 L_i$,  obtain  
\begin{equation}
    h_\rmc(\ep) \Ca \bigg(\frac{\Gamma^{(4)}_{i,q}}{2\ep} - \frac{\GammaLoop_{i,q}}{2\ep} \bigg) 
    = \frac{\Ca}{2} \Big[\gamma_{q}^{24,(1)}(L_i) - \gamma_{q}^{{\rm 1L},(1)}(L_i) \Big] + \order{\ep} \,,
\end{equation}
where $\Ltildei = \log(2E_i/\mu)$. We note that an analogous expansion is also valid for the second term on the \rhs\ \eq\eqref{eq_SigmaDU_ub_c}.
Similarly, we expand the remaining objects on the \rhs\ of \eq\eqref{eq_SigmaDU_ub_c}, and we find
\begin{align}
    & \calX_{i,q} = \frac{\Ca}{2} \Big[\gamma_{z,q \to gq}^{22,(1)} - \gamma_{z,q \to gq}^{24,(1)} - 2L_i \gamma_{z,q \to gq}^{22,(0)} \Big] + \order{\ep} \,, \notag \\
    & \GFSR{i}{f_i \hspace{9.5mm}}{z,f_i \to f_i g} = \T_{f_i}^2 \Big[\gamma_{z,f_i \to f_i g}^{42,(1)}(L_i) - \gamma_{z,f_i \to f_i g}^{22,(1)}(L_i) \Big] + \order{\ep} \,, 
    \label{eq_expansions_prova} \\
    & H_{i,f_i \to \tildef_i q} = \Cf \Big[\gamma_{z,q \to gq}^{24,(1)} - \gamma_{z,q \to gq}^{22,(1)} \Big] + \order{\ep} \notag \,,
\end{align}
while all remaining quantities can be approximated as $\gamma_{x}^{2k,(0)}$.
Thus, \eq\eqref{eq_SigmaDU_ub_c} reduces to a simple expression involving coefficients $\gamma_{x}^{2k,(m)}$, $\gamma_{x}^{42,(m)}$, and $\gamma_{x}^{{\rm 1L},(m)}$ with $m=0,1$, which vanishes at $\order{\ep^{-2}}$.
We observe that the same argument can be repeated for the gluon leg.

Before concluding this section, we briefly discuss  the simple  poles present in the single-boosted and unboosted double-unresolved contributions, which have been  omitted from the preceding discussions.  Similar to the all-gluon case in  Ref.~\cite{Devoto:2023rpv}, their cancellation 
is difficult to trace 
by eye,  but is entirely straightforward to achieve  using  computer algebra. We stress that our ability to absorb the many $1/\ep$ poles into finite structures, such as $\ITot$, simplifies the analysis of the simple poles   dramatically,  since it 
reduces the bulk of it to a 
discussion of \emph{independent} pole cancellations for  each 
of the external legs. 

To cancel the simple poles, we require the integrated triple-collinear subtraction terms, which we take from Ref.~\cite{Delto:2019asp}. However, we need to slightly modify the results of this reference for two reasons. The first is to take into account the damping factors, which in Ref.~\cite{Devoto:2023rpv} were simplified by the symmetry of the final state. The second is that the calculation in Ref.~\cite{Delto:2019asp} assumed that all potentially-unresolved final-state partons are energy-ordered, whereas we impose this ordering only for final states that admit a singular double-soft  limit, and do not require it for e.g.\ $(\Fp_q\Sp_g)$. The changes are relatively minor, as they affect only the strongly-ordered collinear limits. We include updated expressions for the integrated triple-collinear subtraction terms in the ancillary file \usefulfunctions\ (see the last four rows of Table~\ref{table_appendix_ancillary_file}).
 
Upon making these changes and including all sources of simple poles, we are able to confirm the cancellation of all poles, and hence obtain a finite remainder which we present in the following section.

\section{Finite remainder of the NNLO contribution to the cross section}
\label{sec_final_result}

The goal of this section is to present 
the expression for $\dsigmahat_{\gqBorn}^\NNLO$ in \eq\eqref{eq_dsigmahat_NNLO_starting_point} which is free of infrared singularities and, therefore, can be computed numerically.
The main motivations for showing  such a result are its compactness and iterative structure, as well as its proximity to the all-gluon case presented in Ref.~\cite{Devoto:2023rpv}.   
These features, together with the rather general nature of the discussion in the preceding 
sections, give us confidence that a similar expression for \emph{arbitrary} processes at hadron colliders can be derived in the near future. 

To present the result, we split $\dsigmahat_\NNLO$ into fully-resolved, single-unresolved, and double-unresolved contributions, and write
\begin{equation}
    2s_{\inF\inS} \, \dsigmahat^{\gqBorn}_{\NNLO}
    = 2s_{\inF\inS} \big[\dsigmahat^\gqBorn_\fr + \dsigmahat^\gqBorn_\su + \dsigmahat^\gqBorn_\du \big] \,.
    \label{eq_final_result_procgq}
\end{equation}
The first term on the \rhs\ of \eqref{eq_final_result_procgq} is the finite, fully-regulated contribution.
It corresponds to 
\begin{equation}
    2s_{\inF\inS} \, \dsigmahat_\fr^\gqBorn  
    =
    \sum_{n=4}^{6} \sum_{(\Fp\Sp)} \proj \SigmaFR^\genBorn[(\Fp\Sp)]_\RR \,,
    \label{eq_dsigmahat_FR_procgq_final}
\end{equation}
where the quantity $\SigmaFR^\genBorn[(\Fp\Sp)]_\RR $ is defined in \eq\eqref{eq_SigmaFR_gqBorn}.

The second term on the \rhs\ of \eq\eqref{eq_final_result_procgq} is the finite remainder of the single-unresolved contribution.
We write is as the sum of four contributions
\begin{equation}
    2s_{\inF\inS} \, \dsigmahat_\su^\gqBorn = \sum_{i=1}^{4} \proj \SigmaSU^{\fin,(i)}|_\gqBorn \,,
    \label{eq_dsigmahat_procgq_SU_final}
\end{equation}
where in
\begin{align}
    & \begin{aligned}
        &\; \SigmaSU^{\fin,(1)}|_\gqBorn
        = \asbr \, \llint \ONLO^{(\inF,\Fp)} \partFuncACsp{\inF} 
        \log\Big(\frac{\eta_{\inF \Fp}}2\Big) \Delta^{(\Fp)} \big[\PAP_{gg} \conv \FRlo{\FgqprocNLO}{\Fp} \big] \rrint \\
        & + \asbr \llint \ONLO^{(\inS,\Fp)} \,  \partFuncACsp{\inS} \, \log\Big(\frac{\eta_{\inS \Fp}}2\Big) \Delta^{(\Fp)} \big[\FRlo{\FgqprocNLO}{\Fp} \conv \PAP_{qq}\big]\rrint 
        - \sum_{i \in \HP} \, \asbr \Big\langle \ONLO^{(i,\Fp)}  \, \partFuncACsp{i} \\
        & \times \log \Big(\frac{\eta_{i\Fp}}2\Big) \,\Delta^{(\Fp)} \Big[\big(\gamma_i + 2\T_i^2 L_i(E_\Fp) \big) \FLMlo{\FgqprocNLO}{\Fp_g} + (\gamma_i + 2\T_i^2 L_i) \FLMlo{\FgqprocNLO}{\Fp_q} \Big] \Big\rangle  \,,
    \end{aligned}
    \label{eq_SigmaSU_fin_1_final} \\ & \notag \\
    %------------------------------------------------------------------------------------------------------------------
    & \begin{aligned}
        \SigmaSU^{\fin,(2)}|_\gqBorn
        \eqdef & \, 
        - \sum_{i \in \HP} \, \asbr \Lint \ONLO^{(i,\Fp)} \, \partFuncBD{i} \, 
        \log \left ( \frac{\eta_{i \Fp}}{4 (1-\eta_{i \Fp})} \right ) \Delta^{(\Fp)} \Big[\gamma_g \FLMlo{\FgqprocNLO}{\Fp_g} \\
        & + (\gamma_q + 2 L_\Fp \T_q^2) \FLMlo{\FgqprocNLO}{\Fp_q} \Big]\Rint \,, 
    \end{aligned}
    \label{eq_SigmaSU_fin_2_final} \\ & \notag \\
    %-------------------------------------------------------------------------------------------------------------
    & \begin{aligned}
        \SigmaSU^{\fin,(3)}|_\gqBorn
        \eqdef & \, 
        \sum_{i\in\HP} \,\frac{\asbr}{2} \, \gamma_{\bot, g}^{22} \,  \lint \ONLO^{(i,\Fp)} \, \partFuncBD{i} (r_i^\mu r_i^\nu + g^{\mu\nu}) \Delta^{(\Fp)} \FLMunmunu{\FgqprocNLO}{\Fp_g} \rint \\
        & + \sum_{i\in\HP} \,\frac{\asbr}{2} \, \gamma_{\bot, g }^{22, \rmr} \, \lint \ONLO^{(i,\Fp)} \, \partFuncBD{i} \Delta^{(\Fp)} \FLMlo{\FgqprocNLO}{\Fp_g} \rint \,,
    \end{aligned} 
    \label{eq_SigmaSU_fin_3_final} \\ & \notag \\
    %------------------------------------------------------------------------------------------------------------------
    & \begin{aligned}
        &\; \SigmaSU^{\fin,(4)}|_\gqBorn
        = 
        \lint \ONLO^{(\Fp)} \, \Delta^{(\Fp)} \big(\FRVfinlo{\FgqprocNLO}{\Fp_g} + \FRVfinlo{\FgqprocNLO}{\Fp_q}\big) \rint  \\
        & + \asbr \lint \ONLO^{(\Fp)} \, \Delta^{(\Fp)} \big[\ITot^{(0)}(E_\Fp) \colorprod \FLMlo{\FgqprocNLO}{\Fp_g} 
        + \ITot^{(0)}(\Emax) \colorprod \FLMlo{\FgqprocNLO}{\Fp_q} \big] \rint 
        \label{eq_SigmaSU_fin4_def} \\ 
        & + \asbr \lint \ONLO^{(\Fp)} \, \Delta^{(\Fp)} \big[\PNLO_{gg} \conv \FRlo{\FgqprocNLO}{\Fp} \big] \rint 
        + \asbr \lint \ONLO^{(\Fp)} \, \Delta^{(\Fp)} \big[\FRlo{\FgqprocNLO}{\Fp} \conv \PNLO_{qq}\big] \rint \,. 
    \end{aligned}
\end{align}
The definitions of quantities that appear in the above formulas can be found in 
Eqs \eqref{eq_SigmaSU_RR_genBorn_double_soft} and \eqref{eq_SigmaSU_RR_genBorn_nodouble_soft}, as well as in \Sec\ref{sec:single_unr}. 
Here, we note that the objects $\gamma_{\bot,g}^{22}$ and $\gamma_{\bot,g}^{22, \rmr}$ are given in \eq\eqref{eq_gamma_perp_ep_0}, and their expressions are reported in the ancillary file \finalresult\ provided with this paper, see Table \ref{table_ancillary_file}.
The quantity $\ITot^{(0)}$ is the $\order{\ep^0}$ expansion coefficient of the IR-finite operator $\ITot(\ep)$; it can be found in \eq\eqref{eq_IT0_appendix}.
The splitting functions $\PAP_{\alpha\alpha}$ are defined in \eq\eqref{Eq_PAP_0_definition_appendix}, while the splitting functions $\PNLO_{\alpha\alpha}$ are reported in \finalresult, see Table \ref{table_ancillary_file}.
Finally, we remind the reader that the argument $E_\Fp$ in $L_i(E_\Fp)$ and $\ITot^{(0)}(E_\Fp)$ indicates that $\Emax$ must be replaced with $E_\Fp$. 

The third term on the \rhs\ of \eq\eqref{eq_final_result_procgq} corresponds to the finite remainder of the double-unresolved contribution.
It is convenient to write it as the sum  of the following quantities
\begin{equation}
    \dsigmahat_\du^\gqBorn 
    = \dsigmahat_\du^{\gqBorn, \db}
    + \dsigmahat_\du^{\gqBorn, \inFsb}
    + \dsigmahat_\du^{\gqBorn, \inSsb}
    + \dsigmahat_\du^{\gqBorn, \el} \,,
\end{equation}
where $\dsigmahat_\du^{\gqBorn, \db}$ contains all the \emph{double-boosted} contributions, $\dsigmahat_\du^{\gqBorn, \inFsb}$ and $\dsigmahat_\du^{\gqBorn, \inSsb}$ contain all the contributions that are \emph{single-boosted} on leg $\inF$ and leg $\inS$, respectively, and $\dsigmahat_\du^{\gqBorn, \el}$ contains all the \emph{elastic} contributions with the kinematics of the Born process. 
We present each contribution separately, using several functions that we collect in \finalresult.

The double-boosted contribution is described by the very simple expression (see \eq\eqref{eq_double_boosted_expansion})
\begin{equation}
   2s_{\inF\inS} \, \dsigmahat_{\gqBorn, \db}^\du 
   = \asbr^2 \lint \PNLO_{gg} \conv \FLM^\gqBorn \conv \PNLO_{qq} \rint \,.
    \label{eq_dsigmahat_DU_db_procgq_final}
\end{equation} 
Expressions for the single-boosted contributions, related
to the emissions off the incoming partons $\inF$ and $\inS$,  are  slightly more complex. We write them as
\begin{equation}
\begin{split}
    2s_{\inF\inS} \, \dsigmahat_\du^{\gqBorn, \inFsb}
    = &\; \asbr^2 \Big[
    \lint \PNLO_{gg} \conv \big[\ITot^{(0)} \colorprod \FLM^\gqBorn \big] \rint 
    + \lint \calPW_{gg} \conv \big[\Wacfin{\inF} \colorprod \FLM^\gqBorn \big] \rrint \\
    & + \lint \PNNLO_{gg} \conv \FLM^\gqBorn \rint \Big]
    +  \asbr \lint \PNLO_{gg} \conv \FLVfin^\gqBorn \rint \,,
\end{split}
\label{eq_dsigmahat_DU_inFsb_procgq_final}
\end{equation}
and
\begin{equation}
\begin{split}
    2s_{\inF\inS} \, \dsigmahat_\du^{\gqBorn, \inSsb}
    = &\; \asbr^2 \Big[
    \lint \big[\ITot^{(0)} \colorprod \FLM^\gqBorn \big] \conv \PNLO_{qq} \rint 
    + \lint \big[\Wacfin{\inS} \colorprod \FLM^\gqBorn \big] \conv \calPW_{qq} \rint \\
    & +  \lint \FLM^\gqBorn \conv \PNNLO_{qq} \rint \Big] 
    +   \asbr \lint \FLVfin^\gqBorn \conv \PNLO_{qq} \rint \,.
\end{split}
\label{eq_dsigmahat_DU_inSsb_procgq_final}
\end{equation}
\begin{table}[!p] 
    \centering
    \caption{List of functions collected in the ancillary file \finalresult.  
    The first column provides the names of the functions; the second specifies the index range $i,j$; the third indicates the equation where they appear; the fourth provides their names in \finalresult.  
    For brevity, in the second block of the table, $\calP_{gg}^{\, ...}$ and $\calP_{qq}^{\, ...}$ are encoded in $\calP_{xx}^{\, ...}$. To extract the $gg$ or $qq$ splittings, call \texttt{Pgg...[z,En]} or \texttt{Pqq...[z,En]}.  }
    
    \begin{tabular}{
    >{\centering\arraybackslash}m{26mm}  
    >{\centering\arraybackslash}m{30mm} 
    >{\centering\arraybackslash}m{34mm} 
    >{\centering\arraybackslash}m{44.5mm}
    }
        \hline
        \multicolumn{4}{c}{\textbf{Functions collected in the ancillary file \finalresult}} \\
        \hline\hline
            \emph{Function}  
            & \emph{Interval of} $i,j$ 
            & \emph{Ref.} 
            &  \emph{Name in the ancillary file} 
            \\ \hline
        %%%%%%%%%%%%%%%%%%%%%%%%%%%%%%%%%%%%%%%%%%%%%%%%%%%%%%%%%%%%%
        % SPIN-CORRELATED QUANTITIES
        %%%%%%%%%%%%%%%%%%%%%%%%%%%%%%%%%%%%%%%%%%%%%%%%%%%%%%%%%%%%%
        \multicolumn{4}{c}{\emph{Integrals in spin-correlations}} \\ \hline
            $\gamma_{\bot,g}^{22}$    
            & / 
            & \eq\eqref{eq_gamma_perp_ep_0}
            & \texttt{\textgamma gPerp}  
            \\
            %%%%%%%%%%%%%%%%%%%%%%%%%%%%%%%%%%%%%%%%%%%%%%%%%%%%%%%%%%%%%
            $\gamma_{\bot,g}^{22, \rmr}$    
            & / 
            & \eq\eqref{eq_gamma_perp_ep_0}
            & \texttt{\textgamma gPerpR}  
            \\
            %%%%%%%%%%%%%%%%%%%%%%%%%%%%%%%%%%%%%%%%%%%%%%%%%%%%%%%%%%%%%
            $\delta^{(0)}$    
            & / 
            & \eq\eqref{eq_dsigmahat_DU_el_procgq_final}
            & \texttt{\textdelta zero}  
            \\
            %%%%%%%%%%%%%%%%%%%%%%%%%%%%%%%%%%%%%%%%%%%%%%%%%%%%%%%%%%%%%
            $\delta^{\perp,(0)}$    
            & / 
            & \eq\eqref{eq_dsigmahat_DU_el_procgq_final}
            & \texttt{\textdelta Perpzero}  
            \\ \hline
        %%%%%%%%%%%%%%%%%%%%%%%%%%%%%%%%%%%%%%%%%%%%%%%%%%%%%%%%%%%%%
        % SPLITTING FUNCTIONS
        %%%%%%%%%%%%%%%%%%%%%%%%%%%%%%%%%%%%%%%%%%%%%%%%%%%%%%%%%%%%%
        \multicolumn{4}{c}{\emph{Splitting functions}} \\ \hline
            $\PNLO_{xx}(z,E_i)$           
            &  $i\in\{\inF,\inS\}$ 
            & Eqs (\ref{eq_SigmaSU_fin4_def}, \ref{eq_dsigmahat_DU_db_procgq_final} -- \ref{eq_dsigmahat_DU_inSsb_procgq_final})
            & \texttt{PxxNLO[z,En]}
            \\
            %%%%%%%%%%%%%%%%%%%%%%%%%%%%%%%%%%%%%%%%%%%%%%%%%%%%%%%%%%%%%
            $\calPW_{xx}(z,E_i)$           
            &  $i\in\{\inF,\inS\}$ 
            & Eqs (\ref{eq_dsigmahat_DU_inFsb_procgq_final}, \ref{eq_dsigmahat_DU_inSsb_procgq_final})
            & \texttt{PxxW[z,En]}
            \\
            %%%%%%%%%%%%%%%%%%%%%%%%%%%%%%%%%%%%%%%%%%%%%%%%%%%%%%%%%%%%%
            $\PNNLO_{xx}(z,E_i)$           
            &  $i\in\{\inF,\inS\}$ 
            & Eqs (\ref{eq_dsigmahat_DU_inFsb_procgq_final}, \ref{eq_dsigmahat_DU_inSsb_procgq_final})
            & \texttt{PxxNNLO[z,En]}
            \\ \hline
        %%%%%%%%%%%%%%%%%%%%%%%%%%%%%%%%%%%%%%%%%%%%%%%%%%%%%%%%%%%%%
        % ELASTIC FUNCTIONS 
        %%%%%%%%%%%%%%%%%%%%%%%%%%%%%%%%%%%%%%%%%%%%%%%%%%%%%%%%%%%%%
        \multicolumn{4}{c}{\emph{Elastic functions}} \\ \hline
            $\gamma_{z,g \to gg}^{\calW}(L_i)$     
            &  $i\in\HPfg$ 
            & \eq\eqref{eq_dsigmahat_DU_el_procgq_final}
            & \texttt{\textgamma WgTOgg[En]}
            \\
            %%%%%%%%%%%%%%%%%%%%%%%%%%%%%%%%%%%%%%%%%%%%%%%%%%%%%%%%%%%%%
            $\gamma_{z,q \to qg}^{\calW}(L_i)$     
            &  $i\in\HPfq$ 
            & \eq\eqref{eq_dsigmahat_DU_el_procgq_final}
            & \texttt{\textgamma WqTOqg[En]}
            \\
            %%%%%%%%%%%%%%%%%%%%%%%%%%%%%%%%%%%%%%%%%%%%%%%%%%%%%%%%%%%%%
            $D_g^{\rm ISR}(E_i)$     
            &  $i\in\{\inF,\inS\}$ 
            & \eq\eqref{eq_Iuncfin_def}
            & \texttt{DgISR[En]}
            \\
            %%%%%%%%%%%%%%%%%%%%%%%%%%%%%%%%%%%%%%%%%%%%%%%%%%%%%%%%%%%%%
            $D_q^{\rm ISR}(E_i)$     
            & $i\in\{\inF,\inS\}$ 
            & \eq\eqref{eq_Iuncfin_def}
            & \texttt{DqISR[En]}
            \\
            %%%%%%%%%%%%%%%%%%%%%%%%%%%%%%%%%%%%%%%%%%%%%%%%%%%%%%%%%%%%%
            $D_g^{\rm FSR}(E_i)$    
            & $i\in\HPfg$ 
            & \eq\eqref{eq_Iuncfin_def}
            & \texttt{DgFSR[En]}  
            \\
            %%%%%%%%%%%%%%%%%%%%%%%%%%%%%%%%%%%%%%%%%%%%%%%%%%%%%%%%%%%%%
            $D_q^{\rm FSR}(E_i)$    
            & $i\in\HPfq$ 
            & \eq\eqref{eq_Iuncfin_def}
            & \texttt{DqFSR[En]}
            \\ \hline
        %%%%%%%%%%%%%%%%%%%%%%%%%%%%%%%%%%%%%%%%%%%%%%%%%%%%%%%%%%%%%
        % DOUBLE SOFT 
        %%%%%%%%%%%%%%%%%%%%%%%%%%%%%%%%%%%%%%%%%%%%%%%%%%%%%%%%%%%%%
        \multicolumn{4}{c}{\emph{Double-soft finite remainders}} \\ \hline
            $\big(S_{gg,T^2}^{\rm fin}\big)_{ij}$ 
            & $i,j\in\HP, ~ i\neq j$  
            & \eq\eqref{eq_dsigmahat_DU_el_procgq_final}
            & \texttt{SggT2fin[i,j]}
            \\
            %%%%%%%%%%%%%%%%%%%%%%%%%%%%%%%%%%%%%%%%%%%%%%%%%%%%%%%%%%%%%
            $\big(S_{q\qb,T^2}^{\rm fin}\big)_{ij}$
            & $i,j\in\HP, ~ i\neq j$   
            & \eq\eqref{eq_dsigmahat_DU_el_procgq_final}
            & \texttt{SqqbT2fin[i,j]}
            \\
            %%%%%%%%%%%%%%%%%%%%%%%%%%%%%%%%%%%%%%%%%%%%%%%%%%%%%%%%%%%%%
            \noalign{\vspace{1mm}} \hline
    \end{tabular}
    \label{table_ancillary_file}
\end{table}

\noindent The partition-dependent operators $\Wacfin{i}$, with $i = \inF,\inS$, are  given below in \eq\eqref{eq_Wbdfin_def_appendix_final_result}.
Furthermore, $\calPW_{\alpha\alpha}$ and the NNLO splitting functions $\PNNLO_{\alpha\alpha}$ are also reported in the ancillary file \finalresult\ (cf.\ Table \ref{table_ancillary_file}).

Finally, the elastic contribution reads
\begin{equation} 
\begin{split}
    2s_{\inF\inS} \, \dsigmahat_{\gqBorn, \el}^\du 
    & = \asbr^2 \lint\big[\Iccfin + I_{\rm tri}^{\rm fin} + \Iuncfin \big] \colorprod \FLM^\gqBorn \rint  \\
    & + \asbr^2 \sum_{i \in \HP} \llint \Big[\theta_\HPf \gamma_{z,f_i \to f_i g}^{\calW} \, \Wacfin{i} + \delta^{(0)}\Wbdfin{i} + \delta^{\perp,(0)} \Wr{i}\Big] \colorprod \FLM^\gqBorn \rrint \\ 
    & + \asbr^2 \sum_{\inotj} \lint \big[\big(S_{gg,T^2}^{\rm fin}\big)_{ij} + \nf \big(S_{q\qb,T^2}^{\rm fin}\big)_{ij}\big] (\T_i \cdot \T_j) \colorprod \FLM^\gqBorn \rint \\
    & + \asbr \lint \ITot^{(0)} \colorprod \FLVfin^\gqBorn \rint
    + \lint \FLVfinsq^\gqBorn \rint 
    + \lint \FVVfin^\gqBorn \rint \,.
\end{split}
\label{eq_dsigmahat_DU_el_procgq_final}
\end{equation} 
In the first line of \eq\eqref{eq_dsigmahat_DU_el_procgq_final}, we have the combination of quartic and double color-correlated contributions in $\Iccfin$, a triple color-correlated component $\Itrifin$, and a color-uncorrelated part $\Iuncfin$.
The first of these is particularly simple, and reads
\begin{align}
    &\; \Iccfin 
    = \frac{1}{2} \big(\ITot^{(0)}\big)^2 + K \ITot^{(0)} + \beta_0 \bigg[\ITot^{(1)} + \ISofttilde^{(1)} - 2\ISoft^{(1)} + \frac{\pi^2}{24}\IVirt^{(-1)} +2 \ISoft^{(0)} \log{2} \notag \\
    & +\ISoft^{(-1)} \left(4\Lmax \log{2} - 6 \log^2{2} - \frac{\pi^2}{2} \right) \bigg]
    + \Ca \bigg[ \ISoft^{(-1)} \bigg(\frac{1975}{108} - \frac{17 \zeta_3}{4} -\frac{2}{3} \pi^2 \log 2\bigg)
    \label{eq_Iccfin_def} \\
    & + \left( \ISoft^{(0)} + 2 \Lmax \ISoft^{(-1)} \right)  \left(\frac{\pi^2}{3}-\frac{131}{36}\right)\bigg] 
    + \nf \TR \bigg[ \frac{23}{18} \left( \ISoft^{(0)} + 2 \Lmax \ISoft^{(-1)} \right) -\frac{331}{54} \ISoft^{(-1)}  \bigg] \,, \notag
\end{align}
where $K$ is defined in \eq\eqref{eq:constants_VV}, and $\ISoft^{(n)}$, $\ISofttilde^{(n)}$, $\IVirt^{(n)}$, $\ITot^{(n)}$ are the coefficients of the $n$-th power in the $\ep$-expansion of the corresponding operators (cf.\ Appendix \ref{appendix_sec_operator_defn}).
We note that \eq\eqref{eq_Iccfin_def} generalizes Eq.~(I.8) in Ref.~\cite{Devoto:2023rpv}, which was computed for the process $q\qb \to \colsing + \Ng\, g$ with $\nf=0$, to the more general case of the $\gqBorn$ process in \eq\eqref{eq_NLO_gq_LO_process}, where terms proportional to $\nf$ have also been included.  
The definition of the operator $\Itrifin$, which contains triple color-correlated components, remains identical to the one in  Eq.~(I.9) of Ref.~\cite{Devoto:2023rpv}. 
It contains the most complicated functions appearing in the final result, namely classical and generalized polylogarithms (GPLs) up to weight three, whose arguments can be algebraic functions of  the kinematic variables.
The operator $\Iuncfin$ collects color-uncorrelated contributions. We write it as the sum of two terms, one that only depends on the color charges of each external parton, and another, named $D_{f_i}(E_i)$, whose functional form depends on the type of parton  and  whether it is in the initial or the final state. We write
\begin{equation}
\begin{split}
    &\; I_{\rm unc}^{\rm fin} 
    = \sum_{i\in\HP} \T_i^2 \bigg\{  \Ca \bigg[\bigg(\frac{2 \pi^2}{3} - \frac{131}{18} + \frac{22}{3}\log{2}\bigg) \Lmax^2 -\frac{935}{72}\zeta_3 +\frac{9607}{324} - \frac{11}{3} \log^3{2}  \\
    & + \bigg(\frac{1433}{108} - 8\zeta_3 - \frac{11 \pi^2}{6}\bigg) \log{2} - \frac{199 \pi^4}{1440} - \frac{21 \pi^2}{32} + \left(\frac{143}{36}-\frac{\pi ^2}{3}\right) \log ^22\bigg]  \\
    & + \nf \TR \bigg[\bigg(\frac{23}{9} - \frac{8}{3}\log{2}\bigg) \Lmax^2 +\frac{85}{18}\zeta_3 -\frac{746}{81} + \frac{4}{3} \log^3{2} 
    + \bigg( \frac{2 \pi^2}{3}-\frac{67}{27}\bigg) \log{2} \\
    & - \frac{47}{18} \log ^22 + \frac{5 \pi^2}{72} \bigg]
    \bigg\} + \sum_{i \in \HP} D_{f_i}(E_i) \,,
\end{split}
\label{eq_Iuncfin_def}
\end{equation}
where the functions $D_{f_i}$ are reported in the ancillary file \finalresult, see Table \ref{table_ancillary_file}.\footnote{In Table \ref{table_ancillary_file}, we label these functions as $D_{f_i}^{\rm ISR}$ and $D_{f_i}^{\rm FSR}$ if they refer to \emph{initial-state radiation} or \emph{final-state radiation}, respectively.}
The function $I_{\rm unc}^{\rm fin}$ in \eq\eqref{eq_Iuncfin_def} is a generalization of a similar quantity  in Ref.~\cite{Devoto:2023rpv} (see \eq(I.12) therein).

In the second line of \eq\eqref{eq_dsigmahat_DU_el_procgq_final}, we have $\theta_\HPf = 1$ if $i \in \HPf$ and $\theta_\HPf = 0$ otherwise.
The quantities $\delta^{(0)}$, $\delta^{\perp,(0)}$, and $\gamma_{z,f_i \to f_i g}^{\calW}$ are reported in \finalresult, while the partition-dependent operators $\Wacfin{i}$, $\Wbdfin{i}$, $\Wr{i}$ are defined as\footnote{We note that the cancellation of poles, discussed in the previous two sections, occurs independently of the exact expressions of the partition functions, provided they obey certain defining conditions, specified in e.g.\ Ref.~\cite{Devoto:2023rpv}.} 
\begin{align}
    & \Wacfin{i} \eqdef \sum_{\knotl} \int \frac{\rmd \Omega_\Fp^{(3)}}{2\pi} \, \oC_{i \Fp} \left[\log(\eta_{i \Fp}/2) \frac{\rho_{kl}}{\rho_{k\Fp} \rho_{l\Fp}} \, \partFuncACsp{i}  \right] (\T_k \cdot \T_l) \,,
    % \label{eq_Wacfin_def_appendix_final_result} \\
    \notag \\
    %------------------------------------------------------------------------------------------------------------
    & \Wbdfin{i} \eqdef \sum_{\knotl} \int \frac{\rmd \Omega_\Fp^{(3)}}{2\pi} \, \oC_{i\Fp}\bigg[ \log \left(
    \frac{\eta_{i\Fp}}{1-\eta_{i\Fp}}
    \right) \frac{\rho_{kl}}{\rho_{k\Fp} \, \rho_{l\Fp}} \,  \partFuncBD{i} \bigg] (\T_k \cdot \T_l) \,.
    \label{eq_Wbdfin_def_appendix_final_result} \\
    %------------------------------------------------------------------------------------------------------------
    & \Wr{i} \eqdef \sum_{k,l \in \HP} \int \frac{\rmd \Omega_\Fp^{(3)}}{2\pi} \, \partFuncBD{i} \, (r_i^\mu r_i^\nu + g^{\mu\nu}) \frac{n_{k,\mu} n_{l,\nu}}{(n_k \cdot n_\Fp) (n_l \cdot n_\Fp)} \, (\T_k \cdot \T_l) \,, \notag
    % \label{eq_calW_r_i_def_final_result}
\end{align}
where a light-like vector $n_i^\mu$ is defined through the equation $p_i^\mu = E_i \, n_i^\mu$, and $\knotl$ means $k,l \in \HP$, with $k\neq l$.
The term in the third line of \eq\eqref{eq_dsigmahat_DU_el_procgq_final} refers to the finite remainder of the double-soft integrated subtraction term (cf.\ \eq\eqref{eq:DoubSoft_exp}), and it can be extracted from Ref.~\cite{Caola:2018pxp}.
The explicit expressions of $\big(S_{gg,T^2}^{\rm fin}\big)_{ij}$ and $\big(S_{q\qb,T^2}^{\rm fin}\big)_{ij}$, which are functions of $\eta_{ij}$, are given in \finalresult, see Table \ref{table_ancillary_file}. 
Finally, $\FLVfinsq^\gqBorn$ and $\FVVfin^\gqBorn$ are the process-dependent finite remainders of the one-loop squared and two-loop virtual amplitudes, respectively.

\section{Conclusions}
\label{sec_conclusions}

In this paper we discussed the calculation of the integrated 
NNLO QCD subtraction terms 
for the process $gq \to \colsing + \Ng\,g + q$, where the number of hard gluons $\Ng$ is 
\emph{a parameter} and $X$ is an arbitrary colorless final state. We worked  
in the context of the nested soft-collinear subtraction scheme \cite{Caola:2017dug} and followed  
closely a similar study of the all-gluon final state reported 
in Ref.~\cite{Devoto:2023rpv}. 

We have found that the approach 
described in Ref.~\cite{Devoto:2023rpv}
is sufficiently robust, and can be applied to  a complex final state that contains  both quarks and gluons.  
In doing so, we have seen that physical quantities such as collinear anomalous dimensions arise when different singular configurations are properly combined before the subtraction terms are integrated over the unresolved phase space -- a feature  that is important for making computations not only more feasible but also physically transparent.  
Finally, since 
i) soft limits at NNLO depend  only on  the color charges of hard partons,  ii) the  collinear singularities factorize on the external parton legs and iii) an improved  understanding of 
the interplay of the different collinear limits has been achieved in this paper,
we believe that it 
should be straightforward to extend    these results to processes  with an arbitrary number of jets at a hadron collider.

%-----------------------------------------------------
%                     ACKNOWLEDGMENTS
%-----------------------------------------------------

\acknowledgments
We are grateful to Max Delto for discussions concerning integrated triple-collinear subtraction terms. 
The research of K.M. and D.M.T. is supported by the Deutsche Forschungsgemeinschaft (DFG, German Research Foundation) under grant no.\ 396021762 - TRR 257. 
The research of M.T. is supported by ``Karlsruhe School of Elementary and Astroparticle Physics'' (KSETA) through a grant from ``Deutscher Akademischer Austauschdienst'' (DAAD).
F.D. and C.S-S. wish to thank the CERN Theoretical Physics Department for hospitality while part of this work was carried out.  
R.R.\ is partially supported by the Italian Ministry of Universities and Research (MUR) through grant PRIN2022BCXSW9. 
The work of R.R.\ was performed in part at Aspen Center for Physics, which is supported by National Science Foundation grant PHY-2210452. 
This research was supported in part by grant NSF PHY-2309135 to the Kavli Institute for Theoretical Physics (KITP).

%-----------------------------------------------------
%                     APPENDIX
%-----------------------------------------------------

\appendix
\newpage
\section{Collection of relevant constants, functions and operators}
\label{sec:Splitting}

In this Appendix, we collect and define quantities that are used in the main text of this paper.  
We supplement the material presented here with an ancillary file, \usefulfunctions, which contains the $\ep$-series expansions of the functions listed in Table \ref{table_appendix_ancillary_file}.
To simplify various formulas, we use the following notation
\begin{equation}
\begin{gathered}
    \barz = 1 - z \,, 
    \qquad
    \calD_n(z) = \left[\frac{\log^n(1-z)}{1-z}\right]_+ \,,
    \\
    \Ltildei = \log \left(\frac{2 E_i}{\mu}\right) \,, 
    \qquad 
    L_i = \log \left(\frac{\Emax}{E_i} \right) \,, 
    \qquad 
    \Lmax = \log \left(\frac{2\Emax}{\mu} \right) \,.
\end{gathered}
\label{eq_zbar_Li_Ltildei_defs}
\end{equation}
\begin{longtable}[!h]{
    >{\centering\arraybackslash}m{30mm}  
    >{\centering\arraybackslash}m{30mm} 
    >{\centering\arraybackslash}m{30mm} 
    >{\centering\arraybackslash}m{44.5mm}
    }
    %%%%%%%%%%%%%%%%%%%%%%%%%%%%%%%%%%%%%%%%%%%%%%%%%%%%%%%%%%%%%
    %                          CAPTION
    %%%%%%%%%%%%%%%%%%%%%%%%%%%%%%%%%%%%%%%%%%%%%%%%%%%%%%%%%%%%%
    \caption{
    List of functions collected in the ancillary file \usefulfunctions.  
    The first column provides the names of the functions; the second specifies the index range $i$; the third indicates the equation where they appear; the fourth provides their names in \usefulfunctions.  
    For brevity, $\calP_{qq}$, $\calP_{qg}$, $\calP_{gq}$, and $\calP_{gg}$ are encoded in $\calP_{xy}$.  
    To extract a specific configuration -- e.g., the fifth row -- call \texttt{Pqq[z,k,En]}, \texttt{Pqg[z,k,En]}, etc.  
    This convention applies to all splittings in the second and third blocks of the table.  
    For final-state anomalous dimensions, replace \texttt{wTOxy} with \texttt{qTOqg}, \texttt{qTOgq}, \texttt{gTOqq}, or \texttt{gTOgg} to find the corresponding functions in \usefulfunctions.  
    The last four functions refer to triple-collinear contributions; for further details, see the two paragraphs below \eq\eqref{eq_expansions_prova}.     } 
    \label{table_appendix_ancillary_file} \\
    %%%%%%%%%%%%%%%%%%%%%%%%%%%%%%%%%%%%%%%%%%%%%%%%%%%%%%%%%%%%%
    %                          TABLE
    %%%%%%%%%%%%%%%%%%%%%%%%%%%%%%%%%%%%%%%%%%%%%%%%%%%%%%%%%%%%%
    \hline
    \multicolumn{4}{c}{\textbf{Functions collected in the ancillary file \usefulfunctions}} \\
    \hline\hline
        \emph{Function}  
        & \emph{Interval of} $i$ 
        & \emph{Ref.} 
        &  \emph{Name in the ancillary file} 
        \\ \hline \hline 
    %%%%%%%%%%%%%%%%%%%%%%%%%%%%%%%%%%%%%%%%%%%%%%%%%%%%%%%%%%%%%
    % SPIN-CORRELATED QUANTITIES
    %%%%%%%%%%%%%%%%%%%%%%%%%%%%%%%%%%%%%%%%%%%%%%%%%%%%%%%%%%%%%
    \multicolumn{4}{c}{\emph{Integrals in spin-correlations}} \\ \hline
        $\delta_g(\ep)$    
        & / 
        & \eq\eqref{eq_deltam_deltamperp_def}
        & \texttt{\textdelta g[\textepsilon]}  
        \\
        %%%%%%%%%%%%%%%%%%%%%%%%%%%%%%%%%%%%%%%%%%%%%%%%%%%%%%%%%%%%%
        $\delta_q(\ep)$    
        & / 
        & \eq\eqref{eq_deltam_deltamperp_def}
        & \texttt{\textdelta q[\textepsilon]}  
        \\
        %%%%%%%%%%%%%%%%%%%%%%%%%%%%%%%%%%%%%%%%%%%%%%%%%%%%%%%%%%%%%
        $\delta_g^\perp(\ep)$    
        & / 
        & \eq\eqref{eq_deltam_deltamperp_def}
        & \texttt{\textdelta gPerp[\textepsilon]}  
        \\
        %%%%%%%%%%%%%%%%%%%%%%%%%%%%%%%%%%%%%%%%%%%%%%%%%%%%%%%%%%%%%
        $\delta_q^\perp(\ep)$    
        & / 
        & \eq\eqref{eq_deltam_deltamperp_def}
        & \texttt{\textdelta qPerp[\textepsilon]}  
        \\ \hline
    %%%%%%%%%%%%%%%%%%%%%%%%%%%%%%%%%%%%%%%%%%%%%%%%%%%%%%%%%%%%%
    % SPLITTING FUNCTIONS - TREE LEVEL
    %%%%%%%%%%%%%%%%%%%%%%%%%%%%%%%%%%%%%%%%%%%%%%%%%%%%%%%%%%%%%
    \multicolumn{4}{c}{\emph{Tree-level splitting functions}} \\ \hline
        $\calP_{xy}^{(k)}(z,E_i)$    
        & $i\in \{\inF,\inS\}$ 
        & \eq\eqref{eq_P_ab_k_definition_appendix}
        & \texttt{Pxy[z,k,En]}  
        \\
        %%%%%%%%%%%%%%%%%%%%%%%%%%%%%%%%%%%%%%%%%%%%%%%%%%%%%%%%%%%%%
        $\calP_{xy}^{(k),\text{gen}}(z,E_i)$    
        & $i\in \{\inF,\inS\}$ 
        & \eq\eqref{Eq:Paa_GEN_definition_k_general_appendix}
        & \texttt{PxyGEN[z,k,En]}  
        \\ \hline
    %%%%%%%%%%%%%%%%%%%%%%%%%%%%%%%%%%%%%%%%%%%%%%%%%%%%%%%%%%%%%
    % SPLITTING FUNCTIONS - 1 LOOP
    %%%%%%%%%%%%%%%%%%%%%%%%%%%%%%%%%%%%%%%%%%%%%%%%%%%%%%%%%%%%%
    \multicolumn{4}{c}{\emph{One-loop splitting functions}} \\ \hline
        $\PAPone_{gg}(z)$    
        & $i\in \{\inF,\inS\}$ 
        & /
        & \texttt{PAPggOneL[z]}  
        \\
        %%%%%%%%%%%%%%%%%%%%%%%%%%%%%%%%%%%%%%%%%%%%%%%%%%%%%%%%%%%%%
        $\PAPone_{qq,\NStilde}(z)$    
        & $i\in \{\inF,\inS\}$ 
        & /
        & \texttt{PAPqqOneLns[z]}  
        \\
        %%%%%%%%%%%%%%%%%%%%%%%%%%%%%%%%%%%%%%%%%%%%%%%%%%%%%%%%%%%%%
        $P_{xy,\rmi}^{\rm 1L}(z)$    
        & $i\in \{\inF,\inS\}$ 
        & \eq\eqref{eq_P_ab_k_1L_definition}
        & \texttt{PxyOneLISR[z]}  
        \\
        %%%%%%%%%%%%%%%%%%%%%%%%%%%%%%%%%%%%%%%%%%%%%%%%%%%%%%%%%%%%%
        $P_{xy}^{\rm 1L}(z)$    
        & $i\in \HPf$ 
        & \eq\eqref{eq_gamma_nk_1L_appendix}
        & \texttt{PxyOneLFSR[z]}  
        \\
        %%%%%%%%%%%%%%%%%%%%%%%%%%%%%%%%%%%%%%%%%%%%%%%%%%%%%%%%%%%%%
        $\calP_{xy}^{\rm 1L}(z,E_i)$    
        & $i\in \{\inF,\inS\}$ 
        & \eq\eqref{eq_P_ab_k_1L_definition}
        & \texttt{PxyOneL[z,En]}  
        \\
        %%%%%%%%%%%%%%%%%%%%%%%%%%%%%%%%%%%%%%%%%%%%%%%%%%%%%%%%%%%%%
        $\calP_{xy}^{\rm 1L, gen}(z,E_i)$    
        & $i\in \{\inF,\inS\}$ 
        & \eq\eqref{eq_Paa_1L_GEN_definition}
        & \texttt{PxyOneLGEN[z,En]}  
        \\ \hline
    %%%%%%%%%%%%%%%%%%%%%%%%%%%%%%%%%%%%%%%%%%%%%%%%%%%%%%%%%%%%%
    % ANOMALOUS DIMENSIONS - TREE LEVEL
    %%%%%%%%%%%%%%%%%%%%%%%%%%%%%%%%%%%%%%%%%%%%%%%%%%%%%%%%%%%%%
    \multicolumn{4}{c}{\emph{Tree-level anomalous dimensions}} \\ \hline
        $\Gamma_{i, g}^{(k)}$    
        & $i\in \{\inF,\inS\}$ 
        & \eq\eqref{Eq:Gamma_1_2_definition_k_general_appendix}
        & \texttt{\textGamma gISR[k,En]} 
        \\
        %%%%%%%%%%%%%%%%%%%%%%%%%%%%%%%%%%%%%%%%%%%%%%%%%%%%%%%%%%%%%
        $\Gamma_{i, q}^{(k)}$    
        & $i\in \{\inF,\inS\}$ 
        & \eq\eqref{Eq:Gamma_1_2_definition_k_general_appendix}
        & \texttt{\textGamma qISR[k,En]} 
        \\
        %%%%%%%%%%%%%%%%%%%%%%%%%%%%%%%%%%%%%%%%%%%%%%%%%%%%%%%%%%%%%
        $\gamma_{z, w \to xy}^{nk}(L_i)$    
        & $i\in \HPf$ 
        & \eq\eqref{eq_usual_anom_dimens_defs}
        & \texttt{\textgamma wTOxy[n,k,En]} 
        \\
        %%%%%%%%%%%%%%%%%%%%%%%%%%%%%%%%%%%%%%%%%%%%%%%%%%%%%%%%%%%%%
        $\Gamma_{i, w \to xy}^{(k)}$    
        & $i\in \HPf$ 
        & \eq\eqref{eq_Gamma_a_to_bc_defs_appendix}
        & \texttt{\textGamma wTOxy[k,En]} 
        \\
        %%%%%%%%%%%%%%%%%%%%%%%%%%%%%%%%%%%%%%%%%%%%%%%%%%%%%%%%%%%%%
        $\Gamma_{i, g}^{(k)}$    
        & $i\in \HPf$ 
        & \eq\eqref{eq_FSR_generalized_anom_dim_Gamma_q_g}
        & \texttt{\textGamma gFSR[k,En]} 
        \\
        %%%%%%%%%%%%%%%%%%%%%%%%%%%%%%%%%%%%%%%%%%%%%%%%%%%%%%%%%%%%%
        $\Gamma_{i, q}^{(k)}$    
        & $i\in \HPf$ 
        & \eq\eqref{eq_FSR_generalized_anom_dim_Gamma_q_g}
        & \texttt{\textGamma qFSR[k,En]} 
        \\ \hline
    %%%%%%%%%%%%%%%%%%%%%%%%%%%%%%%%%%%%%%%%%%%%%%%%%%%%%%%%%%%%%
    % ANOMALOUS DIMENSIONS - 1 LOOP
    %%%%%%%%%%%%%%%%%%%%%%%%%%%%%%%%%%%%%%%%%%%%%%%%%%%%%%%%%%%%%
    \multicolumn{4}{c}{\emph{One-loop anomalous dimensions}} \\ \hline
        $\Gamma_{i, g}^{\rm 1L}$    
        & $i\in \{\inF,\inS\}$ 
        & \eq\eqref{eq_Gamma_1L_ISR_definition}
        & \texttt{\textGamma OneLgISR[En]} 
        \\
        %%%%%%%%%%%%%%%%%%%%%%%%%%%%%%%%%%%%%%%%%%%%%%%%%%%%%%%%%%%%%
        $\Gamma_{i, q}^{\rm 1L}$    
        & $i\in \{\inF,\inS\}$ 
        & \eq\eqref{eq_Gamma_1L_ISR_definition}
        & \texttt{\textGamma OneLqISR[En]} 
        \\
        %%%%%%%%%%%%%%%%%%%%%%%%%%%%%%%%%%%%%%%%%%%%%%%%%%%%%%%%%%%%%
        $\gamma_{z, w \to xy}^{33,\rm 1L}(L_i)$    
        & $i\in \HPf$ 
        & \eq\eqref{eq_gamma_nk_1L_appendix}
        & \texttt{\textgamma OneLwTOxy[En]} 
        \\
        %%%%%%%%%%%%%%%%%%%%%%%%%%%%%%%%%%%%%%%%%%%%%%%%%%%%%%%%%%%%%
        $\Gamma_{i, w \to xy}^{\rm 1L}$    
        & $i\in \HPf$ 
        & \eq\eqref{eq_Gamma_1L_FSR_definition}
        & \texttt{\textGamma OneLwTOxy[En]} 
        \\ 
        %%%%%%%%%%%%%%%%%%%%%%%%%%%%%%%%%%%%%%%%%%%%%%%%%%%%%%%%%%%%%
        $\Gamma_{i, g}^{\rm 1L}$    
        & $i\in \HPf$ 
        & \eq\eqref{eq_Gamma_1L_q_def_appendix}
        & \texttt{\textGamma OneLgFSR[En]} 
        \\
        %%%%%%%%%%%%%%%%%%%%%%%%%%%%%%%%%%%%%%%%%%%%%%%%%%%%%%%%%%%%%
        $\Gamma_{i, q}^{\rm 1L}$    
        & $i\in \HPf$ 
        & \eq\eqref{eq_Gamma_1L_q_def_appendix}
        & \texttt{\textGamma OneLqFSR[En]} 
        \\ \hline
    %%%%%%%%%%%%%%%%%%%%%%%%%%%%%%%%%%%%%%%%%%%%%%%%%%%%%%%%%%%%%
    % TRIPLE COLLINEAR
    %%%%%%%%%%%%%%%%%%%%%%%%%%%%%%%%%%%%%%%%%%%%%%%%%%%%%%%%%%%%%
    \multicolumn{4}{c}{\emph{Triple-collinear splitting functions}} \\ \hline
        ${\rm TC}_g^{\rm ISR}(z,E_i)$    
        & $i\in \{\inF,\inS\}$ 
        & /
        & \texttt{TCgISR[z,En]}  
        \\
        %%%%%%%%%%%%%%%%%%%%%%%%%%%%%%%%%%%%%%%%%%%%%%%%%%%%%%%%%%%%%
        ${\rm TC}_q^{\rm ISR}(z,E_i)$    
        & $i\in \{\inF,\inS\}$ 
        & /
        & \texttt{TCqISR[z,En]}  
        \\
        %%%%%%%%%%%%%%%%%%%%%%%%%%%%%%%%%%%%%%%%%%%%%%%%%%%%%%%%%%%%%
        ${\rm TC}_g^{\rm FSR}(E_i)$    
        & $i\in\HPfg$ 
        & /
        & \texttt{TCgFSR[En]}  
        \\
        %%%%%%%%%%%%%%%%%%%%%%%%%%%%%%%%%%%%%%%%%%%%%%%%%%%%%%%%%%%%%
        ${\rm TC}_q^{\rm FSR}(E_i)$    
        & $i\in\HPfq$ 
        & /
        & \texttt{TCqFSR[En]}  
        \\
        %%%%%%%%%%%%%%%%%%%%%%%%%%%%%%%%%%%%%%%%%%%%%%%%%%%%%%%%%%%%%
        \hline
\end{longtable} 
\bigskip
%------------------------------------------------------------------------------------------------

\subsection{Useful constants}
\label{app_subsec_useful_constants}
This section 
generalizes 
Appendix A.1 of Ref.~\cite{Devoto:2023rpv} where many  useful constants have been defined. 
We denote the color-charge operators with $\T_i$. 
Squares of these operators  are the Casimir operators of the corresponding representations of the QCD gauge group SU(3), i.e.
\begin{equation}
\begin{split}
    \T_q^2 = \T_\qb^2 = \Cf = \frac{\Nc^2-1}{2\Nc} \,,  \qquad 
    \T_g^2 = \Ca  = \Nc \,,
\end{split}
\end{equation}
where $\Nc$ is the number of colors. 
Quark and gluon anomalous dimensions read
\begin{equation}
    \gamma_q = \frac{3}{2} \Cf \,,
    \qquad
    \gamma_g = \frac{11}{6} \, \Ca - \frac{2}{3} \TR \nf \,,
    \label{eq_appendix_anomalous_dimensions}
\end{equation}
where $\TR = 1/2$, and $\nf$ is the number of massless quark flavors. 

The strong coupling is renormalized in the $\overline{\rm MS}$ scheme. The relation between the bare and renormalized coupling constants reads  
\begin{align}
\label{eq:as-renorm}
    \gsb^2
    = \gs^2 \Sep \mu^{2\ep} \left[1 - \amu \frac{\beta_0}{\ep} 
    + \left(\amu\right)^{\!\! 2} \left(\frac{\beta_0^2}{\ep^2} 
    - \frac{\beta_1}{2\ep} \right)
    + \mathcal{O}(\as^3) \right] \,,
\end{align}
where $\Sep = (4\pi)^{-\ep} e^{\ep\EulerGamma}$, and
\begin{equation}
    \beta_0 = \frac{11}{6}\Ca - \frac{2}{3}\TR\nf  \,,
    \qquad
    \beta_1 = \frac{ 17}{6} \Ca^2 - \frac{5}{3}\Ca\TR\nf - \Cf\TR\nf \,.
\end{equation}
We note that, at leading order, the gluon anomalous dimension and the coefficient of the QCD $\beta$-function coincide,  $\gamma_g = \beta_0$.
Furthermore, it is  convenient to define the following quantity
\begin{align}
  \label{eq:asbr-def}
  \asbr
  \equiv{}&
  \frac{ \as(\mu) }{ 2\pi }
  \frac{e^{\ep\EulerGamma} }{ \Gamma(1-\ep) } 
  \,,
\end{align}
related to the coupling constant. 

In the main body of this paper, we have used $\ep$-dependent constants
\begin{align}
    & N_\ep^{(b,d)} = \frac{\Gamma(1-\ep) \, \Gamma(1+2\ep)}{\Gamma(1+\ep)} 
    = 1+\frac{\pi^2}3\ep^2 + \order{\ep^3} \,, 
    \notag \\
    %---------------------------------------------------
    & N_{\Fp \parallel \Sp}(\ep) = 2^{2\ep} \frac{\Gamma(1+2\ep) \Gamma(1-2\ep)}{\Gamma(1+\ep) \Gamma(1-\ep)} 
    = 1 +2 \ep \log 2 +\frac{1}{2} \ep^2 \big(\pi ^2 +4 \log^2 2\big) + \order{\ep^3} \,, 
    \label{eq:normalisation} \\
    %---------------------------------------------------
    & N_c(\ep) = - \frac{\Gamma(1-\ep) \Gamma(1-2\ep)}{\Gamma(1-3\ep)} + \frac{2\Gamma^2(1-\ep)}{\Gamma(1-2\ep)}
    = 1 + \order{\ep^3} \,, \notag \\
    %---------------------------------------------------
    & \Nscbd = 2^{2\ep} \frac{\Gamma(1+2\ep) \Gamma^3(1-2\ep)}{\Gamma(1-3\ep)\Gamma(1+\ep)\Gamma^2(1-\ep)} 
    = 1 +2 \ep \log 2 +\frac{1}{3}\ep^2 \big(\pi^2 + 6\log^2 2\big) + \order{\ep^3} \,, \notag 
\end{align}
which originate from the definition of certain angular integrals. 

Singular parts of double-virtual amplitudes, written in Ref.~\cite{Catani:1998bh}, require the following quantities 
\begin{align}
    K & = \left( \frac{67}{18} - \frac{\pi^2}{6} \right) \Ca  - \frac{10}{9} \TR \nf \,,
    \notag \\
    c_\ep & = \frac{e^{-\ep \gamma_E} \Gamma(1-2\ep)}{\Gamma(1-\ep)} = 1+ \frac{\pi^2}{4} \, \ep^2 + \frac73 \zeta_3 \, \ep^3 + \mathcal{O}(\ep^4) \,,
    \label{eq:constants_VV} \\
    \cTilde & = \frac{\Gamma(1-\ep)}{e^{\ep \EulerGamma}} = 1 + \frac{\pi^2}{12} \ep^2 + \order{\ep^3} \,. \notag 
\end{align}
To parametrize  integrated double-soft contributions, we have introduced
\begin{equation}
\begin{split}
    c_1(\ep) & = 1 + \left(\frac{\pi^2}{6} - \frac{32}{9}\right)\ep^2 + \left(\frac{217}{27} - \frac{137}{9}\log2 - 22 \log^2 2 + \frac{11 \zeta_3}{2}\right)\ep^3 \,, \\
    %----------------------------------------------------
    c_2(\ep) & = 1 + \frac{\pi^2}{3} \ep^2  \,, \\
    %----------------------------------------------------
    c_3(\ep) & = 4 \log 2+ 8 \ep \log^2 2 \,, \\
    %----------------------------------------------------
    c_4(\ep) & = - \frac{13}{6} + \bigg(\frac{125}{18} - \frac{35}{3} \log2 - 12 \log^2{2} \bigg) \,\ep \,.
\end{split}
\label{eq:constants_SS}
\end{equation}
We emphasize that the above expressions are \emph{exact} and do not contain higher powers 
of $\ep$.
To describe  soft and collinear limits of the real-virtual contributions, it is convenient to define the following quantities
\begin{equation}
\begin{split}
    A_K(\ep) & = \frac{\Gamma^3(1+\ep) \, \Gamma^5(1-\ep)}{\Gamma(1+2\ep) \, \Gamma^2(1-2\ep)} = 1 - \frac{\pi^2}{3} \ep^2 + \order{\ep^3} \,, \\
    %---------------------------------------------------
    h_\rmc(\ep) & =
    \frac{\Gamma^2(1-2\ep) \Gamma(1+\ep)}{\Gamma(1-3\ep)} = 1 + \order{\ep^3} \,.
    \label{eq:constants_RV}
\end{split}
\end{equation}

%%%%%%%%%%%%%%%%%%%%%%%%%%%%%%%%%%%%%%%%%%%%%%%%%%%%%%%%%%%%%%%%%%%%%%%%%%%%%%%%%%%%%%%%%%%%%%%%%%%%%%%%%%%%%%%%%%%%%%%%%%%%%%%%%%%%%%%%%%%%%%%%%%%%%%%%%%%%%%%%%%%%%%%%

\subsection{Tree-level, one-loop and Altarelli-Parisi splitting functions}
\label{appendix_subsec_AP_splittings}

\begin{figure}[t]
    \centering
    \begin{align*}
    %%%%%%%%%%%%%%%%%%%%%%%%%%%%%%%%%%%%%%%%%%%%%%%%%%%%%%%%%%%%%%%%%%%%%%%%%
    % FIRST FIGURE
    %%%%%%%%%%%%%%%%%%%%%%%%%%%%%%%%%%%%%%%%%%%%%%%%%%%%%%%%%%%%%%%%%%%%%%%%%
    {\rm FSR:} \quad & C_{i\Fp} \, \Delta^{(\Fp)} \left|
    \begin{tikzpicture}[baseline=(c.base)]
    \begin{feynman}
        \vertex (ic);
        \vertex [blob, right=0.5cm of ic] (c) {};
        \vertex [right=1.3cm of c] (i1);
        \vertex [right=1.9cm of i1] (i2);
        \vertex [above=0.7cm of i2] (i3);
        \vertex [above=0.125cm of ic] (f2);
        \vertex [above=0.25cm of f2] (f21);
        \vertex [above=0.25cm of f21] (f22);
        \vertex [below=0.125cm of ic] (f3);
        \vertex [below=0.25cm of f3] (f31);
        \vertex [below=0.25cm of f31] (f32);
        \diagram{
            (c) -- [red, plain, edge label'=$p_{[i\Fp]}$] (i1) -- [red, plain, edge label'=$p_i$] (i2);
            (i1) -- [red, plain, edge label={$p_\Fp \sim \zb\, p_{[i\Fp]}$}, sloped] (i3);
            (c) -- [plain] (f2);
            (c) -- [plain] (f21);
            (c) -- [plain] (f22);
            (c) -- [plain] (f3);
            (c) -- [plain] (f31);
            (c) -- [plain] (f32);
        };
    \end{feynman} 
    \end{tikzpicture} 
    \right|^2
    \xrightarrow{\theta_{i\Fp} \to \, 0}  \frac{\gsb^2 \mu_0^{2\ep}}{p_i \cdot p_\Fp} \, z P_{f_{[i\Fp]} f_i}(z) \times
    \left|
    \begin{tikzpicture}[baseline=(c.base)]
    \begin{feynman}
        \vertex (ic);
        \vertex [blob, right=0.5cm of ic] (c) {};
        \vertex [right=1.3cm of c] (i1);
        \vertex [above=0.125cm of ic] (f2);
        \vertex [above=0.25cm of f2] (f21);
        \vertex [above=0.25cm of f21] (f22);
        \vertex [below=0.125cm of ic] (f3);
        \vertex [below=0.25cm of f3] (f31);
        \vertex [below=0.25cm of f31] (f32);
        \diagram{
            (c) -- [red, plain, edge label'=$p_{[i\Fp]}$] (i1); 
            (c) -- [plain] (f2);
            (c) -- [plain] (f21);
            (c) -- [plain] (f22);
            (c) -- [plain] (f3);
            (c) -- [plain] (f31);
            (c) -- [plain] (f32);
        };
    \end{feynman} 
    \end{tikzpicture}
    \right|^2 \\
    %%%%%%%%%%%%%%%%%%%%%%%%%%%%%%%%%%%%%%%%%%%%%%%%%%%%%%%%%%%%%%%%%%%%%%%%%
    % SECOND FIGURE
    %%%%%%%%%%%%%%%%%%%%%%%%%%%%%%%%%%%%%%%%%%%%%%%%%%%%%%%%%%%%%%%%%%%%%%%%%
    {\rm ISR:} \quad & C_{i\Fp} \, \Delta^{(\Fp)} \left|
    \begin{tikzpicture}[baseline=(c.base)]
    \begin{feynman}
        \vertex (ic);
        \vertex [blob, right=1.2cm of ic] (c) {};
        \vertex [left=1.2cm of ic] (i1);
        \vertex [above right=0.7cm and 1.2cm of ic] (i2);
        \vertex [right=0.8cm of c] (fc);
        \vertex [above=0.125cm of fc] (f2);
        \vertex [above=0.25cm of f2] (f21);
        \vertex [above=0.25cm of f21] (f22);
        \vertex [below=0.125cm of fc] (f3);
        \vertex [below=0.25cm of f3] (f31);
        \vertex [below=0.25cm of f31] (f32);
        \diagram{
            (i1) -- [red, plain, edge label'=$p_i$] (ic) -- [red, plain, edge label'=$p_{[i\Fp]}$] (c);
            (i2) -- [red, plain, edge label={$p_\Fp$}, sloped] (ic);
            (c) -- [plain] (f2);
            (c) -- [plain] (f21);
            (c) -- [plain] (f22);
            (c) -- [plain] (f3);
            (c) -- [plain] (f31);
            (c) -- [plain] (f32);
        };
    \end{feynman} 
    \end{tikzpicture} 
    \right|^2
    \xrightarrow{\theta_{i\Fp} \to \, 0} \frac{\gsb^2 \mu_0^{2\ep}}{p_i \cdot p_\Fp} \, P_{f_{[i\Fp]} f_i, \rmi}(z) \times \frac{1}{z}
    \left|
    \begin{tikzpicture}[baseline=(c.base)]
    \begin{feynman}
        \vertex (ic);
        \vertex [blob, right=1.2cm of ic] (c) {};
        \vertex [right=0.8cm of c] (fc);
        \vertex [above=0.125cm of fc] (f2);
        \vertex [above=0.25cm of f2] (f21);
        \vertex [above=0.25cm of f21] (f22);
        \vertex [below=0.125cm of fc] (f3);
        \vertex [below=0.25cm of f3] (f31);
        \vertex [below=0.25cm of f31] (f32);
        \diagram{
            (ic) -- [red, plain, edge label'=$p_{[i\Fp]}$] (c);
            (c) -- [plain] (f2);
            (c) -- [plain] (f21);
            (c) -- [plain] (f22);
            (c) -- [plain] (f3);
            (c) -- [plain] (f31);
            (c) -- [plain] (f32);
        };
    \end{feynman} 
    \end{tikzpicture}
    \right|^2 
    \end{align*} 
    %%%%%%%%%%%%%%%%%%%%%%%%%%%%%%%%%%%%%%%%%%%%%%%%%%%%%%%%%%%%%%%%%%%%%%%%%
    % CAPTION
    %%%%%%%%%%%%%%%%%%%%%%%%%%%%%%%%%%%%%%%%%%%%%%%%%%%%%%%%%%%%%%%%%%%%%%%%%
    \caption{Graphical representation of the convention adopted for collinear factorization.  
    The top figure describes the final-state splitting process $[i\Fp]^* \to i(z) + \Fp(\zb)$, where $z=1-E_\Fp/E_{[i\Fp]}$, and $\zb = 1-z$.  
    The bottom figure corresponds to the initial-state splitting process $i \to [i\Fp]^* + \Fp$, with $z=1-E_\Fp/E_i$.   
    We note that in the case of FSR, the action of $C_{i\Fp}$ on $\Delta^{(\Fp)}$ produces an extra factor of $z$, which does not occur in ISR. }
    \label{fig_collinear_factorization_FSR_ISR}
\end{figure}

Consider the final-state splitting process $[i\Fp]^* \to i(z) + \Fp(1-z)$, where $i$ and $\Fp$ are two partons with flavors $f_i$ and $f_\Fp$, respectively.  
The variable $z=1-E_\Fp/E_{[i\Fp]}$ represents the energy fraction carried by parton $i$ in the collinear splitting, while parton $\Fp$ takes the energy fraction $1-z$.  
As for $[i\Fp]$, it is the clustered ``mother'' parton.  
We denote the spin-averaged final-state splitting functions as $P_{f_{[i\Fp]}f_i}(z)$ (cf.\ Figure \ref{fig_collinear_factorization_FSR_ISR}, above).  
These functions read  
\begin{equation}
\begin{split}
    P_{qq}(z) & \eqdef \Cf \left[\frac{1+z^2}{1-z} - \ep(1-z)\right] \,, \\ 
    P_{qg}(z) & \eqdef \Cf \left[\frac{1+(1-z)^2}{z} - \ep z\right] \equiv P_{qq}(1-z) \,, \\
    P_{gq}(z) & \eqdef \TR \left[1 - \frac{2z(1-z)}{1-\ep}\right] \,, \\
    P_{gg}(z) & \eqdef 2\Ca \left[\frac{z}{1-z} + \frac{1-z}{z} + z(1-z)\right] \,.
\end{split}
\label{eq_final_state_splitting_functions_defs_appendix}
\end{equation}

Next, we consider the initial-state splitting process $i \to [i\Fp]^* + \Fp$, where $i$ is now the incoming parton, $[i\Fp]$ is the parton that enters the hard scattering, and $\Fp$ is an outgoing parton. 
The variable $z$ corresponds to $z=1-E_\Fp/E_i$.
In this case, we denote the initial-state splitting function, averaged over initial-state color and polarizations, as $P_{[i\Fp]i,\rmi}(z)$ (cf.\ Figure \ref{fig_collinear_factorization_FSR_ISR}, below).  
These functions are related to the splittings in \eq\eqref{eq_final_state_splitting_functions_defs_appendix} by the following relations
\begin{align}
    P_{qq,\rm i}(z) & \eqdef -z P_{qq}(1/z) \equiv P_{qq}(z) \,, \notag \\
    P_{qg,\rm i}(z) & \eqdef \left[\frac{2N_c}{2(1-\ep) (N_c^2-1)}\right] z P_{qg}(1/z) \equiv P_{gq}(z) \,, 
    \label{eq_initial_state_splitting_functions_defs_appendix} \\
    P_{gq,\rm i}(z) & \eqdef \left[\frac{2(1-\ep) (N_c^2-1)}{2N_c}\right] z P_{gq}(1/z) \equiv P_{qg}(z) \,,  \notag \\
    P_{gg,\rm i}(z) & \eqdef -z P_{gg}(1/z) \equiv P_{gg}(z) \,. \notag
\end{align}

We observe that the notation just described for initial- and final-state splittings, and illustrated in Figure \ref{fig_collinear_factorization_FSR_ISR}, applies to all tree-level and one-loop splittings used in this article.  
Therefore, it must be applied to all functions defined in the remainder of this appendix, as well as those collected in \finalresult\ (cf.\ Table \ref{table_ancillary_file}) and \usefulfunctions\ (cf.\ Table \ref{table_appendix_ancillary_file}).  

Finally, we report the conventional leading-order Altarelli-Parisi splitting functions that we use throughout the paper.
They read \cite{Ellis:1996mzs}
\begin{align}
    \PAP_{qq}(z) & \eqdef \Cf \left[2\calD_0(z) - (1+z) + \frac{3}{2} \delta(1-z) \right] \, , 
    \notag
    \\
    \PAP_{qg}(z) & \eqdef \TR \Big[(1-z)^2 + z^2\Big] \, , 
    \label{Eq_PAP_0_definition_appendix}
    \\
    \PAP_{gq}(z) & \eqdef \Cf \left[\frac{1 + (1-z)^2}{z}\right] \, , 
    \notag
    \\
    \PAP_{gg}(z) & \eqdef 2\Ca \left[\calD_0(z) + z(1-z) + \frac{1}{z} - 2 \right] + \beta_0 \delta(1-z) \,.
    \notag
\end{align}
We also require the  one-loop Altarelli-Parisi splitting functions; they  can be found in \Sec4.3 of Ref.~\cite{Ellis:1996mzs}.
They are also given in the ancillary file \usefulfunctions, see Table \ref{table_appendix_ancillary_file}.
We point out that, for the $qq$ splitting, we need the \emph{non-singlet} splitting function from which the analog of the interference contribution of identical quarks has been subtracted.

%%%%%%%%%%%%%%%%%%%%%%%%%%%%%%%%%%%%%%%%%%%%%%%%%%%%%%%%%%%%%%%%%%%%%%%%%%%%%%%%%%%%%%%%%%%%%%%%%%%%%%%%%%%%%%%%%%%%%%%%%%%%%%%%%%%%%%%%%%%%%%%%%%%%%%%%%%%%%%%%%%%%%%%%%%%%%%%%%%%%%%%%%%%%%%%%%%%%%%%%

\subsection{Generalized splitting functions  and anomalous dimensions}
\label{appendix_sec_generalized_splitting}

\subsubsection{Tree-level}
\label{subsubsec_app_tree_level}

Phase-space remnants and 
damping factors introduce 
additional factors to integrands over splitting functions that we require.
For this reason, it is convenient to define generalized splitting functions and their integrals.  

For the initial-state splitting where parton $\Fp$ becomes collinear to an initial-state parton $\inF$ of flavor $f_\inF$, the following integral appears 
\begin{equation}
    E_a^{-k\ep} \int_{0}^{1}  \dz\,  (1-z)^{-k \ep} P_{f_{[\inF\Fp]} f_\inF,\rmi}(z) \, F(z) \,.
\end{equation}
The initial-state splitting functions $P_{\alpha \beta, \rm i}$ are given in \eq\eqref{eq_initial_state_splitting_functions_defs_appendix}, and the function $F(z)$ typically contains the relevant matrix element squared. 
Such integrals may contain infrared divergences which need to be extracted.
To facilitate this, we define a new splitting function 
\begin{equation}
    \calP_{f_{[\inF\Fp]} f_\inF}^{(k)}(z,E_\inF) 
    \eqdef
    \oS_z \Big[(1-z)^{-k \ep} P_{f_{[\inF \Fp]} f_\inF, \rmi}(z)\Big]
    - 2 \delta_{f_{[\inF \Fp]} f_\inF} \, \T_{f_{[\inF \Fp]}}^2 \frac{1 - e^{-k \ep L_\inF}}{k \ep}\delta(1-z) \,.
\label{eq_P_ab_k_definition_appendix}
\end{equation}
Here, $\oS_z = \iden - S_z$, where $S_z$ subtracts the soft singularity at $z =1$. 
The relevant powers of phase-space factors $(1-z)^{-\ep}$ are $k=2$ at NLO, and $k=4$ at NNLO. 
Explicit expressions for the functions $\calP_{\alpha\beta}^{(k)}$ in \eq\eqref{eq_P_ab_k_definition_appendix} are provided in the ancillary file \usefulfunctions, see Table \ref{table_appendix_ancillary_file}.
We note that we omit the superscript for $k=2$ when writing the generalized splitting functions and their integrals to simplify the notation. 
Furthermore, we point out that functions $\calP_{\alpha \beta}^{(k)}(z,E_\inF)$ in \eq\eqref{eq_P_ab_k_definition_appendix} are, in fact,  energy-independent if $\alpha \not= \beta$. 

For the manipulations described in the main body of the paper, it is 
convenient to rewrite the generalized  splitting functions as follows
\begin{equation}
   - \bigg[\left(\frac{2E_\inF}{\mu}\right)^{\!\!-2\ep} \frac{\Gamma^2(1-\ep)}{\Gamma(1-2\ep)}\bigg]^{\frac{k}{2}} \calP_{\alpha \beta}^{(k)}(z,E_\inF) 
   = \Gamma_{\inF, \alpha}^{(k)} \,
   \delta_{\alpha \beta} \; \delta(1-z) + \calP_{\alpha \beta}^{(k),\text{gen}}(z,E_\inF) \,,
   \label{eq_Pab_k_relation_with_Gamma_a_and_Pab_GEN_k_appendix}
\end{equation}
with
\begin{align}
    \Gamma_{\inF, \alpha}^{(k)} 
    & \eqdef \bigg[\left(\frac{2E_\inF}{\mu}\right)^{\!\!-2\ep} \frac{\Gamma^2(1-\ep)}{\Gamma(1-2\ep)}\bigg]^{\frac{k}{2}} \left[\gamma_{\alpha} + 2\T_{\alpha}^2 \frac{1 - e^{-k\ep L_\inF}}{k \ep}\right] \,, 
    \label{Eq:Gamma_1_2_definition_k_general_appendix} \\
    %---------------------------------
    \calP_{\alpha \beta}^{(k),\text{gen}}(z,E_\inF)
    & \eqdef \bigg[\left(\frac{2E_\inF}{\mu}\right)^{\!\!-2\ep} \frac{\Gamma^2(1-\ep)}{\Gamma(1-2\ep)}\bigg]^{\frac{k}{2}} \left[- \PAP_{\alpha \beta}(z) + \ep \,  \calP_{\alpha \beta}^{(k),\text{fin}}(z)\right] \,. 
    \label{Eq:Paa_GEN_definition_k_general_appendix}
\end{align}
Here, $\PAP_{\alpha \beta}$ are the Altarelli-Parisi splitting functions given in \eq\eqref{Eq_PAP_0_definition_appendix}, while $\calP_{\alpha \beta}^{(k),\text{fin}}(z)$ corresponds to the $\ep$-expansion of $-\calP_{\alpha\beta}^{(k)}(z,E_\inF = \Emax)$ starting with $\order{\ep}$. 
Explicit expressions for quantities that appear in  Eqs \eqref{Eq:Gamma_1_2_definition_k_general_appendix} and \eqref{Eq:Paa_GEN_definition_k_general_appendix} are provided in \usefulfunctions, see Table \ref{table_appendix_ancillary_file}.

We continue with the discussion of the final-state splittings. Then, if   the  final-state parton $\Fp$ becomes collinear to the final-state parton  $i$ of flavour $f_i$, the generalized final-state anomalous dimension reads
\begin{equation}
    \Gamma_{i, f_{[i\Fp]} \to f_i f_\Fp}^{(k)} 
    \eqdef 
    \bigg[\left(\frac{2E_i}{\mu}\right)^{\!\!-2\ep} \frac{\Gamma^2(1-\ep)}{\Gamma(1-2\ep)}\bigg]^{\frac{k}{2}} \gamma_{z,f_{[i\Fp]} \to f_i f_\Fp}^{2k}(L_i) \,,
    \label{eq_Gamma_a_to_bc_defs_appendix}
\end{equation}
where 
\begin{equation}
\begin{split}
    \gamma_{g(z), f_{[i\Fp]} \to f_i f_\Fp}^{nk}(L_i) 
    \eqdef & - \int_{0}^{1} \dz \, \oS_z \Big[z^{-n\ep}(1-z)^{-k\ep} g(z) \,  P_{f_{[i\Fp]} f_i}(z)\Big] \\
    & + 2 \delta_{f_{[i\Fp]} f_i} \, \T^2_{f_{[i\Fp]}} \frac{1 - e^{-k\ep L_i}}{k\ep} g(1) \,.
\end{split}
\label{eq_usual_anom_dimens_defs}
\end{equation}
The  splitting functions $P_{\alpha \beta}$ can be found in \eq\eqref{eq_final_state_splitting_functions_defs_appendix}, while the explicit expressions for the quantities defined in Eqs \eqref{eq_Gamma_a_to_bc_defs_appendix} and \eqref{eq_usual_anom_dimens_defs} are provided in \usefulfunctions, see Table \ref{table_appendix_ancillary_file}. 

Finally, we define ``physical'' combinations of quark and gluon collinear anomalous dimensions 
\begin{equation}
    \Gamma_{i,q}^{(k)} \eqdef \Gamma_{i, q \to q g}^{(k)} + \Gamma_{i, q \to g q}^{(k)} \,,
    \qquad
    \Gamma_{i,g}^{(k)} \eqdef \Gamma_{i, g \to g g}^{(k)} + 2\nf\, \Gamma_{i, g \to q\qb}^{(k)} \,,
    \label{eq_FSR_generalized_anom_dim_Gamma_q_g}
\end{equation}
which appear in the operators $\IColl$ and $\ITot$.

%%%%%%%%%%%%%%%%%%%%%%%%%%%%%%%%%%%%%%%%%%%%%%%%%%%%%%%%%%%%%%%%%%%%%%%%%%%%%%%%%%%%%%%%%%%%%%%%%%%%%%%%%%%%%%%%%%%%%%%%%%%%%%%%%%%%%%%%%%%%%%%%%%%%%%%%%%%%%%%%%%%%%%%%

\subsubsection{Spin-correlations}

To describe the spin-correlated contributions  arising from sectors $\theta^{(b)}$ and $\theta^{(d)}$, we require the following
integrals
\begin{equation}
\begin{split}
    \delta_\Fp(\ep) 
    & \eqdef \frac{N_\ep^{(b,d)} \Emax^{4\ep}}{2} 
    \int_{\Emax}^{2\Emax} \frac{\rmd E_{[\Fp \Sp]}}{E_{[\Fp \Sp]}^{1+4\ep }} \int_{1-\xi}^{\xi} \dz\, [z(1-z)]^{-2\ep} \big[P_{g\Fp}(z) + \ep \, P_{g\Fp}^{\perp,\rmr}(z) \big] \,, \\
    \delta_\Fp^{\perp}(\ep) 
    & \eqdef \frac{N_\ep^{(b,d)} \Emax^{4\ep}}{2} 
    \int_{\Emax}^{2\Emax} \frac{\rmd E_{[\Fp \Sp]}}{E_{[\Fp \Sp]}^{1+4\ep }} \int_{1-\xi}^{\xi} \dz\, [z(1-z)]^{-2\ep} P_{g\Fp}^\perp(z) \,,
\end{split}
\label{eq_deltam_deltamperp_def}
\end{equation}
where $\xi = \Emax/E_{[\Fp \Sp]}$, 
$\xa$  can be either a quark or a gluon, and 
the transversal splitting functions are defined as follows
\begin{align}
    \begin{aligned}
        P_{gg}^{\bot}(z) & \eqdef 4 \Ca (1-\ep) \, z (1-z) \,,
        &\quad
        P_{gg}^{\bot, \rmr}(z) & \eqdef 2 \Ca\, z (1-z) (1 - 2 \ep) \,,
        \\
        %----------------------------------------------------------------------------------------
        P_{gq}^{\bot}(z) & \eqdef - 4 \TR \, z (1-z) \,,
        &\quad
        P_{gq}^{\bot, \rmr}(z) & \eqdef - 2 \TR \, z (1-z) \frac{1-2\ep}{1-\ep} \,.
    \end{aligned}
    \label{eq_P_gg_Pgq_perp_defs}
\end{align}
The $\ep$-expansions of the $\delta$-integrals in \eq\eqref{eq_deltam_deltamperp_def} are reported in \usefulfunctions, see Table \ref{table_appendix_ancillary_file}.
Note that we combine the gluon and quark components as
\begin{equation}
    \delta(\ep) \eqdef \delta_g(\ep) + 2\nf\,\delta_q(\ep) \,,
    \qquad
    \delta^\perp(\ep) \eqdef \delta_g^\perp(\ep) + 2\nf\,\delta_q^\perp(\ep) \,.
    \label{eq_delta_ep_delta_perp_defs}
\end{equation}

Additionally, we find it convenient to introduce  the  following notation for the  integrals 
\begin{equation}
    \gamma_{\bot, g\to \Fp\Sp}^{22} \eqdef - \int_0^1 \rmd z \; \frac{P_{g\Fp}^\bot(z)}{ [ z (1-z) ]^{2\ep}} 
     \,,
    \qquad 
    \gamma_{\bot, g\to \Fp\Sp}^{22, \rmr} \eqdef
    - \int_{0}^1 \rmd z \; \frac{P_{g\Fp}^{\bot, \rmr}(z)}{[ z (1-z) ]^{2\ep}}
     \,,
     \label{eq:gamma_tilde}
\end{equation}
where $(\Fp\Sp) = (gg)$ or $(q\qb)$.  ``Physical'' combinations of anomalous dimensions are defined as follows: 
\begin{equation}
    \gamma_{\bot, g}^{22} \eqdef \gamma_{\bot, g\to gg}^{22} + 2\nf\,\gamma_{\bot, g\to q\qb}^{22} \,,
    \qquad
    \gamma_{\bot, g}^{22,\rmr} \eqdef \gamma_{\bot, g\to gg}^{22,\rmr} + 2\nf\,\gamma_{\bot, g\to q\qb}^{22,\rmr} \,.
\label{eq_gamma_perp_g_def}
\end{equation}
In the $\ep=0$ limit, they read   
\begin{equation}
    \gamma_{\bot,g}^{22} = - \frac23 (\Ca - 2\nf \TR) \,,
    \qquad
    \gamma_{\bot,g}^{22, \rmr} = - \frac13 (\Ca - 2\nf \TR) \,,
\label{eq_gamma_perp_ep_0}
\end{equation}  
and are reported in \finalresult, see Table~\ref{table_ancillary_file}.

%%%%%%%%%%%%%%%%%%%%%%%%%%%%%%%%%%%%%%%%%%%%%%%%%%%%%%%%%%%%%%%%%%%%%%%%%%%%%%%%%%%%%%%%%%%%%%%%%%%%%%%%%%%%%%%%%%%%%%%%%%%%%%%%%%%%%%%%%%%%%%%%%%%%%%%%%%%%%%%%%%%%%%%%

\subsubsection{One-loop}

Similar to the discussion of the generalized tree-level splitting functions 
and anomalous dimensions in \Sec\ref{subsubsec_app_tree_level}, we need to define generalized splitting functions for the one-loop case.  For the initial state splitting, when a parton  $\Fp$ becomes collinear to an initial-state parton $\inF$ with flavor $f_\inF$, we define 
\begin{equation}
\begin{split}
\calP_{f_{[\inF \Fp]} f_\inF}^{(k),\mathrm{1L}}(z,E_\inF) 
   \eqdef &\; \oS_z \Big[(1-z)^{-k \ep} P_{f_{[\inF \Fp]} f_\inF, \rmi}^{\text{1L}}(z) \Big]  \\
    & + 2 \delta_{f_{[\inF \Fp]} f_\inF} \Ca \T_{f_{[\inF\Fp]}}^2 \, \frac{1- e^{-(2+k) \ep L_\inF}}{(2+k) \ep^2} \pi \cot(\pi \ep) \delta(1-z)  \,.
\end{split}
\label{eq_P_ab_k_1L_definition}
\end{equation}
The splitting functions $P_{\alpha\beta,\rm i}^{\rm 1L}$ can be obtained from Refs \cite{Bern:1999ry, Kosower:1999rx, Campbell:1999ah}, and they are collected in \usefulfunctions, see Table \ref{table_appendix_ancillary_file}.
In analogy with \eq\eqref{eq_Pab_k_relation_with_Gamma_a_and_Pab_GEN_k_appendix}, we find it convenient to rewrite the splitting function $\calP_{\alpha\beta}^{(k),\mathrm{1L}}$ as follows
\begin{equation}
    \frac{\ep^2}{C_x} \bigg[\left(\frac{2E_\inF}{\mu}\right)^{\!\!-2\ep} \frac{\Gamma^2(1-\ep)}{\Gamma(1-2\ep)}\bigg]^{k} \calP_{\alpha\beta}^{(k),\mathrm{1L}}(z,E_\inF) 
    \eqdef  
    \Gamma_{\inF, \alpha}^{(k),\mathrm{1L}} \delta_{\alpha\beta} \delta(1-z) + \calP_{\alpha\beta}^{(k),\mathrm{1L,gen}}(z,E_\inF) \,,
    \label{eq_relation_Pab1L_vs_Pab1L_gen_appendix}
\end{equation}
where $C_x = 2\Cf - \Ca$ if $\alpha\beta = gq$, and $C_x = \Ca$ otherwise. Furthermore, we define 
\begin{align}
    & \Gamma_{\inF, \alpha}^{(k),\mathrm{1L}} 
    \eqdef 
    \bigg[\left(\frac{2E_\inF}{\mu}\right)^{\!\!-2\ep} \frac{\Gamma^2(1-\ep)}{\Gamma(1-2\ep)}\bigg]^{k} 
    \left[\gamma_{\alpha} + 2 \T_{\alpha}^2 \, \frac{1- e^{-(2+k) \ep L_\inF}}{(2+k)} \pi \frac{\cos(\pi \ep)}{\sin(\pi \ep)} \right] \,, \label{eq_Gamma_1L_ISR_definition} 
    \\
    & \calP^{(k),1\rmL,\gen}_{\alpha\beta}(z,E_\inF) \eqdef \bigg[\left(\frac{2E_\inF}{\mu}\right)^{\!\!-2\ep} \frac{\Gamma^2(1-\ep)}{\Gamma(1-2\ep)}\bigg]^{k} 
    \Big[-\PAP_{\alpha\beta}(z) + \ep \, \calP^{(k),1\rmL,\fin}_{\alpha\beta}(z)\Big] \,.
    \label{eq_Paa_1L_GEN_definition}
\end{align}
In \eq\eqref{eq_Paa_1L_GEN_definition}, $\calP^{(k),1\rmL,\fin}_{\alpha\beta}(z)$ corresponds to the $\ep$-expansion of
\begin{equation}
    \frac{\ep^2}{C_x} \, \calP_{\alpha\beta}^{(k),\mathrm{1L}}(z,E_\inF = \Emax) \,,
\end{equation}
starting from $\order{\ep}$.
Note that, although the above definitions are provided for any value of $k$, we set $k=2$ in the real-virtual contributions.
The explicit expression of the functions in Eqs \eqref{eq_P_ab_k_1L_definition}, \eqref{eq_Gamma_1L_ISR_definition}, \eqref{eq_Paa_1L_GEN_definition}, with $k=2$, are provided in \usefulfunctions, see Table \ref{table_appendix_ancillary_file}.

If the unresolved final-state parton $\Fp$ goes collinear to a final-state parton, say $i$ of flavor $f_i$, the generalized final-state anomalous dimensions read
\begin{equation}
    \Gamma_{i,f_{[i\Fp]} \to f_i f_\Fp}^{(k), \mathrm{1L}} 
    \eqdef 
    - \left[\left(\frac{2E_i}{\mu}\right)^{\!\!-2\ep} \frac{\Gamma^2(1-\ep)}{\Gamma(1-2\ep)}\right]^{k} \frac{\ep^2 \cos(\pi\ep)}{C_x} \, \gamma_{z,f_{[i\Fp]} \to f_i f_\Fp}^{3(k+1),\mathrm{1L}}(L_i) \,,
\label{eq_Gamma_1L_FSR_definition}
\end{equation}
where $C_x = 2\Cf-\Ca$ for the $g\to q\qb$ splitting, and $C_x = \Ca$ otherwise.
The  one-loop anomalous dimensions $\gamma_{g(z),f_{[i\Fp]} \to f_i f_\Fp}^{n(k+1), \rm 1L}$ that appear in \eq\eqref{eq_Gamma_1L_FSR_definition} are defined as
\begin{equation}
\begin{split}
    \gamma_{g(z),f_{[i\Fp]} \to f_i f_\Fp}^{n(k+1), \rm 1L}(L_i) \eqdef & - \int_{0}^{1} \dz\, \oS_z \left[z^{-n \ep} (1-z)^{-(k+1)\ep} g(z) \, P_{f_{[i\Fp]} f_i}^{\text{1L}}(z)\right]  
    \\
    & - 2 \delta_{f_{[i\Fp]} f_i} \Ca \T^2_{f_{[i\Fp]}} \, \frac{1- e^{-(2+k) \ep L_i}}{(2+k)} \frac{\pi}{\ep^2 \sin(\pi \ep)} g(1) \,.
\end{split}
\label{eq_gamma_nk_1L_appendix}
\end{equation}
The quantities defined in Eqs \eqref{eq_Gamma_1L_FSR_definition} and \eqref{eq_gamma_nk_1L_appendix}, with $n=3$ and $k=2$, are provided in \usefulfunctions, together with  the one-loop final-state splitting functions $P_{\alpha\beta}^{\rm 1L}$ (see Table \ref{table_appendix_ancillary_file}).
We note that, similar to the expressions in \eq\eqref{eq_FSR_generalized_anom_dim_Gamma_q_g}, we define the one-loop case as follows
\begin{equation}
    \Gamma_{i,q}^{\mathrm{1L}} \eqdef \Gamma_{i, q \to q g}^{\mathrm{1L}} + \Gamma_{i, q \to g q}^{\mathrm{1L}} \,,
    \qquad
    \Gamma_{i,g}^{\mathrm{1L}} \eqdef \Gamma_{i, g \to g g}^{\mathrm{1L}} + 2\nf \Gamma_{i, g \to q\qb}^{\mathrm{1L}} \,.
    \label{eq_Gamma_1L_q_def_appendix}
\end{equation}

%%%%%%%%%%%%%%%%%%%%%%%%%%%%%%%%%%%%%%%%%%%%%%%%%%%%%%%%%%%%%%%%%%%%%%%%%%%%%%%%%%%%%%%%%%%%%%%%%%%%%%%%%%%%%%%%%%%%%%%%%%%%%%%%%%%%%%%%%%%%%%%%%%%%%%%%%%%%%%%%%%%%%%%%

\subsection{Definitions  and relevant properties of $I$-operators}
\label{appendix_sec_operator_defn}

In Ref.~\cite{Devoto:2023rpv},  we have introduced 
virtual, soft, and collinear operators, and we employ similar operators in this paper as well. The virtual operator is defined as
\begin{equation}
    \IVirt(\ep) \eqdef \ICatbar(\ep) + \ICatbar^\dagger(\ep) \,,
    \label{eq_IVirt_def_appendix}
\end{equation}
where \cite{Catani:1998bh} \textcolor{red}{}
\begin{equation}
    \ICatbar(\ep) 
    \eqdef \frac{1}{2} \sum_{\inotj} \frac{\calV_i^\sing(\ep)}{\T_i^2} \, \T_i\cdot\T_j \left(\frac{\musq}{2p_i\cdot p_j}\right)^{\!\! \ep} e^{i \pi\lambda_{ij} \ep} \,,
    \qquad \calV_i^\sing(\ep) = \frac{\T_i^2}{\ep^2} + \frac{\gamma_i}{\ep} \,.
    \label{eq_ICatbar_def_appendix}
\end{equation}
In \eq\eqref{eq_ICatbar_def_appendix}, $\gamma_i$ are the anomalous dimensions that can be found in \eqref{eq_appendix_anomalous_dimensions}, the sum goes over the unordered pairs of external particles $i,j$ with $i \ne j$, 
and $\lambda_{ij}$ are constants which are equal to 1 if both $i$ and $j$ are incoming or outgoing, and to $0$ otherwise.
The soft operators which appear in the double-real and in the real-virtual corrections read
\begin{equation}
\begin{split}
    \ISoft(\ep) & \eqdef - \frac{(2\Emax/\mu)^{-2\ep}}{\ep^2} \sum_{\inotj} \eta_{ij}^{-\ep} K_{ij} \, (\T_i \cdot \T_j) \,,
    \\
    \ISofttilde(2\ep) & \eqdef - \frac{(2\Emax/\mu)^{-4\ep}}{(2\ep)^2} \sum_{\inotj} \eta_{ij}^{-2\ep} \tildeK_{ij} \, (\T_i \cdot \T_j) \,,
\end{split}
\label{eq_ISoft_definition_appendix}
\end{equation}
where $\eta_{ij} = (1 - \cos\theta_{ij})/2$, and 
\begin{equation}
\begin{split}
    K_{ij} & \eqdef \frac{\Gamma^2(1-\ep)}{\Gamma(1-2\ep)} \eta_{ij}^{1+\ep} \hypF(1,1,1-\ep,1-\eta_{ij}) \,, 
    \\
    \tildeK_{ij} &\eqdef \frac{\Gamma^2(1-2\ep)}{\Gamma(1-4\ep)} \,  \eta_{ij}^{1+3\ep} \hypF(1+\ep, 1+\ep, 1-\ep, 1-\eta_{ij}) \,.
\end{split}
\label{eq_K_Ktilde_def_appendix}
\end{equation}
A useful relation between  $\ISoft$ and $\ISofttilde$ is 
\begin{equation}
    \ISofttilde(2\ep) = \ISoft(2\ep) + \order{\ep} \,.
    \label{eq_relation_between_ISofttilde_and_ISoft_appendix}
\end{equation}
The Laurent expansions of  $\IVirt(\ep)$, $\ISoft(\ep)$, and $\ISofttilde(2\ep)$ can be found in Appendix A.5 of Ref.~\cite{Devoto:2023rpv}. 

The collinear operators which appear in the double-real and in the real-virtual corrections read, respectively,
\begin{equation}
    \IColl^{(k)}(\ep) \eqdef \sum_{i\in\HP} \frac{\Gamma_{i,f_i}^{(k)}}{\ep \,( k/2)} \,,
    \qquad
    \IColltilde(2\ep) \eqdef \sum_{i \in \HP} \frac{\Gamma_{i,f_i}^{\rm 1L}}{2\ep} \,,
\label{eq_IColl_IColltilde_def_appendix}
\end{equation}
where generalized initial- and final-state anomalous dimensions are given in  Eqs \eqref{Eq:Gamma_1_2_definition_k_general_appendix} and \eqref{eq_Gamma_a_to_bc_defs_appendix} for $\IColl^{(k)}$, and in Eqs \eqref{eq_Gamma_1L_ISR_definition} and \eqref{eq_Gamma_1L_FSR_definition} for $\IColltilde$. 
We note that the following relations hold
\begin{equation}
    \IColltilde(2\ep) = \IColl(2\ep) + \order{\ep^0} \,, 
    \qquad
    \IColl^{(4)}(\ep) = \IColl(2\ep) + \order{\ep^0} \,. 
\label{eq_relation_between_IColltilde_and_IColl_appendix}
\end{equation}

Finally, the $\ep$-finite operator $\ITot$ 
is defined as follows
\begin{equation}
    \ITot(\ep) \eqdef \IVirt(\ep) + \ISoft(\ep) + \IColl(\ep) \,.
\end{equation}
Its expansion in powers of $\ep$ reads
\begin{equation}
    \ITot(\ep) = \sum_{n=0}^\infty \ep^n \ITot^{(n)} \,,
\end{equation}
where the $n=0,1$ coefficients are given by
\begin{align}
    \ITot^{(0)} = & -\sum_{\inotj} (\T_i \cdot \T_j) \bigg[
    \left( 2\Lmax +\frac12 \log\eta_{ij} \right) \log \eta_{ij} -\frac12 L_{ij} \left(L_{ij} + \frac{2 \gamma_i}{\T_i^2}\right) \notag \\
    & +\Li_2(1-\eta_{ij}) + \frac{\pi^2}{2} \lambda_{ij} \bigg] + \sum_{i\in\HP} \T_i^2 \bigg[2\Lmax^2 - \frac{\pi^2}{6} - \Big(2\Ltildei\gamma_{i}^{22,(0)} - \gamma_{i}^{22,(1)} \Big) \frac{\theta_\HPf}{\T_i^2} 
    \label{eq_IT0_appendix} \\
    & - 2\bigg(L_i^2 + 2L_i \Ltildei + \Ltildei \frac{\gamma_i}{\T_i^2} \bigg) \bar{\theta}_\HPf \bigg] \,, \notag \\ & \notag \\
%%%%%%%%%%%%%%%%%%%%%%%%%%%%%%%%%%%%%%%%%%%%%%%%%%%%%%%%%%%%%%%%%%%%%%%%%%%%%%%%%%%%%%%%%%%%%%%%%%%%%%%%%%%%%%%%%%%%%%%%%%%%%%%%%%%%%%%%%%%%%%%%%%%%%%%%%%%%%%%%%%%%%%%%%%%%%%%%%%%%%%%%%%%%%%%%%%%%%%%%%%%%%%%%%%%%%%%%%%%%%%%%%%%%%%%%%%%%%%%%%%%%%%%%%%%%%%%%%%%%%%%%%%%%%%%%%%%%%%%%%%%%%%%%%%%%%%%%%%%%%%%%%%%%%
    \ITot^{(1)} = & \sum_{\inotj} (\T_i \cdot \T_j) \bigg[\frac{1}{6} \big(L_{ij}^3 + \log^3\eta_{ij}\big) + 2\Lmax^2 \log\eta_{ij} -\frac{\pi^2}{2} \lambda_{ij} L_{ij} \notag \\
    & + \bigg(\Lmax - \frac{1}{2} \log(1-\eta_{ij})\bigg)\log^2\eta_{ij} + 2\Lmax \, \Li_2(1-\eta_{ij}) - \Li_3(\eta_{ij}) \notag \\
    & - \Li_3(1-\eta_{ij}) + \frac{\gamma_i}{2\T_i^2} \big(L_{ij}^2 - \pi^2 \lambda_{ij}\big) \bigg] 
    + \sum_{i\in\HP} \T_i^2 \bigg\{- \frac{4}{3} \Lmax^3 + \frac{\pi^2}{3}\Lmax - 3 \zeta_3 
    \label{eq_IT1_appendix} \\
    & + \bigg[\bigg(2\Ltildei^2 - \frac{\pi^2}{6}\bigg) \gamma_{i}^{22,(0)} - 2\Ltildei\gamma_{i}^{22,(1)} + \gamma_{i}^{22,(2)}\bigg] \frac{\theta_\HPf}{\T_i^2}
    + \bigg[\frac{4}{3} L_i^3 + 4 L_i^2 \Ltildei + 4 L_i \Ltildei^2 \notag \\
    & - \frac{\pi^2}{3} L_i + \frac{\gamma_i}{\T_i^2} \bigg(2\Ltildei^2 - \frac{\pi^2}{6}\bigg) \bigg] \bar{\theta}_\HPf \bigg\} \,. \notag
\end{align} 
To write the above equations, we have used
\begin{equation}
    \theta_\HPf \eqdef 
    \begin{cases}
        1 & \text{if } i \in \HPf \\
        0 & \text{otherwise} \,,
    \end{cases} 
    \qquad\qquad
    \bar{\theta}_\HPf \eqdef 1 - \theta_\HPf \,, 
\end{equation}
together with the expansion coefficients of the collinear anomalous dimensions $\gamma_i^{22}(L_i)$, defined as 
\begin{equation}
    \gamma_i^{22}(L_i) = \sum_{n=0}^{\infty} \ep^n \, \gamma_{i}^{22,(n)}(L_i) \,,
    \qquad
    i = g,q \,. 
\end{equation}
We note that in Eqs \eqref{eq_IT0_appendix} and \eqref{eq_IT1_appendix}, the dependence of the coefficients $\gamma_{i}^{22,(n)}$ on the logarithms $L_i$ is  not shown explicitly.

%%%%%%%%%%%%%%%%%%%%%%%%%%%%%%%%%%%%%%%%%%%%%%%%%%%%%%%%%%%%%%%%%%%%%%%%%%%%%%%%%%%%%%%%%%%%%%%%%%%%%%%%%%%%%%%%%%%%%%%%%%%%%%%%%%%%%%%%%%%%%%%%%%%%%%%%%%%%%%%%%%%%%%%%%%%%%%%%%%%%%%%%%%%%%%%%%%%%%%%%%%%%%%%%%%%%%%%%%%%%%%%%%%%%%%%%%%%%%%%%%%%%%%%%%%%%%%%%%%%%%%%%%%%%%%%%%%%%%%%%%%%%%%%%

\subsection{Double-unresolved contributions}

In this section, we define  the  double-unresolved terms in Eqs \eqref{eq_SigmaDU_HP_sym_main_formula} 
and \eqref{eq_SigmaDU_genSet_unsym_main_formula} which were left unspecified  in the main body of the paper. 
These contributions are  at most of $\order{\ep^{-1}}$.
We start from \eq\eqref{eq_SigmaDU_HP_sym_main_formula}, which refers to the $(\Fp\Sp) \in \DS$ case (cf.\ \eq\eqref{eq_DS_def}).
We write\footnote{The numbering scheme for contributions to $\SigmaDU|_\genBorn$ and $\SigmaDU^{\rm fin}|_\genBorn$ is kept consistent  with the definitions introduced in Ref.~\cite{Devoto:2023rpv}.}
\begin{equation}
    \SigmaDU^{\genBorn,\rest}[(\Fp\Sp) \in \DS]_\RR
    = 
    \SigmaDU^{\rm rdc}|_{\genBorn} 
    + \SigmaDU^{(2)}|_{\genBorn} 
    + \SigmaDU^{(8)}|_{\genBorn} 
    + \sum_{i=1}^{5} \SigmaDU^{\fin,(i)}|_{\genBorn} \,,
    \label{eq_SigmaDUrest_genBorn_ds}
\end{equation}
with
\begin{align}
    \SigmaDU^{\rm rdc}|_{\genBorn} = &\; \frac{1}{2} \sum_{i \in \HP} \llint \big[2(\eta_{i\Sp}/2)^{-\ep} -1\big] \calS(\Fp,\Sp) \, C_{i \Sp} C_{i \Fp} \Delta^{(\Fp\Sp)} \FLMlo{\genBorn}{\Fp,\Sp}\rrint \,,
    \label{eq_SigmaDU_rdc_genBorn_ds} \\
    %%%%%%%%%%%%%%%%%%%%%%%%%%%%%%%%%%%%%%%%%%%%%%%%%%%%%%%%%%%%%%%%%%%%%%%%%%%%%%%%%%%%%%%%%%%%%%%%%%
    \SigmaDU^{(2)}|_\genBorn = &\; \lint \oS_{\Fp \Sp} \oS_{\Sp} \Omega_2 \Delta^{(\Fp \Sp)} \THmn \FLMlo{\genBorn}{\Fp,\Sp} \rint \,, 
    \label{eq_SigmaDU_2_genBorn_ds} \\
    %%%%%%%%%%%%%%%%%%%%%%%%%%%%%%%%%%%%%%%%%%%%%%%%%%%%%%%%%%%%%%%%%%%%%%%%%%%%%%%%%%%%%%%%%%%%%%%%%%
    \SigmaDU^{(8)}|_\genBorn = &\; \sum_{i\in\HP} \, \asbr\frac{N_\ep^{(b,d)}}{2} \llint \oS_{[\Fp\Sp]} C_{i[\Fp\Sp]} (E_{[\Fp\Sp]}/\mu)^{-2\ep} \sigma_{i[\Fp\Sp]}^{-\ep} \Big[\gamma_{\bot, g \to \Fp\Sp}^{22} (r_i^\mu r_i^\nu + g^{\mu\nu}) \notag \\
    & - \gamma_{\bot, g\to \Fp\Sp}^{22, \rmr} \, g^{\mu\nu} \Big] \Delta^{(\Fp)} \FLMlomunu{\genBorn}{[\Fp\Sp]} \rrint \,,
    \label{eq_SigmaDU_8_genBorn_ds}
    \\ & \notag \\
    %%%%%%%%%%%%%%%%%%%%%%%%%%%%%%%%%%%%%%%%%%%%%%%%%%%%%%%%%%%%%%%%%%%%%%%%%%%%%%%%%%%%%%%%%%%%%%%%%%
    \SigmaDU^{\fin, (1)}|_\genBorn = & - \bigg[\left(\frac{\Gamma(1-2\ep)}{\Gamma^2(1-\ep)}\right)^{\! 2} - 1\bigg] \sum_{\inotj} \lint \oS_{\Sp} C_{j \Sp} C_{i \Fp} \Delta^{(\Fp \Sp)} \, \THmn \FLMlo{\genBorn}{\Fp,\Sp} \rint \,, 
    \label{eq_SigmaDU_fin1_genBorn_ds} \\
    %%%%%%%%%%%%%%%%%%%%%%%%%%%%%%%%%%%%%%%%%%%%%%%%%%%%%%%%%%%%%%%%%%%%%%%%%%%%%%%%%%%%%%%%%%%%%%%%%%
    \SigmaDU^{\fin, (2)}|_\genBorn = &\; \sum_{i\in\HP} \llint \big[(\eta_{i\Fp}/2)^{-\ep} -1\big] \oC_{i\Fp} S_\Fp \lint \oS_\Sp C_{i\Sp} \, \partFuncNNLO{i}{i} \Delta^{(\Fp \Sp)} \THmn \FLMlo{\genBorn}{\Fp,\Sp} \rint \rrint \,, \\
    %%%%%%%%%%%%%%%%%%%%%%%%%%%%%%%%%%%%%%%%%%%%%%%%%%%%%%%%%%%%%%%%%%%%%%%%%%%%%%%%%%%%%%%%%%%%%%%%%%
    \SigmaDU^{\fin, (3)}|_\genBorn = & \left[\frac{\Gamma(1-2\ep)}{\Gamma^2(1-\ep)} - 1\right] \Lint \calS(\Fp,\Sp) \bigg[\sum_{\inotj} C_{j\Sp} C_{i\Fp} + \sum_{i\in\HP} (\eta_{i\Sp}/2)^{-\ep} C_{i\Sp} C_{i\Fp} \bigg] \notag \\
    & \times \Delta^{(\Fp \Sp)}\FLMlo{\genBorn}{\Fp,\Sp}\Rint \,, 
    \label{eq_SigmaDU_fin3_genBorn_ds}\\
    %%%%%%%%%%%%%%%%%%%%%%%%%%%%%%%%%%%%%%%%%%%%%%%%%%%%%%%%%%%%%%%%%%%%%%%%%%%%%%%%%%%%%%%%%%%%%%%%%%
    \SigmaDU^{\fin, (4)}|_\genBorn = &\; \asbr^2 \, 2^{2\ep} \, \delta_\Fp(\ep) \left(\frac{2\Emax}{\mu}\right)^{\!\!-4\ep} \sum_{i\in\HP} \lint \Wbdfin{i} \colorprod \FLM^\genBorn \rint \,, 
    \label{eq_SigmaDU_fin4_genBorn_ds} \\
    %%%%%%%%%%%%%%%%%%%%%%%%%%%%%%%%%%%%%%%%%%%%%%%%%%%%%%%%%%%%%%%%%%%%%%%%%%%%%%%%%%%%%%%%%%%%%%%%%%
    \SigmaDU^{\fin, (5)}|_\genBorn = &\; \asbr^2 \, 
    \delta_\Fp^{\perp}(\ep) \left(\frac{E_{\rm max}}{\mu}\right)^{\!\!-4\ep} 
    \sum_{i\in\HP} \lint \Wr{i}  \colorprod \FLM^{\genBorn} \rint \,.
    \label{eq_SigmaDU_fin5_genBorn_ds}
\end{align}
The  terms in Eqs (\ref{eq_SigmaDU_rdc_genBorn_ds} -- \ref{eq_SigmaDU_8_genBorn_ds}) are  $\order{\ep^{-1}}$, whereas the terms in Eqs (\ref{eq_SigmaDU_fin1_genBorn_ds} -- \ref{eq_SigmaDU_fin5_genBorn_ds}) are $\ep$-finite.
In Eqs \eqref{eq_SigmaDU_rdc_genBorn_ds} and \eqref{eq_SigmaDU_fin3_genBorn_ds}, the operator $\calS(\Fp,\Sp)$ is defined as 
\begin{equation}
    \calS(\Fp,\Sp) \eqdef 
    \begin{cases}
        \oS_\Sp (\iden  - S_\Fp \THnm)  + S_\Sp \oS_\Fp \THnm \,,  & \text{if } (\Fp\Sp) = (gg) \,, \\
        \iden \,, & \text{if } (\Fp\Sp) =(q_i\qb_i) \,.
    \end{cases} 
\end{equation}
In \eq\eqref{eq_SigmaDU_2_genBorn_ds}, the operator $\Omega_2$ is the triple-collinear operator defined in \eq(D.6) of Ref.~\cite{Devoto:2023rpv}.
The triple-collinear splittings are reported in \usefulfunctions, see Table~\ref{table_appendix_ancillary_file}.
In \eq\eqref{eq_SigmaDU_8_genBorn_ds}, we used $\sigma_{ij} = \eta_{ij}/(1-\eta_{ij})$, the anomalous dimensions $\gamma_{\bot, g \to \Fp\Sp}^{22}$ and $\gamma_{\bot, g \to \Fp\Sp}^{22,\rmr}$  defined in \eq\eqref{eq:gamma_tilde}, and the vector $r_i^\mu$ specified in Appendix E of Ref.~\cite{Devoto:2023rpv}.
The quantities $\delta_\Fp(\ep)$ and $\delta_\Fp^\perp(\ep)$ appearing in Eqs (\ref{eq_SigmaDU_fin4_genBorn_ds} -- \ref{eq_SigmaDU_fin5_genBorn_ds})  are defined in \eq\eqref{eq_deltam_deltamperp_def}, while the two partition-dependent operators $\Wbdfin{i}$ and $\Wr{i}$ are defined in \eq\eqref{eq_Wbdfin_def_appendix_final_result}. 

For  the double-unresolved terms appearing in \eq\eqref{eq_SigmaDU_genSet_unsym_main_formula}, which refers to $(\Fp\Sp) \in \noDS$ case, we write 
\begin{equation}
    \SigmaDU^{\genBorn,\rest}[(\Fp\Sp) \in \noDS]_\RR
    = \SigmaDU^{\rm rdc}|_{\genBorn} 
    + \SigmaDU^{(2)}|_{\genBorn} 
    + \sum \limits_{i=1,3} \SigmaDU^{\fin,(i)}|_{\genBorn} \,,
    \label{eq_SigmaDUrest_genBorn_nods}
\end{equation}
where
\begin{align}
    \SigmaDU^{\rm rdc}|_{\genBorn} = &\; \frac{1}{2}\sum_{i \in \HP} \lint \big[2(\eta_{i\Sp}/2)^{-\ep} - 1\big] \oS_\Sp \big( C_{i\Fp} C_{i\Sp} + C_{i\Sp} C_{i\Fp} \big) \Delta^{(\Fp \Sp)} \FLMlo{\genBorn}{\Fp, \Sp} \rint \,,
    \label{eq_SigmaDU_rdc_genBorn_nods} \\
    %%%%%%%%%%%%%%%%%%%%%%%%%%%%%%%%%%%%%%%%%%%%%%%%%%%%%%%%%%%%%%%%%%%%%%%%%%%%%%%%%%%%%%%%%%%%%%%%%%
    \SigmaDU^{(2)}|_{\genBorn} = &\; \lint \oS_{\Sp} \Omega_2 \Delta^{(\Fp \Sp)} \FLMlo{\genBorn}{\Fp,\Sp} \rint \,, 
    \label{eq_SigmaDU_2_genBorn_nods} \\ & \notag \\
    %%%%%%%%%%%%%%%%%%%%%%%%%%%%%%%%%%%%%%%%%%%%%%%%%%%%%%%%%%%%%%%%%%%%%%%%%%%%%%%%%%%%%%%%%%%%%%%%%%
    \SigmaDU^{\fin, (1)}|_{\genBorn} = & - \bigg[\left(\frac{\Gamma(1-2\ep)}{\Gamma^2(1-\ep)}\right)^{\! 2} - 1\bigg] \sum_{\inotj} \lint \oS_{\Sp} C_{j \Sp} C_{i \Fp} \Delta^{(\Fp \Sp)} \FLMlo{\genBorn}{\Fp,\Sp} \rint \,, 
    \label{eq_SigmaDU_fin1_genBorn_nods} \\
    %%%%%%%%%%%%%%%%%%%%%%%%%%%%%%%%%%%%%%%%%%%%%%%%%%%%%%%%%%%%%%%%%%%%%%%%%%%%%%%%%%%%%%%%%%%%%%%%%%
    \SigmaDU^{\fin, (3)}|_{\genBorn} = & \left[\frac{\Gamma(1-2\ep)}{\Gamma^2(1-\ep)} - 1\right] \Lint \oS_\Sp \bigg[2\sum_{\inotj} C_{j\Sp} C_{i\Fp} + \sum_{i\in\HP} (\eta_{i\Sp}/2)^{-\ep} \big(C_{i\Sp} C_{i\Fp} \notag \\
    & + C_{i\Fp} C_{i\Sp}\big) \bigg] \Delta^{(\Fp \Sp)}\FLMlo{\genBorn}{\Fp,\Sp}\Rint \,.
    \label{eq_SigmaDU_fin3_genBorn_nods}
\end{align}
The terms in Eqs (\ref{eq_SigmaDU_rdc_genBorn_nods} -- \ref{eq_SigmaDU_2_genBorn_nods}) are  $\order{\ep^{-1}}$, and those in Eqs (\ref{eq_SigmaDU_fin1_genBorn_nods} -- \ref{eq_SigmaDU_fin3_genBorn_nods}) are $\ep$-finite.

% \input{Appendix/B_Final_result}

%-----------------------------------------------------
%                     BIBLIOGRAPHY
%-----------------------------------------------------

\newpage
\bibliographystyle{JHEP}
\bibliography{biblio.bib}

\end{document}